\long\def\equalign
#1\end#2{\def\\{\cr\noalign{\smallskip}}
    \refstepcounter{equation}$$ \vcenter{\tabskip=5pt
    \halign{\hfil$\displaystyle{##}$&&$\displaystyle{{}##}$\hfil\cr
    #1 \cr}}\end{#2}}

\def\ua1{$\hbox{U}_A$(1)}
\def\uv1{$\hbox{U}_V$(1)}
\def\sua2{$\hbox{SU}_A$(2)}
\def\suv2{$\hbox{SU}_V$(2)}

\def\pmb#1{\mbox{\boldmath #1}}

\def\boldtau{\pmb{$\tau$}}

\input psfig

\def\xslide#1#2#3#4#5#6{\centerline{\psfig
{figure=#1,height=#2,bbllx=#3bp,bblly=#4bp,bburx=#5bp,bbury=#6bp,clip=}}}

\documentstyle[11pt]{report}
\begin{document}


\newpage

\title{{\Large REPORT No. 1739/PH } \\ $\mbox{}$ \\ $\mbox{}$ \\
       {\Large \bf DESCRIPTION OF HOT COMPRESSED HADRONIC MATTER BASED ON
                  AN EFFECTIVE CHIRAL LAGRANGIAN}}

\author{\\ $\mbox{}$ \\ $\mbox{}$ \\ $\mbox{}$ \\ 
        {\Large Wojciech Florkowski} \\ $\mbox{}$ \\}


\maketitle

\newpage
\thispagestyle{empty}
$\mbox{}$

\vspace{2cm}
\centerline{\underline{\bf ACKNOWLEDGMENTS}}
\vspace{2cm}

The results presented in this report have been obtained during the
joint work with other physicists in different research and educational
institutes. I would like to thank them all for the help in the
preparation of this work. However, I particularly appreciate the
assistance of my closest collaborators Wojciech Broniowski, Ji\v
r\'{\i} Dolej\v s\'{\i}, Bengt Friman, J\"org H\"ufner, Sandy
Klevansky and Ludwig Neise. 

\bigskip

I am most grateful to Prof. Jan Kwieci\'nski for his encouragement
and continuous interest in this work. I also thank all my
colleagues from the Theory Department of the H. Niewodnicza\'nski 
Institute of Nuclear Physics for stimulating and friendly atmosphere.

\bigskip

I would like to thank Maciek A. Nowak  for clarifying discussions 
concerning chiral symmetry and for pointing my attention to
the lithographs by M.C. Escher. I am also indebted to Pengfei Zhuang
for illuminating discussions on the transport theory.

\bigskip

I am obliged to H. Niewodnicza\'nski Institute of Nuclear
Physics, Gesellschaft f\"ur Schwerionenforschung (GSI) and the Institute
of Theoretical Physics of the Heidelberg University for supporting
me with the comfortable work conditions.  

\bigskip

The research presented below has been supported by the Polish
State Committee for Scientific Research (KBN) under Grants No.
2.0204.91.01 and 2 P03B 188 09, and also by the Stiftung f\"ur
Deutsch-Polnische Zusammenarbeit project 1522/94/LN. 

\renewcommand{\headheight}{-10mm}  
\setcounter{page}{2}

\tableofcontents

\part{\bf OVERVIEW}

\chapter{\bf Introduction}
\label{chapt:introduction}

It is now commonly accepted that {\it quantum chromodynamics} (QCD) is
the underlying theory of strong interactions \cite{QCD}. Consequently,
one tries to understand all hadronic phenomena in terms of quark-gluon
processes. To large extent, the success of such a program has been
already achieved in the {\it deep inelastic lepton-hadron collisions}
\cite{DIS}; due to the asymptotic freedom of QCD, in the high energy
regime one can apply perturbation theory. On the other hand,
description of low energetic processes in terms of QCD is very
difficult, since in this case the theory exhibits a complicated
nonperturbative structure, responsible for such important phenomena
like, e.g., confinement or spontaneous breaking of chiral
symmetry. The most fundamental approach to low energetic QCD (energies
smaller or comparable to the typical hadron mass, $\sim$ 1 GeV) is
based on the {\it lattice simulations} \cite{LQCD}. However, due to
the problems with, e.g., finite size effects or proper inclusion of
fermions, such calculations are still not fully satisfactory. In this
situation, the important role is played by various {\it effective
theories} which maintain the basic features of QCD but, at the same
time, they are much easier to deal with.  One of such theories is the
Nambu -- Jona-Lasinio (NJL) model.  The NJL model was introduced
already in the early sixties as a theory of interacting nucleons
\cite{NJL}. Later it was reformulated in terms of quark degrees of
freedom. Numerous calculations demonstrate the success of the model in
describing hadronic data. In this report we are going to discuss the
properties of hot compressed hadronic matter as described by this
model.

The two very important characteristics of QCD are: the nonperturbative
structure of its ground state, characterized by a nonvanishing value
of the quark condensate, and the appearance of light pseudoscalar
particles, which are identified with the (quasi) Goldstone bosons of
the spontaneously broken chiral symmetry. There is another feature of
QCD which concerns hot (dense) systems. Due to the asymptotic freedom,
at high temperature (density) the quarks and gluons are not confined
to hadrons anymore but form the so-called {\it quark-gluon plasma}
(QGP).  Moreover, one expects that during such a deconfinement phase
transition the chiral symmetry is additionally restored \cite{HMO96}.
The successful effective theory of QCD should incorporate these
properties.

An essential feature of the NJL model is its chiral invariance,
spontaneously broken in the true ground state of the theory. This
phenomenon leads directly to the appearance of the Goldstone bosons.
The relative simplicity of the model allows for explicit calculations
in this case. The model includes also the chiral symmetry restoration
phase transition. This fact resembles us the case of QCD where an
analogous phase transition is expected. In consequence, the NJL model
seems to be a very interesting tool to study the temperature and
density dependence of various physical quantities.

There are also other interesting properties of the NJL model, which
deserve some comments. The addition of small (a few MeV) current quark 
masses to the NJL Lagrangian breaks explicitly chiral
invariance.  In this case, the latter becomes only an approximate
symmetry.  Moreover, the difference between various current quark
masses leads to the mass splittings among the mesons (baryons) with
different flavour content. We note that this type of behaviour is
typical for QCD, where the small current quark masses play a similar
role. The NJL model allows us also to understand the success of the
constituent quark picture.  Due to the strong four-fermion pointlike
interaction, the light (current) quarks appearing in the NJL
Lagrangian gain large (constituent) masses. In this way, we can find
connections between the concepts of the spontaneous breaking of the
chiral symmetry and the concepts of the nonrelativistic quark model.

The shortcomings of the NJL model are: i) the pointlike
character of the quark-antiquark interaction causes that the theory is
not renormalized, ii) the NJL type of the quark interaction does not
explain the confinement, moreover iii) there are no gluons in the
model.  The fact that the model is not renormalizable is not so
important as long as we do not want to treat it as a fundamental
theory. The lack of confinement is a more serious problem. This may
lead to unphysical processes like, e.g., decays of heavy mesons into
quark-antiquark pairs. The treatment of such cases in the NJL model
must be done with a special care. Similarly, the fact that there are
no gluons as dynamic degrees of freedom may lead to problems,
especially, if one tries to apply the model at relatively high
energies. Because of these disadvantages the model can be easily
criticized. Nevertheless, in our opinion, the NJL model is, in spite
of its deficiencies, a very interesting tool for studying phenomena
related to the chiral symmetry and the chiral phase transition.

Through the use of the NJL model one can evaluate the temperature (density)
dependence of various physical quantities like, e.g., the quark
condensate (constituent quark mass) \cite{HK85,BMZ87b,FB96}, the pion decay
constant \cite{BMZ87b,HK87}, the meson masses
\cite{HK85,HK87,BMZ87a,BM88a,BM88b,TK91}, the meson-meson and quark-meson
coupling constants \cite{HK87,BM88a}. In the framework of the model,
the calculations of the critical temperature and the critical baryon
chemical potential have been performed \cite{HK85,BMZ87b}, and the
order of the chiral symmetry restoration phase transition was studied
\cite{AY89}. Moreover, the thermodynamic properties of the quark-meson
plasma were investigated \cite{HKZV94,ZHK94}. Many results obtained
from the NJL model (e.g., the temperature dependence of the quark-number
susceptibility \cite{K91}, or of the meson screening masses \cite{FF94a})
are in good qualitative agreement with the lattice simulations of QCD.

In contrast to the abundant studies of the equilibrium situations, the
investigations of the non-equilibrium cases in the NJL model are
relatively rare. Initially, a transport theory for the model has been
derived in the framework of the Keldysh closed-time-path formalism
combined with the effective action method \cite{ZW92}. Later, another
derivation (in the mean-field approximation) has been presented
\cite{FHKN96}, which leads to the kinetic equations whose form is
explicitly chirally invariant and which include the spin dynamics. A
certain class of solutions of the classical transport equations found
in \cite{ZW92} has been studied in both \cite{AA95} and \cite{F94}.
The aim of a series of publications \cite{CS} was the evaluation of
the in-medium cross sections.  Such cross sections can be used in the
phenomenological kinetic equations describing non-equilibrium
phenomena in hadronic matter.  In particular, the singular cross
sections (giving the critical scattering at the chiral phase
transition) were used in the rate equations describing the
hadronization of the quark plasma \cite{DFH95}.

\newpage
In this report we give the review of the recent results obtained in
the NJL model, describing the properties of hot compressed hadronic
matter.  The first large class of problems concerns the behaviour of
static meson correlation functions. In particular, this includes the
investigation of the {\it screening of meson fields} at finite
temperature or density.  Another wide range of problems presented in
our report concerns the {\it formulation of the transport theory} for
the NJL model and its applications to the description of high energy
nuclear collisions. Recently, several reviews of the NJL model have
been published \cite{NJLR}. The aim of the present article is not to
give yet another wide and complete review, but rather to summarize a
series of the new results. Our report is based on the following
original articles:

\begin{itemize}

\item[I] W. Florkowski and B. L. Friman: {\it Spatial Dependence
of Meson Correlation Functions at High Temperature},  
Zeit. f\"ur Phys. {\bf A347} (1994) 271-276 \cite{FF94b}, 

\item[II] W. Florkowski and B. L. Friman: {\it Meson             
Screening Masses in the Nambu -- Jona-Lasinio Model},              
Acta Phys. Pol. {\bf B25} (1994) 49-71 \cite{FF94a},

\item[III] W. Florkowski and B. L. Friman: {\it Oscillations
of the Static Meson Fields at Finite Baryon Density}, Nucl. Phys. 
{\bf A} in print \cite{FF97},

\item[IV] W. Florkowski and W. Broniowski: {\it Melting of the
Quark Condensate in the NJL Model with Meson Loops}, Phys. Lett. 
{\bf B386} (1996) 62-64 \cite{FB96},

\item[V] W. Florkowski, J. H\"ufner, S.P. Klevansky and
L. Neise: {\it Chirally Invariant Transport Equations for
Quark Matter}, Ann. Phys. (NY) {\bf 245} (1996) 445-463 \cite{FHKN96},

\item[VI] W. Florkowski: {\it Large Time-Scale Fluctuations        
of the Quark Condensate at High Temperature}, Phys. Rev. {\bf C50}
(1994) 3069-3078 \cite{F94},                                                   

\item[VII] J. Dolej\v s\'{\i}, W. Florkowski, and J. H\"ufner:
{\it Critical Scattering at the Chiral Phase Transition and
Low-$p_T$ Enhancement of Mesons in Ultra-Relativistic Heavy-Ion
Collisions}, Phys. Lett. {\bf B349} (1995) 18-22 \cite{DFH95}.
\end{itemize}

Let us now outline the organization of this paper. Since our studies
within the NJL model have been motivated by the physics of the
relativistic heavy-ion collisions, Chapter \ref{chapt:HENC} serves as
a brief survey of high-energy nuclear reactions. This gives us a
general background for the discussion of the properties of hot and
dense hadronic matter.  Chapter \ref{chapt:NJL} is the introduction to
the NJL model, where the basic concepts are presented and the
connections between different formulations of the model are
clarified. In Chapters \ref{chapt:imagin_time} --- \ref{chapt:GBMF}
[EQUILIBRIUM ENSEMBLES] we discuss the physical systems in
thermodynamic equilibrium, whereas in Chapters \ref{chapt:real_time}
--- \ref{chapt:critical} [NON-EQUILIBRIUM PHENOMENA] the cases
out of equilibrium are described. In order to make our report more
self-contained, Chapters \ref{chapt:imagin_time} and
\ref{chapt:real_time} are the short introductions to the
imaginary-time and real-time formalisms.  (We note that the
imaginary-time formalism is usually used to describe the equilibrium
ensembles, whereas the real-time formalism is used to study the
systems out of equilibrium. We follow this tendency here.)  More
details about the structure of this article can be found in the Table
of Contents.

\chapter{\bf High Energy Nuclear Collisions} 
\label{chapt:HENC}

\section{Relativistic Collisions}

In the {\it relativistic heavy-ion collisions}, the energies of the
colliding nuclei are of the order of a few GeV per nucleon. In this
energy range, the basic properties of the nuclear equation of state
can be tested, which has important astrophysical relevance to neutron
stars and supernova explosions. Here one encounters many interesting
and well established phenomena like, e.g, collective flows or
subthreshold production of particles. In general, during heavy-ion
collisions, large systems of hot and dense hadronic matter are
produced (central collisions of symmetric heavy ions at 1 GeV per
nucleon are likely to yield about 3 times normal nuclear matter
density).  The particles inside such a fireball do not propagate
completely freely: their Compton wavelength may be comparable with
their mean free path. In this situation, we expect that some of the
particle properties (e.g., their masses, widths or coupling constants)
can be changed. These {\it in-medium modifications} can lead to the
experimentally observed phenomena.  For example, the change of the
$\rho$ meson mass in dense matter can influence the measured dilepton
spectrum \cite{HADES}. Nowadays, one attempts to connect in-medium
modifications of hadron properties with the partial restoration of
chiral symmetry. If this idea turns out to be correct, we can treat
the change of hadron properties as a signature of the rearrangement of
the QCD vacuum.

The experiments with relativistic heavy-ions are intensely carried
out in many places, for example, at the Gesellschaft f\"ur
Schwerionenforschung in Darmstadt (GSI), at the Grand Accelerateur
d'Ions Lourds in Caen (GANIL), at the Joint Institute for Nuclear
Research in Dubna (JINR), and at the Lawrence Berkeley Laboratory in
Berkeley (LBL).

\section{Ultra-Relativistic Collisions}

If the beam energies are larger (the energy per nucleon exceeds 10
GeV) we talk usually about the {\it ultra-relativistic heavy-ion
collisions}.  The ultimate goal of such collisions is the observation
of the phase transition from hadronic matter to the quark-gluon plasma
\cite{QGP,QM90,QM91,QM93}.  The search for signatures of this phase 
transition is one of the most challenging problems of high-energy
nuclear physics, both from the experimental and theoretical points of
view. Several plasma signatures have been proposed so far. They can be
grouped into the following categories: photons and lepton pairs,
strangeness and antibaryon enhancement, $J/\psi$ suppression,
transverse flow and thermodynamic variables measuring equation of
state (for a complete review see \cite{S93}).

The first experiments with ultra-relativistic heavy-ion collisions
took place at the Brookhaven National Laboratory (BNL) and CERN in
1986.  The Alternating Gradient Synchrotron (AGS) accelerated beams up
to ${}^{20}$Si at 14.5 GeV per nucleon. At CERN, the Super Proton
Synchrotron (SPS) accelerated ${}^{16}$O at 60 and 200 GeV per nucleon
in 1986 and ${}^{32}$S at 200 GeV per nucleon in 1987. In 1990 a
long-term project on heavy-ion physics was realized at CERN with
several weeks of ${}^{32}$S beams. Initially, no special heavy-ion
machines were constructed for ultra-relativistic heavy ion collisions,
but rather old existing accelerators had been upgraded.  Only now, the
completely new experiments take place at CERN with ${}^{208}$Pb
beams. These are for the first time really ``heavy'' ions, providing
large volumes and lifetimes of the reaction zone.  The future of the
field is connected with the construction of the colliders: RHIC at BNL
(Au on Au reactions at $\sqrt{s}$ = 200 GeV per nucleon), and LHC at
CERN (Pb on Pb reactions at $\sqrt{s}$ = 6.3 TeV per nucleon).

The present experimental evidence indicates that in ultra-relativistic
heavy-ion collisions an extended and very dense system of hadronic
matter is indeed formed. It differs in many aspects from the systems
formed in elementary hadron-hadron reactions. However, it has also
become clear that our search for QGP will not be an easy
exercise. Although a number of ``plasma signals'' have been observed,
at the same time the conventional theoretical models have been
improved to such a level of the agreement with the data, that it does
not require any radically new physics. Nevertheless, the physics of
the ultra-relativistic heavy-ion collisions only now comes into its
mature age. With Pb on Pb collisions at LHC, which offer the initial
energy density 50 to 100 times larger than that of normal nuclear
matter, the ``new physics'' of one kind or another should certainly
appear.

\section{Theoretical Concepts}

The physics of the (ultra)relativistic heavy-ion collisions is an
interdisciplinary field which combines different methods and ideas
from the particle and nuclear physics. Since we deal with 
macroscopic objects containing a large number of particles, the methods
of {\it statistical physics} and {\it thermodynamics} are especially
useful.  In fact, many estimates are done on the basis of purely
thermodynamic considerations. Nevertheless, the systems produced in
these collisions are not static. The need for the dynamical description
involves rich applications of {\it hydrodynamics}. Furthermore, since
the matter produced in high energy nuclear collisions lives only for a
short while, it is natural to expect that its space-time evolution
proceeds far away from equilibrium. Consequently, there exists a
growing interest in applying and developing {\it transport theories}
which are suitable for the description of non-equilibrium
processes. For the relativistic heavy-ion collisions one uses
transport equations of the BUU (Boltzmann-Uehling-Uhlenbeck) type,
which evolve phase-space densities of nucleons, pions and other
hadrons. With the ultra-relativistic collisions in mind, a quark-gluon
kinetic theory based directly on QCD has been formulated
\cite{H83,EGV86}. However this approach, with a few exceptions
\cite{BCDF88}, is still far from practical applications.

To a large extent the successful description of ultra-relativistic
reactions has been obtained on the basis of the {\it microscopic
Monte-Carlo simulations}. These models are in some sense an
extrapolation of low energy hadron-hadron string models \cite{AGIS83}.
At energies of 200 GeV per nucleon or higher, the density of strings
becomes high and they start to overlap. Consequently, the interaction
and fusion of strings must be taken into account in most cases
\cite{W92}. For extremely high energies, the concept of strings and
flux tubes can be even abandoned and replaced by the picture of a
parton cascade developing in the perturbative vacuum \cite{GM92}.

Last but not at least, the physics of the ultra-relativistic
heavy-ion collisions has triggered the fast development of the
{\it quantum theory of fields in and out of equilibrium}. Using
the methods of field theory one can study the in-medium
properties of particles. Moreover, this approach allows also for 
the formulation of the kinetic equations satisfied by the particle
distribution functions.

A description of the whole complexity of high-energy nuclear
collisions in terms of a simple model is impossible. In particular,
effective chiral theories cannot explain the mechanism of all
processes happening at different energy scales. However, many
important problems discussed in the context of high-energy nuclear
collisions can be successfully analyzed within the NJL model.  Let us
mention a few of them: rearrangement of the QCD vacuum under extreme
conditions, modification of the properties of the particles
propagating through a hot and dense hadronic medium, structure of the
chiral phase transition, and non-equilibrium dynamics of hadronic
matter.  It is an attractive feature of the NJL model that these
problems can be addressed in a unified manner.

\bigskip
\chapter{\bf Nambu -- Jona-Lasinio Model}
\label{chapt:NJL}
\bigskip

This Chapter is the introduction to the NJL model. We define here our
notation and conventions. The relationship between the model and QCD
is claryfied. The presentation starts with the discussion of the
simplest, one-flavour version. Later on, more realistic cases are
considered.

\section{One-Flavour Case}

The simplest variant of the NJL model is its one-flavour version based
on the following form of the Lagrange density

\begin{equalign}
\label{l1}
{\cal L} = {\bar \psi} \left( i\! \not \! \partial 
- m \right) \psi +
G \left[({\bar \psi} \psi)^2 + 
({\bar \psi} i \gamma_5 \psi)^2 \right].
\end{equalign}

\noindent Here $\psi$ is the Dirac field describing quarks, $m$ is the 
current quark mass, and $G$ is the coupling constant \footnote{In our
report we shall always work in 1+3 dimensions.  The formulation of the
theory in 1+1 dimensions has different aspects, since in this case the
model is renormalizable. A 2-dimensional massless theory is commonly
called the Gross-Neveu model \cite{GN74}.}.  The latter has a
dimension (mass)$^{-2}$ and is assumed to be positive (in this case
the forces between quarks and antiquarks are attractive).  The field
$\psi$ can carry additional colour degrees of freedom.  In such a
case, the notation in (\ref{l1}) assumes summation over colour
indices, for example, ${\bar \psi} \psi =
\sum_{a=1}^{N_c}{\bar \psi}^a {\psi}^a$, where $N_c$ is the number of 
colours.

Expression (\ref{l1}) is an effective Lagrangian describing the local
quark-antiquark interaction. The pointlike character of this coupling
is usually explained by the assumption that the gluon degrees of
freedom can be absorbed in the effective quark-antiquark coupling
constant $G$. This type of reduction is considered to be a suitable
approximation to QCD in the low-energy (long wavelength) limit.

In the case $m=0$, Lagrangian (\ref{l1}) is invariant under \ua1 
transformations                                  
\begin{equalign}                                                       
\label{ua1}                                                           
\psi \rightarrow \psi^{\prime} = 
\exp\left(-i\gamma_5 {\chi \over 2}\right)     
\psi.
\end{equalign}                                                         
                                                                       
\noindent Consequently, we find that the {\it axial} current is 
conserved   

\begin{equalign}                                                       
\label{acc}                                                            
\partial_{\mu} A^{\mu}(x) = 0, \,\,\,\,\,                              
A^{\mu}(x) = 
{\bar \psi}(x) \gamma^{\mu} \gamma_5 \psi(x).
\end{equalign}                                                         

\noindent If the current quark mass does not vanish, $m \not = 0$, the 
equality (\ref{acc}) is not fulfilled anymore. In this case we find

\begin{equalign}
\label{pacc}                                                            
\partial_{\mu} A^{\mu}(x) = 2i m {\bar \psi}(x) \gamma_5 \psi(x).   
\end{equalign}

\noindent Since $m$ is small, Eq. (\ref{pacc}) is usually called the 
{\it partial conservation} of the axial current (PCAC).

For arbitrary $m$ Lagrangian (\ref{l1}) is 
invariant under  \uv1 transformations, 
\begin{equalign}                                                       
\label{uv1}                                                           
\psi \rightarrow \psi^{\prime} = \exp\left(-i{\phi \over 2}\right)\psi, 
\end{equalign}

\noindent which leads to the conservation of the {\it baryon} current, 
\begin{equalign}                                                       
\label{bcc}                                                            
\partial_{\mu} V^{\mu}(x) = 0, \,\,\,\,\,                              
V^{\mu}(x) = 
{\bar \psi}(x) \gamma^{\mu} \psi(x).
\end{equalign}                                                         

\bigskip
\section{Two-Flavour Symmetric Case}
\bigskip

A straightforward generalization of the one-flavour model leads to  the 
following Lagrangian

\begin{equalign}
\label{l2s}
{\cal L} = {\bar \psi} \left( i\! \not \! \partial -  {\hat m} \right) \psi 
+ \sum_{i=0}^3 {G_S \over 2} \left[({\bar \psi} \tau_i \psi)^2 + 
({\bar \psi} i \gamma_5 \tau_i \psi)^2 \right].
\end{equalign}

\noindent In this case, the field $\psi$ has an extra flavour
index and $\tau_i$ are the isospin Pauli matrices (with $\tau_0
= 1$). The quantity ${\hat m}$ is a two by two diagonal matrix containing 
the current quark masses, i.e., ${\hat m} = \hbox{diag}(m_u, m_d)$.

In the chiral limit, ${\hat m} \rightarrow 0$, Lagrangian (\ref{l2s})
is invariant under \ua1 transformations defined by the rule
(\ref{ua1}).  This fact gives the conservation of the axial current,
which we already know from the consideration of the one-flavour model,
see Eq. (\ref{acc}).  Moreover, also for ${\hat m}=0$, Lagrangian
(\ref{l2s}) does not change if we make \sua2 transformations

\begin{equalign}                                                       
\label{sua2}                                                           
\psi \rightarrow \psi^{\prime} = \exp\left(-i\gamma_5 { \pmb{$\tau$} \cdot
\pmb{$\chi$} \over 2}\right) \psi,
\end{equalign}        

\noindent where $\pmb{$\tau$} = (\tau_1,\tau_2,\tau_3)$.
This symmetry leads us to the conservation of the {\it chiral} 
current 
\begin{equalign}                                                       
\label{ccc}                                                            
\partial_{\mu} {\bf A}^{\mu}(x) = 0, \,\,\,\,\,                              
{\bf A}^{\mu}(x) = 
{\bar \psi}(x) \gamma^{\mu} \gamma_5 \pmb{$\tau$} \psi(x).
\end{equalign}                                                         

\noindent Similarly to the one-flavour case, if ${\hat m} \not = 0$
instead of Eq. (\ref{ccc}) one finds

\begin{equalign}
\label{pccc}                                                            
\partial_{\mu} {\bf A}^{\mu}(x) = 
i {\bar \psi}(x) 
\left(\pmb{$\tau$}{\hat m}+{\hat m}\pmb{$\tau$}\right)
\gamma_5 \psi(x).
\end{equalign}

In the case $m_u = m_d = m$, one finds that 
Lagrangian (\ref{l2s}) is invariant under \suv2 transformations
\begin{equalign}                                                       
\label{suv2}                                                           
\psi \rightarrow \psi^{\prime} = \exp\left(-i  { \pmb{$\tau$} \cdot
\pmb{$\phi$} \over 2}\right) \psi,
\end{equalign}                                                

\noindent which gives the conservation of the {\it isospin} current
\begin{equalign}                                                       
\label{icc}                                                            
\partial_{\mu} {\bf V}^{\mu}(x) = 0, \,\,\,\,\,                              
{\bf V}^{\mu}(x) = 
{\bar \psi}(x) \gamma^{\mu} \pmb{$\tau$} \psi(x).
\end{equalign}         

\noindent Finally, for any value of ${\hat m}$, one can check that 
the Lagrange density (\ref{l2s}) is invariant under \uv1 transformations,
such that the baryon current defined by Eq. (\ref{bcc}) is conserved. 

Let us now make a few comments about our terminology. Both \ua1 and
\sua2 symmetries will be later called axial or chiral ones.
Similarly, $A^{\mu}$ and ${\bf A}^{\mu}$ can be called axial or chiral
currents. However, the name axial current will be usually reserved for
$A^{\mu}$, and the name chiral will refer to ${\bf A}^{\mu}$.  In the
analogous way, both baryon and isospin currents, $V^{\mu}$ and ${\bf
V}^{\mu}$, can be shortly called the vector currents. Finally, since
the theory based on Lagrangian (\ref{l2s}) is for ${\hat m}=0$
invariant under both \ua1 and \sua2 symmetry groups, we shall call this 
version of the model symmetric.

Discussing the properties of our two Lagrangians, we have introduced
four types of symmetries, and associated with them four conserved
currents.  It is well known that these symmetries characterize also
the QCD Lagrangian.  In this sense, the NJL model has symmetry
features common with QCD.  Symmetries connected with the unitary
transformations \uv1 and \suv2 have a simple manifestation in
nature. The first one leads to the baryon number conservation, whereas
the second one is responsible for the organization of hadrons into the
isospin multiplets. On the other hand, the axial symmetries \ua1 and
\sua2 do not have any direct realization.  In particular, they do not
lead to any degeneracy of the particle spectra: If the \ua1 symmetry
were naively realized in nature, each hadron would have an opposite
parity partner. Similarly, a naive realization of the \sua2 symmetry
would require that each isospin multiplet has a partner multiplet
which groups the particles with opposite parity. Such a situation,
however, does not take place.

The problem with the \sua2 symmetry is resolved by the assumption that
this symmetry is realized in the Goldstone mode, i.e., it is
spontaneously broken. The phenomenon of the spontaneous symmetry
breaking is always accompanied by the appearance of the so-called
Goldstone bosons. Hence, in the case of \sua2 we expect that there
exist isovector pseudoscalar massless particles. Having in mind the
smallness of the pion mass, we can state that such particles indeed
exist. Speaking more precisely, pions should be regarded as quasi
Goldstone bosons, i.e., particles which appear after the breaking of
an approximate symmetry.

The situation with the \ua1 symmetry looks different. Since we do not
observe any Goldstone bosons which could be associated with \ua1, we
conclude that this symmetry cannot be spontaneously broken. In fact,
as it was shown by t' Hooft \cite{tH76}, the \ua1 symmetry is broken
due to the instanton effects.

The two versions of the NJL model, which have been discussed so far,
exhibit the spontaneous breaking of the \ua1 symmetry. In addition,
the two-flavour symmetric model breaks spontaneously the \sua2
invariance.  Unfortunately, as we have just learnt, the \ua1 symmetry
is not spontaneously broken in the real world. Consequently, these two
models must be still improved, if we want to use them for the
description of real particles.  In fact, this can be easily done by
the modification of Lagrangian (\ref{l2s}). One can add a term to
expression (\ref{l2s}), which simulates instanton effects and breaks
explicitly the \ua1 invariance. At the same time, this term leaves the
\sua2 symmetry intact (including its spontaneous breaking). This
procedure will be described in detail in the next Section.

Although the study of more realistic situations requires the elimination
of the \ua1 symmetry, the two simplest forms of the NJL model can be
still useful.  The point is, that in the discussed versions of NJL
model, the mechanism of the symmetry breaking is the same for both
\ua1 and \sua2 groups. Consequently, in many problems where the
flavour structure is not so important, we can study simplified
theories. In this report, we shall investigate the properties of
particles in hot and dense medium. In such a case, the extra
complexities resulting from the exact consideration of the flavour
content of particles are very often not so important.  Therefore, many
of our investigations will be based on simplified models defined by
Lagrangians (\ref{l1}) and (\ref{l2s}).

\bigskip                                                                       
\section{Two-Flavour Standard Case}
\bigskip

The \ua1 breaking interaction, accounting for the axial anomaly in QCD,
can be described by the following Lagrangian \cite{tH76}

\begin{eqnarray}
\label{l2det}
{\cal L}_A 
&=&  G_A \left\{ 
\hbox{det} \left[{\bar \psi}_i (1+\gamma_5) \psi_j \right] +
\hbox{det} \left[{\bar \psi}_i (1-\gamma_5) \psi_j \right] \right\} 
\nonumber \\
&=&  {G_A  \over 2} \left[
({\bar \psi} \psi)^2 + 
({\bar \psi} i \gamma_5 \pmb{$\tau$} \psi)^2 -
({\bar \psi} \pmb{$\tau$} \psi)^2 -
({\bar \psi} i \gamma_5 \psi)^2 \right],
\end{eqnarray}

\noindent Here det stands for the determinant with respect to
the flavour indices, and $G_A$ is the coupling constant describing the
strength of this interaction. (The last equality in (\ref{l2det}) is
valid only for the two-flavour case.)  One can check that expression
(\ref{l2det}) is invariant under \sua2 transformations. The easiest
way of proving this property is the consideration of an infinitesimal
\sua2 transformation. The formula (\ref{sua2}) for very small
$\pmb{$\chi$}$ (denoted later by $\delta \pmb{$\chi$} $) gives

\begin{equalign}
\label{isu2a1}
{\bar \psi}^{\prime} \psi^{\prime} = {\bar \psi} \psi
- {\bar \psi} i\gamma_5  \pmb{$\tau$} \psi \delta \pmb{$\chi$}, 
\,\,\,\,\,\,\,\, {\bar \psi}^{\prime} i\gamma_5 \psi^{\prime} = 
{\bar \psi} i\gamma_5 \psi
+ {\bar \psi} \pmb{$\tau$} \psi \delta \pmb{$\chi$}, \nonumber
\end{equalign}
\begin{equalign}
\label{isu2a2}
{\bar \psi}^{\prime}  \pmb{$\tau$}\psi^{\prime} = 
{\bar \psi}  \pmb{$\tau$}\psi - {\bar \psi} i\gamma_5 \psi \delta 
\pmb{$\chi$},  \,\,\,\,\,\,\,\,
{\bar \psi}^{\prime} i\gamma_5  \pmb{$\tau$} \psi^{\prime} =
{\bar \psi} i\gamma_5  \pmb{$\tau$} \psi
+ {\bar \psi} \psi \delta \pmb{$\chi$}. \nonumber
\end{equalign}

\noindent One can notice that the first two terms on the right-hand-side
(RHS) of the second line in (\ref{l2det}) and the last two terms (also
in this line) are separately \sua2 invariant. On the other hand,
Lagrangian (\ref{l2det}) is not \ua1 invariant. Thus, adding formula
(\ref{l2det}) to Lagrangian (\ref{l2s}), we obtain a theory which is
only \sua2 invariant.  In the case $G_A = G_S = G$ one finds

\begin{equalign}
\label{l2}
{\cal L} = {\bar \psi} \left( i\! \not \! \partial - {\hat m} \right) \psi 
+ G \left[({\bar \psi} \psi)^2 + 
({\bar \psi} i \gamma_5 \pmb{$\tau$} \psi)^2 \right].
\end{equalign}

The first Lagrangian having exactly the same form as expression
(\ref{l2}) was written by Nambu and Jona-Lasinio (the second paper
published in 1961 \cite{NJL}). Since at that time quarks and gluons
were not known, the field $\psi$ was used to describe nucleons
(protons and neutrons).  However, shortly after its formulation, the
model of Nambu and Jona-Lasinio was abandoned because its
non-fundamental character (reflected mainly in the nonrenormalizability
of the theory) became clear. Later on the model was
reinterpreted in terms of the quark degrees of freedom.

\bigskip
\section{Mean-Field Approximation}
\bigskip

Let us now come back to the discussion of the one-flavour model.
By introducing the fields ${\hat \sigma}$ and ${\hat \pi}$ defined 
as

\begin{equalign}
\label{sp}
{\hat \sigma} = - 2 G \, {\bar \psi} \psi, \,\,\,\,
{\hat \pi} = - 2 G \, {\bar \psi} i \gamma_5 \psi,
\end{equalign}

\noindent we can recast Lagrangian (\ref{l1}) into the form

\begin{equalign}
\label{l1mf}
{\cal L} = {\bar \psi} (i\! \not \! \partial - m) \psi -
{\hat \sigma} \, {\bar \psi} \psi - {\hat \pi} \, {\bar \psi}
i \gamma_5 \psi - {{\hat \sigma}^2 + {\hat \pi}^2 \over 4G}.
\end{equalign}

\noindent The equivalence of (\ref{l1}) and (\ref{l1mf}) becomes
clear if one checks that the variation of (\ref{l1}) with respect
to $\bar \psi$ gives the same equation of motion for $\psi$ as
the variation of (\ref{l1mf}) with respect to ${\bar \psi}, {\hat
\sigma}$ and ${\hat \pi}$, namely

\begin{equalign}
\label{de}
\left[i\! \not\! \partial - m - {\hat \sigma}(x) - i \gamma_5
{\hat \pi}(x) \right]\psi(x) = 0.
\end{equalign}

In the {\it mean-field approximation}, the operators ${\hat \sigma}$ and 
${\hat \pi}$ are replaced by their mean values, namely
\begin{equalign}
\label{mfa}
{\hat \sigma} \rightarrow \sigma =  \langle {\hat \sigma} \rangle
= \hbox{Tr}( {\hat \rho} {\hat \sigma} ), \,\,\,\,\,
{\hat \pi} \rightarrow \pi = \langle {\hat \pi} \rangle
= \hbox{Tr}( {\hat \rho} {\hat \pi} ).
\end{equalign}

\noindent This leads to the self-consistent mean-field equations:

\begin{equalign}
\label{scmfe1}
\left[ i\! \not\! \partial - m - \sigma (x) - 
i \gamma_5 \pi (x) \right]\psi(x) = 0, 
\end{equalign}

\begin{equalign}
\label{scmfe2}
-2\,G\,\hbox{Tr}
\left[{\hat \rho}{\bar \psi}(x) \psi(x) \right] = \sigma(x), 
\end{equalign}

\begin{equalign}
\label{scmfe3}
-2\,G\,\hbox{Tr}
\left[{\hat \rho}{\bar \psi}(x) i\gamma_5 \psi(x) \right] = \pi(x).
\end{equalign}

\noindent In Eqs. (\ref{mfa}) - (\ref{scmfe3})
${\hat \rho}$ is the density operator and Tr denotes the trace over
all physical states of the system. The particular form of ${\hat
\rho}$ is not important for our considerations.  We only note that in
typical situations, ${\hat \rho}$ is either an operator projecting on
the parity invariant ground state or it describes parity invariant
statistical ensembles. In these two cases $\langle {\hat \pi} \rangle
=0$. In more general situations (e.g., when the system
is out of equilibrium) we cannot exclude the case $\langle {\hat \pi}
\rangle \not = 0$ \footnote{Generally speaking, in the space-time regions 
where $\langle {\hat \pi} \rangle \not = 0$ the quark condensate is
chirally rotated from its usual orientation in the isospin space. Such
a piece of wrongly oriented condensate is called the {\it disoriented
chiral condensate} \cite{DCC}. Its production may explain rare events
with a deficit or excess of neutral pions observed in cosmic ray
experiments.}.

The \ua1 transformation of the quark field $\psi$ induces the change of
the fields ${\hat \sigma}$ and ${\hat \pi}$. Altogether, the fields   
$\psi$, ${\hat \sigma}$ and ${\hat \pi}$ obey the following transformation
rule

\begin{eqnarray}                                                       
\label{chi1}                                                           
\psi &\rightarrow& \psi^{\prime} = \exp(-i\gamma_5 {\chi \over 2})     
\psi, \\                                                               
\label{chi2}                                                           
{\hat \sigma} &\rightarrow& {\hat {\sigma^{\prime}}} =                 
{\hat \sigma} \cos \chi - {\hat \pi} \sin \chi, \\                     
\label{chi3}                                                           
{\hat \pi   } &\rightarrow& {\hat {\pi^{\prime}}}    =                 
{\hat \pi   } \cos \chi + {\hat \sigma} \sin \chi.                     
\end{eqnarray}                                                         

\noindent In the vector space spanned by the fields  ${\hat \sigma}$ 
and ${\hat \pi}$, the chiral symmetry manifests itself simply as a
rotation. 

It is also important to realize that in the mean-field
approximation the conservation of the axial current holds only for the
mean value.  Speaking more precisely, in the limit $m=0$ one finds

\begin{equalign}
\label{mfac1}
\partial_{\mu} A^{\mu} (x) = 2 \sigma(x) {\bar \psi}(x) i\gamma_5 
\psi(x) - 2 \pi(x) {\bar \psi}(x) \psi(x),
\end{equalign}

\noindent which after calculation of the trace gives

\begin{equalign}
\label{mfac2}
\partial_{\mu} \langle A^{\mu} (x) \rangle = 
\partial_{\mu} \hbox{Tr} ( {\hat \rho}  A^{\mu} (x) ) = 0.
\end{equalign} 

\part{\bf EQUILIBRIUM ENSEMBLES}

\chapter{\bf Imaginary-Time Formalism}
\label{chapt:imagin_time}

Chapters 5 --- 9 of our report will be devoted to the study of systems
in thermodynamic equilibrium. In order to calculate different physical
quantities at finite temperature or at finite baryon chemical
potential, we shall use the imaginary-time formalism \cite{FW71,K89,M81}.  
In this Chapter, we give a short introduction to this formalism and 
discuss several computational rules used at finite
$T$ or $\mu$.

\section{Temperature Green's Functions}

While studying many particle systems, it is convenient to use the {\it
grand canonical ensemble}. Introducing the operator ${\hat K} = {\hat
H} - \mu {\hat N}$ (${\hat H}$ is the Hamiltonian and ${\hat N}$ is
the baryon number operator) we can define the partition function and
the statistical density operator by the following expressions

\begin{equalign}
\label{Z}
Z = \hbox{Tr} \, e^{-\beta {\hat K}} \equiv e^{-\beta \Omega}
\end{equalign}
and
\begin{equalign}
\label{rho}
{\hat \rho} = Z^{-1} e^{-\beta {\hat K}},
\end{equalign}

\noindent where $\beta$ is the inverse temperature and $\Omega$ is
the thermodynamic potential ($k_B=1$).

The operator ${\hat K}$ can be interpreted as the grand canonical
Hamiltonian. Thus, for each operator in the Schr\"odinger picture,
${\hat O}_S({\bf x})$, we can define its partner in the modified
Heisenberg picture, namely

\begin{equalign}
\label{mhp}
O_K(\tau,{\bf x}) = e^{{\hat K}\tau}{\hat O}_S({\bf x})
e^{-{\hat K}\tau}.
\end{equalign}

\noindent Using this prescription, we introduce the quark 
{\it temperature Green's} function

\begin{equalign}
\label{tgf}
{\cal G}_{ab}(\tau,{\bf x};\tau^{\prime},{\bf x}^{\prime}) 
= - \hbox{Tr} \left\{ {\hat \rho}  {\hat T}_{\tau}
\left[ \psi_{K\,a}(\tau,{\bf x}) 
{\bar \psi}_{K\,b}(\tau^{\prime},{\bf x}^{\prime}) \right] \right\}.
\end{equalign}

\noindent Here $\psi_{K\,a}(\tau,{\bf x})$ is the quark field operator, 
$a$ and $b$ are the indices describing internal degrees of freedom
(spinor, flavour and colour), and ${\hat T}_{\tau}$ is the
$\tau$-ordering operator.  Of course, an analogous expression can be
given for other types of fields.  Therefore, from now on we shall
continue our discussion treating simultaneously the case of fermions
and bosons.

In the typical situations (${\hat H}$= const, the system is uniform)
${\cal G}_{ab}(\tau,{\bf x};\tau^{\prime},{\bf x}^{\prime})$ is a
function of the difference of its arguments, i.e., we can write ${\cal
G}_{ab}(\tau,{\bf x};\tau^{\prime},{\bf x}^{\prime})= {\cal
G}_{ab}(\tau-\tau^{\prime},{\bf x}-{\bf x}^{\prime})$.  Using the
fermion (boson) commutation relations and the cyclic property of the
trace, we find that the Green's function is an anti-periodic
(periodic) function of time

\begin{equalign}
\label{pgf}
{\cal G}_{ab}(\tau-\tau^{\prime},{\bf x}-{\bf x}^{\prime})
= \mp \, {\cal G}_{ab}(\tau-\tau^{\prime}+\beta,
{\bf x}-{\bf x}^{\prime}).
\end{equalign}

\noindent Equation (\ref{pgf}) allows us to write the following 
representation

\begin{equalign}
\label{fs}
{\cal G}_{ab}(\tau,{\bf x}-{\bf x}^{\prime}) = T \sum_n
e^{-i\omega_n\tau} {\cal G}_{ab}(i\omega_n,{\bf x}-{\bf x}^{\prime}),
\end{equalign}

\noindent where

\begin{equalign}
\label{fc}
{\cal G}_{ab}(i\omega_n,{\bf x}-{\bf x}^{\prime}) =
\int_{0}^{\beta} d\tau  e^{i\omega_n\tau}
{\cal G}_{ab}(\tau,{\bf x}-{\bf x}^{\prime}),
\end{equalign}

\noindent and $\omega_n$ are the so-called {\it Matsubara} frequencies:

\begin{equalign}
\label{mf}
\omega_n = (2n+1)\pi T \,\,\,\, \hbox{(for fermions)}, \,\,\,\,\,
\omega_n = 2n \pi T    \,\,\,\, \hbox{(for bosons)}.
\end{equalign}

The temperature Green's functions are very useful for the quantitative 
calculation of the thermodynamic properties of the system. On the other 
hand, in order to study the frequencies and lifetimes of excited states 
at finite temperature, we should study the structure of the
{\it real-time Green's} functions. They are defined by the expression

\begin{equalign}
\label{rtgf}
i{\overline G}_{ab}(t,{\bf x};t^{\prime},{\bf x}^{\prime}) 
=  \hbox{Tr} \left\{ {\hat \rho} T
\left[ 
\psi_{K\,a}(t,{\bf x}) 
{\bar \psi}_{K\,b}(t^{\prime},{\bf x}^{\prime})
\right] \right\},
\end{equalign}

\noindent which is a straightforward generalization of the $T=\mu=0$ case

\begin{equalign}
\label{vgf}
i G_{ab}(t,{\bf x};t^{\prime},{\bf x}^{\prime}) 
=  \langle \hbox{0} | T \left[ \psi_{K\,a}(t,{\bf x}) 
{\bar \psi}_{K\,b}(t^{\prime},{\bf x}^{\prime})
\right]  | \hbox{0} \rangle.
\end{equalign}

\noindent
In Eqs. (\ref{rtgf}) and  (\ref{vgf}) ${\hat T}$ is the $t$-ordering 
operator and  the field operators are written in the standard 
Heisenberg  picture, i.e., the following notation is used here

\begin{equalign}
\label{hp}
O_K(t,{\bf x}) = e^{i{\hat K}t}{\hat O}_S({\bf x}) e^{-i{\hat K}t}. 
\end{equalign}

\noindent We note that the real-time Green's function (\ref{vgf}) is
defined as the expectation value of the operator product in the
ground state $ | \hbox{0} \rangle $, and that the grand canonical
density operator ${\hat \rho}$ is reduced to the projection operator
onto this state in the case  $T=\mu=0$.

Besides the formula (\ref{rtgf}), one employs also the 
{\it retarded} and {\it advanced} Green's functions 

\begin{equalign}
\label{retgf}
i{\overline G}^{\, R}_{ab}(t,{\bf x};t^{\prime},{\bf x}^{\prime}) 
=  \theta(t-t^{\prime}) \hbox{Tr} \left\{ {\hat \rho} 
\left[ 
\psi_{K\,a}(t,{\bf x}), 
{\bar \psi}_{K\,b}(t^{\prime},{\bf x}^{\prime})
\right]_{\pm} \right\}
\end{equalign}
and
\begin{equalign}
\label{advgf}
i{\overline G}^{\, A}_{ab}(t,{\bf x};t^{\prime},{\bf x}^{\prime}) 
=  -\theta(t^{\prime}-t) \hbox{Tr} \left\{ {\hat \rho} 
\left[ 
\psi_{K\,a}(t,{\bf x}), 
{\bar \psi}_{K\,b}(t^{\prime},{\bf x}^{\prime})
\right]_{\pm} \right\},
\end{equalign}

\noindent where $[\,\,,\,\,]_{\pm}$ denotes the anticommutator for fermions
and the commutator for bosons. These two functions have the following
spectral representations

\begin{equalign}
\label{specrr}
{\overline G}^{\, R}_{ab}(\omega,{\bf q}) = \int\limits_{-\infty}^{\infty}
{d\omega^{\prime} \over 2\pi} 
{\rho_{ab}(\omega^{\prime},{\bf q}) \over \omega - \omega^{\prime} 
+ i\epsilon}
\end{equalign}
and
\begin{equalign}
\label{specra}
{\overline G}^{\, A}_{ab}(\omega,{\bf q}) = \int\limits_{-\infty}^{\infty}
{d\omega^{\prime} \over 2\pi} 
{\rho_{ab}(\omega^{\prime},{\bf q}) \over \omega - \omega^{\prime} 
- i\epsilon}.
\end{equalign}

\noindent In Eqs. (\ref{specrr}) and (\ref{specra}) the function
$\rho_{ab}(\omega^{\prime},{\bf q})$ is the spectral density (the same
in the two cases) and $\epsilon$ is an infinitesimally positive
constant.

The crucial point in the imaginary-time formalism is that the
temperature Green's function has the same spectral representation,
i.e., we can write

\begin{equalign}
\label{specrt}
{\cal G}_{ab}(i\omega_n,{\bf q}) = \int\limits_{-\infty}^{\infty}
{d\omega^{\prime} \over 2\pi} 
{\rho_{ab}(\omega^{\prime},{\bf q}) \over i\omega_n - \omega^{\prime} },
\end{equalign}

\noindent where again the same spectral function $\rho_{ab}(\omega^{\prime},
{\bf q})$ appears. Using Eqs. (\ref{specrr}) - (\ref{specrt}) and the
requirement of the correct asymptotics of the Green's function in
infinity, one can show that the temperature Green's function (defined
only at complex and discrete frequencies $i\omega_n$) has a unique
analytic continuation to the whole complex plane, where it coincides
either with ${\overline G}^{\, R}_{ab}(\omega,{\bf q})$ (for upper
half-plane) or ${\overline G}^{\, A}_{ab}(\omega,{\bf q})$ (for lower
half-plane) \cite{BM61}.  This allows us for the calculation of
${\overline G}^{\, R}_{ab}$ and ${\overline G}^{\, A}_{ab}$ from the
knowledge of the temperature Green's function ${\cal
G}_{ab}(i\omega_n,{\bf q})$. Furthermore, knowing ${\overline G}^{\,
R}_{ab}$ and ${\overline G}^{\, A}_{ab}$ we find ${\overline
G}_{ab}(\omega,{\bf q})$ through the relation \cite{FW71}

\begin{equalign}
\label{cra}
{\overline G}_{ab}(\omega,{\bf q}) = [1 \pm e^{-\beta \omega}]^{-1}\,\,
{\overline G}^{\, R}_{ab}(\omega,{\bf q}) +  [1 \pm e^{\beta \omega}]^{-1}
\,\,{\overline G}^{\, A}_{ab}(\omega,{\bf q}).
\end{equalign}

\noindent {\it The importance of the discussed connections between the
temperature and real-time Green's functions is related to the fact
that the Feynman perturbation theory applies only to the temperature
Green's functions; at finite temperature the real-time Green's
functions do not have such a diagrammatic expansion, unless one
introduces an extended Hilbert space} \cite{K89}.  Thus, when studying
excited states at finite temperature we first calculate the
temperature Green's function (in this case we can use all known
techniques for Feynman diagrams), and only afterwards we do the
analytic continuation to real frequencies.

\section{Sums over Frequencies}

The Feynman rules for the temperature Green's functions \cite{KM76} 
coincide with the $T=\mu=0$ rules with the replacement

\begin{equalign}
\label{is}
\int {d^4p \over (2\pi)^4} \longrightarrow
iT \sum_n \int {d^3p\over (2\pi)^3},
\end{equalign}

\noindent where the continuous variable $p^0$ should be replaced
by discrete complex energies \footnote{In this report we consider
the case when the bosonic chemical potential is
zero.}:

\begin{equalign}
p^0 \longrightarrow i\omega_n + \mu = (2n+1)\pi i T + \mu
\,\,\,\, \hbox{(for fermions)}
\end{equalign}
and
\begin{equalign}
p^0 \longrightarrow i\omega_n       = 2n \pi i T    
\,\,\,\, \hbox{(for bosons)}.
\end{equalign}

\begin{figure}[hb]
\label{fig:cps}
\xslide{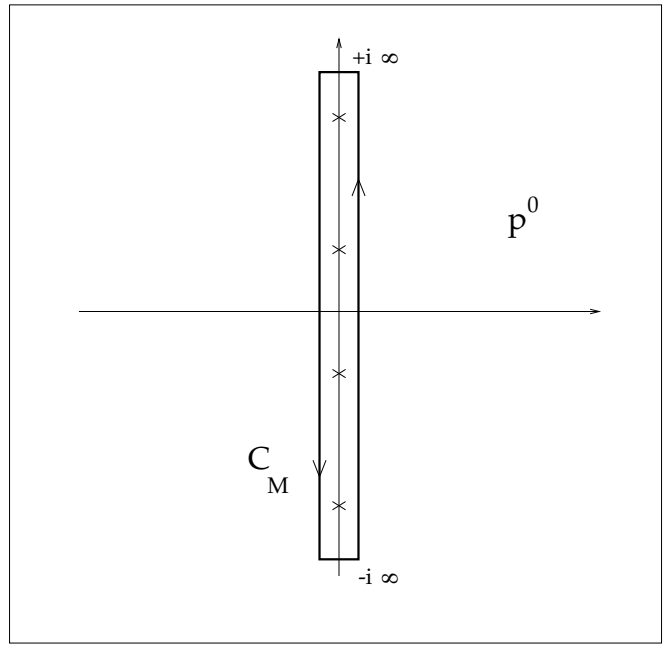}{9cm}{24}{148}{566}{687}
\caption{\small The sum over the Matsubara frequencies can be calculated as
                a contour integral in the complex energy plane.}
\end{figure}

Very often, it is convenient to convert the frequency sums to contour
integrals. In the case of fermions, one can write \cite{FW71}

\begin{equalign}
\label{soff}
iT \sum_n f(i\omega_n+\mu) = - {1 \over 2\pi} \oint\limits_{{\cal C}_M}
{dp^0 f(p^0+\mu) \over \exp(\beta p^0) + 1},
\end{equalign}

\noindent where ${{\cal C}_M}$ denotes the integration contour in the complex
$p^0$ space, see Fig. [4.1]. Changing the integration variable, we
obtain \cite{KM76}

\begin{eqnarray}
\label{soff1}
iT \sum_n f(i\omega_n+\mu) &=& - {1 \over 2\pi} 
\int\limits_{-i\infty + \mu + \epsilon}^{+i\infty + \mu + \epsilon}
{dp^0 f(p^0) \over \exp[\beta (p^0-\mu)] + 1}
- {1 \over 2\pi} 
\int\limits_{-i\infty + \mu - \epsilon}^{+i\infty + \mu - \epsilon}
{dp^0 f(p^0) \over \exp[\beta (\mu-p^0)] + 1} \nonumber \\
&+& {1 \over 2\pi} \oint\limits_{\cal C} dp^0 f(p^0)
+ {1 \over 2\pi} \int\limits_{-i\infty}^{+i\infty} dp^0 f(p^0),
\end{eqnarray}

\noindent where ${\cal C}$ is another integration contour in the complex 
energy plane, shown in Fig. [4.2]. The position of ${\cal C}$ is
fixed by the value of the chemical potential $\mu$. 

\begin{figure}[ht]
\label{fig:mats}
\xslide{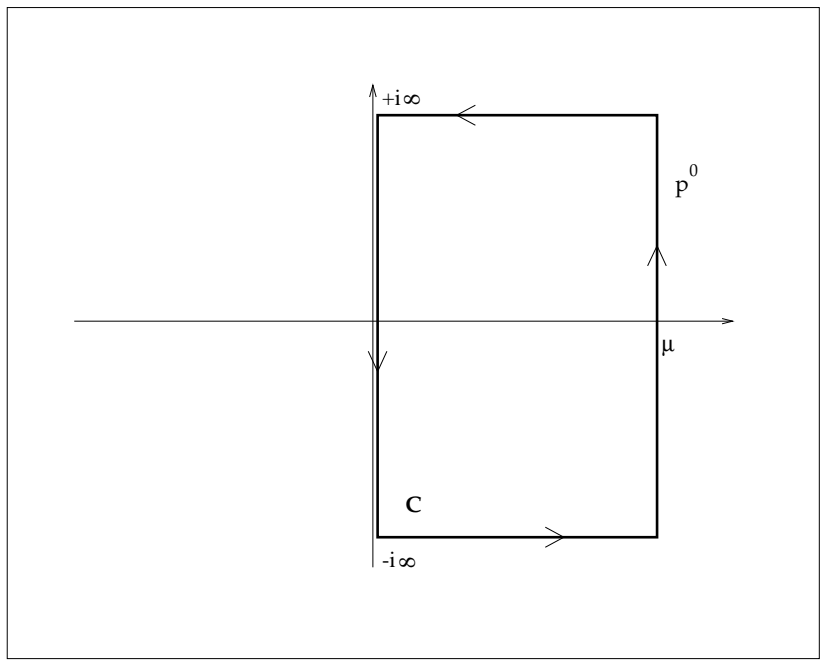}{9cm}{1}{179}{590}{662}
\caption{\small Integration contour in the complex energy plane used
                for the evaluation of the matter parts.}
\end{figure}

There are two special cases which will be particularly interesting for us, 
i.e., the case of vanishing chemical potential and the case when the 
temperature is zero. For $\mu=0$ and $T \not = 0$, Eq. (\ref{soff1})
can be simplified to the form

\begin{equalign}
\label{soff2}
iT \sum_n f(i\omega_n) =  {1 \over 2\pi}
\int\limits_{-i\infty}^{+i\infty} dp^0 f(p^0) 
- {1 \over 2\pi} \int\limits_{-i\infty + \epsilon}^{+i\infty + \epsilon}
{dp^0 \left[f(p^0)+f(-p^0)\right] \over \exp(\beta p^0) + 1},
\end{equalign}   

\noindent whereas in the limit  $T \rightarrow  0$ and $\mu \not = 0$, 
we can  use the rule

\begin{equalign}
\label{soff3}
iT \sum_n f(i\omega_n+\mu) \rightarrow {1 \over 2\pi}
\int\limits_{-i\infty}^{+i\infty} dp^0 f(p^0)
+ {1 \over 2\pi} \oint\limits_{\cal C} dp^0 f(p^0).
\end{equalign}

\noindent In the case when the integrand depends additionally on the 
external frequency, e.g., $f = f(p^0,{\bf p}, \omega)$, the integrals
on the RHS of expressions (\ref{soff2}) and (\ref{soff3}) should be
evaluated for purely imaginary values of $\omega$ and {\it
subsequently} analytically continued to real frequencies. Similar
prescriptions for the conversion of the frequency sum into the contour
integrals can be given for bosons. We skip their presentation here
since our later considerations will be restricted mainly to the
fermionic sums.

The meaning of the decompositions (\ref{soff2}) and (\ref{soff3}) is that
we can separate the so-called {\it vacuum part} from the {\it medium
part}. The vacuum part of a physical quantity does not {\it explicitly}
depend on the occupation of phase space and reduces at $T = \mu = 0$
to its vacuum expectation value. On the other hand, the medium part
depends explicitly on the occupation of phase space and consequently
vanishes in vacuum.

\section{Meson Correlation Functions}

Let us now consider again the special case when $\psi_{K a}(t,{\bf x)}$ 
is the quark field operator. In this situation, we define the meson 
correlation function by the expression

\begin{equalign}
\label{mcf}
\chi_{ab}(t,{\bf x}) = i \hbox{Tr} \left\{ {\hat \rho} {\hat T}
{\hat O}_{K a}(t,{\bf x}) {\hat O}_{K b}(0,{\bf 0}) \right\},
\end{equalign}

\noindent where

\begin{equalign}
\label{opo}
{\hat O}_{K a}(t,{\bf x}) = {\bar \psi}_K(t,{\bf x}) M_a
\psi_K(t,{\bf x}) -  \hbox{Tr} \left\{ {\hat \rho} {\hat T}
{\bar \psi}_K(t,{\bf x}) M_a
\psi_K(t,{\bf x}) \right\}.
\end{equalign}

\noindent The quantities $M_a$ in Eq. (\ref{opo}) are products of the
Dirac spinor matrices $\Gamma_A$ and the Pauli isospin matrices
$\tau_i$ (in the case of the SU(3) flavour group the Pauli matrices
should be replaced by the Gell-Mann matrices $\lambda_a$).
In writing Eq. (\ref{mcf}) we have assumed that the Hamiltonian
is independent of time and that the system is uniform (this allowed
us to make a shift of one of the arguments to zero).

The meson correlation function (\ref{mcf}) describes the 
real-time propagation of a meson which consists of a $q {\overline q}$ 
pair. Similarly to the case of one-particle Green's functions discussed
before, in order to calculate the real-time correlation function 
(\ref{mcf}) we initially calculate the temperature correlation 
function and subsequently do the analytic continuation. The 
temperature meson correlation function is defined as

\begin{equalign}
\label{tmcf}
\chi_{ab}(\tau,{\bf x}) = \hbox{Tr} \left\{ {\hat \rho} {\hat T}_{\tau}
{\hat O}_{K a}(\tau,{\bf x}) {\hat O}_{K b}(0,{\bf 0}) \right\},
\end{equalign}

\noindent where the field operators are written in the modified Heisenberg
picture (\ref{mhp}) (we do not introduce the special notation for the
temperature correlation function, it can be distinguished by its
imaginary time argument $\tau$).

\chapter{\bf Quark Self-Energy and the Gap Equation}
\label{chapt:gap}

In the NJL model, one calculates the quark self-energy from the
self-consistent Schwinger-Dyson equation. This is usually done either
in the {\it Hartree} or {\it Hartree-Fock} approximation. Furthermore,
different regularization schemes can be used to define divergent
integrals.  In order to illustrate these possibilities, we come back
to the separate discussion of the one-flavour, two-flavour symmetric,
and two-flavour standard models. The one-flavour case will be treated
in the Hartree approximation and the 3-dimensional cutoff will be used
as a regulator.  The two-flavour symmetric case will be considered in
the Hartree-Fock approximation and we shall regularize it using a
version of the Pauli-Villars method. Finally, we shall discuss the
standard version of the two-flavour model in the Hartree-Fock
approximation, regularizing it with the help of the Schwinger
proper-time method.

We analyze the differences in the formulation of the theory in more
detail, since several problems of interest for us will be studied
later in different versions of the model --- it often happens that one
version is more suitable for the analysis of a given problem than
other versions. In particular, the in-medium dependence of the dynamic
and screening masses of mesons (Chapters 6 and 7) as well as the
oscillations of the static meson fields at finite baryon density
(Chapter 8) are studied within the two-flavour symmetric model. The
methods going beyond the Hartree-Fock approximation (Chapter 9) are
introduced on the basis of the two-flavour standard model. On the
other hand, the transport theory for the model (Chapter 11) is
constructed for the one-flavour model.

Using the methods introduced in Chapter 4, we calculate the quark
self-energy at finite temperature or density. In this way we can
determine the in-medium dependence of the constituent quark mass.
Illustrative calculations done in different versions of the 
model lead to the same qualitative conclusion: with increasing
$T$ ($\mu$) the constituent mass decreases, and for sufficiently
large temperature (density) it approaches the value of the current
mass.

\section{One-Flavour Case}

In the one-flavour model, using the Hartree approximation we write
the self-consistent Schwinger-Dyson equation in the following
form

\begin{equalign}
\label{qse1f}
\Sigma = 2 i G \int { d^4p \over (2\pi)^4 } 
\left\{ \hbox{tr} \left[ G(p) \right] + 
i \gamma_5  \, \hbox{tr} \left[i \gamma_5 G(p) \right] 
\vphantom{{1\over2}} \right\},
\end{equalign}

\noindent where tr is the trace over spinor (and colour) indices and
$G(p)$ is the quark propagator

\begin{equalign}
\label{prop}
G(p)^{-1} = \not \! p - \Sigma - m + i\epsilon.
\end{equalign}

\begin{figure}[ht]
\label{qse1}
\xslide{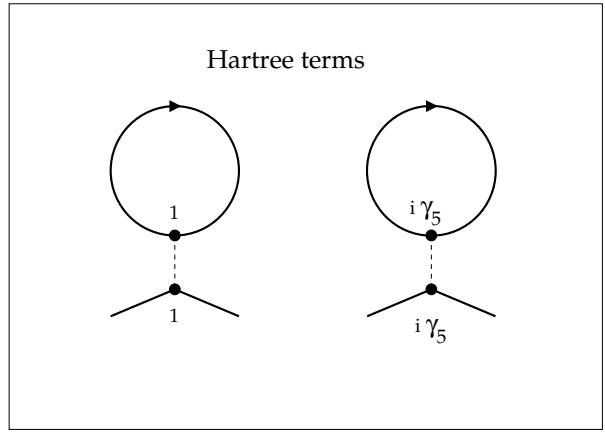}{9cm}{42}{135}{555}{700}
\caption{\small Diagrams used for the calculation of the quark self-energy
         in the Hartree approximation (one-flavour model). 
         The solid line describes the quark propagator, whereas the
         dashed line is introduced to describe the point-like
         quark-antiquark interaction. Only the 
         first term gives a non-zero contribution to the self-energy.}
\end{figure}

\noindent The two terms on the RHS of Eq. (\ref{qse1f})
appear since Lagrangian (\ref{l1}) describes both scalar and
pseudoscalar interactions, see Fig. [\ref{qse1}]. However, parity
implies that the second term in (\ref{qse1f}) vanishes. 

Using the general formula for the quark propagator, Eq. (\ref{vgf}),
we define the {\it quark condensate} through equation

\begin{equalign}
\label{cond}
\langle {\overline q}  q\rangle 
= - i\, \hbox{tr}\, G(x=0^-).
\end{equalign}

\noindent Rewriting Eq. (\ref{cond}) in the momentum space, we find
that

\begin{equalign}
\label{cond1}
\langle {\overline q} q\rangle = 
- 4 N_c \, i \int {d^4 p \over (2\pi)^4}
{\Sigma + m \over p^2 -
(\Sigma + m)^2 + i\epsilon}.
\end{equalign}

\noindent The sum $\Sigma + m$ can be regarded as the {\it constituent}
(effective or dynamic) quark mass $M$, whose origin are the quark-antiquark
interactions.  Eqs.  (\ref{qse1f}) and (\ref{cond1}) lead to the simple
relation

\begin{equalign}
\label{gap1f}
M = m - 2 G \langle {\overline q} q\rangle,
\end{equalign}

\noindent which connects the constituent quark mass directly to the
condensate. Eq. (\ref{gap1f}) is called the {\it gap equation} since
it determines the energy gap in the energy spectrum of one-particle
excitations.

In writing expressions (\ref{qse1f}) - (\ref{gap1f}) we have used the
Feynman rules for $T = \mu = 0$. As it was pointed out in Chapter 4,
the Feynman rules at finite $T$ and $\mu$ have the same
structure. Consequently, we can always start our discussion with the
consideration of the standard Feynman diagrams. Later on, the
expressions valid at finite $T$ and $\mu$ are obtained by employing
the prescriptions: (\ref{is}), (\ref{soff2}) and (\ref{soff3}). In
particular, it turns out that the form of formula (\ref{gap1f}) is
valid also at finite $T$ and $\mu$ provided one writes $ \langle
{\overline q} q\rangle$ as a sum of two contributions

\begin{equalign}
\label{condec}
\langle {\overline q} q\rangle = \langle {\overline q} q\rangle_{\hbox{vac}}
+ \langle {\overline q} q\rangle_{\hbox{med}}.
\end{equalign}
 
For the moment, let us concentrate in more detail on the vacuum 
part of the condensate $\langle {\overline q} q\rangle_{\hbox{vac}}$.  Of
course, it can be evaluated directly from Eq. (\ref{cond1}). For
further use, it is convenient to introduce the function
$I_{1,\hbox{vac}}(M^2)$ defined as

\begin{equalign}
\label{i1vac}
I_{1,\hbox{vac}}(M^2) = 8iN_c \int {d^4p \over (2\pi)^4} {1 \over p^2 - M^2
+ i\epsilon}.
\end{equalign}

\noindent Using Eqs. (\ref{cond1}) and (\ref{i1vac}) we write

\begin{equalign}
\label{condvac}
\langle {\overline q} q\rangle_{\hbox{vac}} = - {M \over 2}
I_{1,\hbox{vac}}(M^2).
\end{equalign}

\noindent Integration over energy in (\ref{i1vac}) leads to the following 
expression
 
\begin{equalign}
\label{i1vac1}
I_{1,\hbox{vac}}(M^2) = 4N_c \int {d^3p \over (2\pi)^3} {1 \over E_p},
\,\,\,\, E_p = \sqrt{ {\bf p}^2+M^2}.
\end{equalign}

\noindent The 3-dimensional integral appearing in Eq. (\ref{i1vac1})
diverges. The simplest way of its regularization is the introduction of a 
cutoff $\Lambda$, confining the region of integration to the sphere
$|{\bf p}| < \Lambda$. In this case one finds

\begin{equalign}
\label{i1vacR3D}
I_{1, \hbox{vac}}^{R, 3D}(M^2) = {N_c \Lambda^2 \over \pi^2}
\left[ \sqrt{1+\left({M\over\Lambda}\right)^2 }
-\left({M\over\Lambda}\right)^2 \ln 
{\Lambda + \sqrt{\Lambda^2+M^2} \over M} \right].
\end{equalign}

\noindent Using now Eqs. (\ref{gap1f}), (\ref{condvac}) and (\ref{i1vacR3D}) 
we obtain the following explicit form of the gap equation

\begin{equalign}
\label{gap3D}
M \left[1 - {m \over M} \right] = M \,  {N_c G \Lambda^2 \over \pi^2}
\left[ \sqrt{1+\left({M\over\Lambda}\right)^2 }
-\left({M\over\Lambda}\right)^2 \ln 
{\Lambda + \sqrt{\Lambda^2+M^2} \over M} \right].
\end{equalign}

It is interesting to look at some consequences of Eq. (\ref{gap3D}).
At first, let us consider the case of the chiral limit $m=0$.  The
expression in the square bracket on the RHS of formula (\ref{gap3D})
is positive and smaller than 1 for $M>0$. Thus, the nontrivial
solution exists only if $ 0 < \pi^2 / (N_c G \Lambda^2) < 1 $. See
Fig. [5.2], where Eq. (\ref{gap3D}) has been plotted for two different
values of $G$. As $G \Lambda^2$ increases over the critical value
$\pi^2 / N_c$, $M$ starts rising from zero. Consequently, the form of
Eq. (\ref{gap3D}) indicates that the force between quarks and
antiquarks must be positive ($G > 0$) and strong enough to cause the
formation of the condensate. In the analogy to the BCS theory of
super-conductivity, one expects that the nontrivial solution of
Eq. (\ref{gap3D}) corresponds to the true ground state of the model
\footnote{In the Hartree approximation the true ground state of
the theory can be described as consisting of the Dirac sea of the
constituent quarks --- {\it complicated} vacuum of strongly
interacting {\it light} current quarks is approximated by a {\it
simple} vacuum of the {\it heavy} constituent quarks.}.  This is
indeed so, and this fact can be verified by performing a variational
calculation (for more details on this subject see the original paper
by Nambu and Jona-Lasinio \cite{NJL} or the last review listed in
\cite{NJLR}). If the current quark mass $m$ does not vanish, one can
find out that there is always a nontrivial solution to
Eq. (\ref{gap3D}).  Moreover, it corresponds to the ground state of
the theory, similarly as in the special case $m=0$.

\begin{figure}[hb]
\label{crg}
\xslide{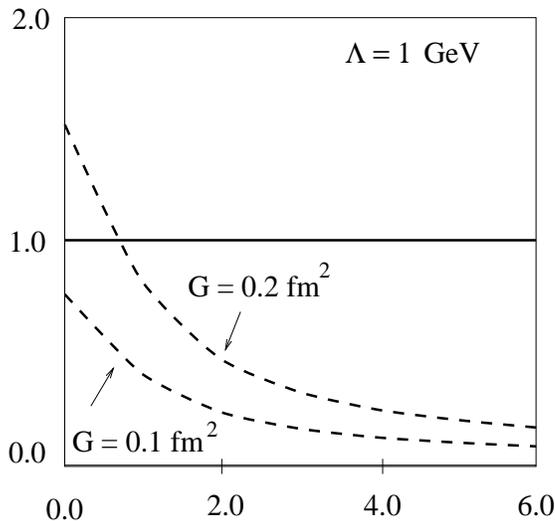}{10cm}{42}{135}{555}{700}
\caption{\small Graphical representation of Eq. (\ref{gap3D}) in the
limiting case $m=0$ and for two different values of the coupling
constant $G$. The common factor $M$, appearing on both sides of
Eq. (\ref{gap3D}), has been omitted. The LHS is represented by the
solid line, whereas the RHS is represented by the two dashed
lines. Only if $G \Lambda^2$ is greater than the critical value 
$\pi^2 / N_c$, there exists a nontrivial solution.}
\end{figure}

Of course, formula (\ref{gap3D}) is valid only for $T=\mu=0$. If the
system is in contact with a heat bath or with a reservoir of
particles, Eq. (\ref{gap3D}) has to be generalized. To find the
in-medium form of the gap equation we calculate the in-medium part of
the condensate $\langle {\overline q} q\rangle_{\hbox{med}}$.  Similarly to
Eq. (\ref{condvac}) we can write

\begin{equalign}
\label{conmed}
\langle {\overline q} q\rangle_{\hbox{med}} = - {M \over 2}
I_{1,\hbox{med}}(M^2).
\end{equalign}

\noindent In the special cases (i.e., for $T>0 \,\hbox{and}\, \mu=0$ 
or for $T=0\, \hbox{and}\, \mu>0$) 
we have:

\begin{eqnarray}
\label{i1tem}
I_{1, \hbox{tem}}(M^2)  &=&  -16 i N_c
\int\limits_{-i\infty+\epsilon}^{i\infty+\epsilon} {dp^0 \over 2\pi}
\int {d^3 p \over (2\pi)^3} {1 \over p_0^2 -{\bf p}^{\, 2} - M^2}
{1 \over e^{p^0/T} + 1} \nonumber \\
&=& - {4 N_c \over \pi^2} \int\limits_{0}^{\infty}
{dp\, p^2 \over E_p} {1 \over e^{E_p/T} + 1}
\end{eqnarray}
and
\begin{eqnarray}
\label{i1mat}
I_{1, \hbox{mat}}(M^2)  &=&  8i N_c
\oint\limits_{\cal C} {dp^0 \over 2\pi} \int {d^3 p \over (2\pi)^3} 
{1 \over p_0^2 -{\bf p}^{\, 2} - M^2} \nonumber \\
&=&  - {2 N_c \over \pi^2} \int\limits_{0}^{\infty}
{dp\, p^2 \over E_p} \theta(\mu-E_p) =
- {2 N_c \over \pi^2} \int\limits_{0}^{\sqrt{\mu^2-M^2}}
{dp\, p^2 \over E_p}.
\end{eqnarray}
The form of Eqs. (\ref{i1tem}) and (\ref{i1mat}) follows from
the application of the replacement rules (\ref{is}), (\ref{soff2})
and (\ref{soff3}) in the expression on the RHS of Eq. (\ref{i1vac}).

It is important to realize that expressions (\ref{i1tem}) and
(\ref{i1mat}) are well defined converging integrals, and they do not
have to be regularized, which is in contrast to the calculation of the
vacuum part (\ref{i1vac}).  Note that the sum $I_{1, \hbox{vac}}(M^2) +
I_{1, \hbox{med}}(M^2)$ should vanish in the limit $T \rightarrow \infty$ or
$\mu \rightarrow \infty$. This condition follows from the general
property of the fermionic Matsubara sums.  Since we have introduced
the finite cutoff in the calculation of $I_{1, \hbox{vac}} (M^2)$, the
discussed condition may not be fulfilled. For instance, at finite
temperature we find the expression

\begin{equalign}
\label{sumi1}
I^{R, 3D}_{1, \hbox{vac}}(M^2) + I_{1, \hbox{tem}}(M^2) = 
{2 N_c \over \pi^2} \left[
\int_0^{\Lambda} {dp \, p^2 \over E_p} -  \int_0^{\infty}
{dp \, p^2 \over E_p} { 2 \over e^{E_p/T} + 1} \right],
\end{equalign}

\noindent which does not vanish at $T \rightarrow \infty$. However, if
the function $I_{1, \hbox{tem}}(M^2)$ is regularized in the same way
as the function $I_{1, \hbox{vac}}(M^2)$, the sum $I^{R, 3D}_{1,
\hbox{vac}}(M^2)+I^{R, 3D}_{1, \hbox{tem}}(M^2)$ has the desired high
temperature limit.  This example leads us to the simple conclusion:
{\it if the cutoffs used for the regularization of the vacuum parts
are finite, the temperature (matter) parts must be additionally
regularized}.  This ensures the correct high temperature (density)
behaviour of the complete expression.

With these conditions in mind, we write the finite temperature gap
equation in the form

\begin{equalign}
\label{gap3Dt}
M \left[ 1 - {m\over M} \right] = M \, {2 N_c G \over \pi^2}
\int\limits_0^{\Lambda} {dp \, p^2 \over E_p} \left[ 1
- {2 \over e^{E_p/T} + 1} \right].
\end{equalign}

\noindent Since the RHS of Eq. (\ref{gap3Dt}) vanishes
in the limit $T \rightarrow \infty$, the constituent quark mass $M$
drops to the current one $m$ (with increasing $T$). In the special
case $m=0$, there exists a critical temperature $T_c$ such that for $T
> T_c$ formula (\ref{gap3Dt}) has only a trivial solution $M=0$.  This
fact implies the restoration of chiral symmetry at high
temperature. See Fig. [5.3], where the temperature dependence of the
constituent quark mass has been plotted.

\begin{figure}[hb]
\label{q3Dtps}
\xslide{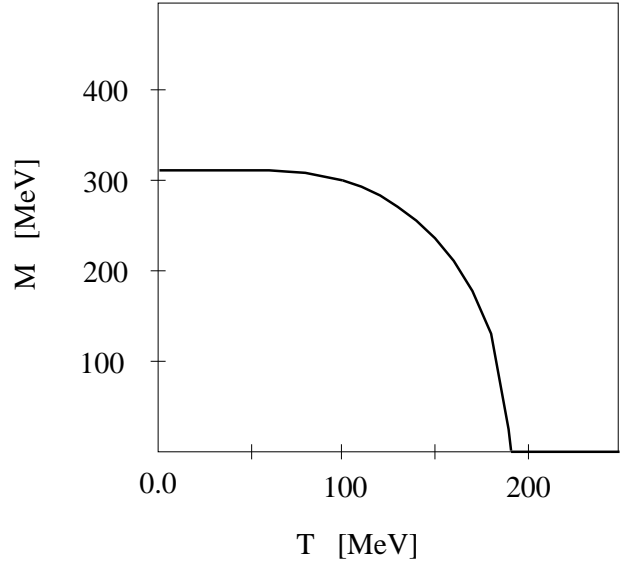}{10cm}{42}{135}{555}{700}
\caption{\small Temperature dependence of the constituent quark mass
$M$ in the limiting case $m=0$.  If temperature is greater than the
critical one (in this case $T_c$ turns out to be 190 MeV) there exists
only a trivial solution to the gap equation (\ref{gap3Dt}), i.e.,
$M=0$ for $T>T_c$. This fact indicates the restoration of chiral
symmetry. In the present calculation the 3-dimensional cutoff
regularization was adopted and we used the following parameters:
$\Lambda$ = 0.65 GeV, $G$ = 0.4 $\hbox{fm}^2$ and $N_c$ = 3 (this gives
$M = M_0$ = 313 MeV at $T=0$).}
\end{figure}

In the analogous way, for finite densities but $T=0$, one finds

\begin{equalign}
\label{gap3Dd1}
M \left[ 1 - {m\over M} \right] = M \, {2 N_c G \over \pi^2}
\int\limits_{p_F}^{\Lambda} {dp \, p^2 \over E_p},
\end{equalign}

\noindent 
where $p_F$ is the Fermi momentum of quarks, i.e., $p_F = \sqrt{\mu^2
- M^2}$. Calculating explicitly the integral on the RHS of expression
(\ref{gap3Dd1}) one finds

\begin{equalign}
\label{gap3Dd2}
M \left[ 1 - {m\over M} \right] & = M \, {N_c G \Lambda^2 \over \pi^2}
\left\{ \sqrt{1+\left({M\over\Lambda}\right)^2 }
-\left({M\over\Lambda}\right)^2 \ln 
{\Lambda + \sqrt{\Lambda^2+M^2} \over M} \right. \\
& \left. - \sqrt{1+\left({M\over p_F}\right)^2 }
+ \left({M\over p_F}\right)^2 \ln 
{p_F + \sqrt{p_F^2+M^2} \over M} \right\}.
\end{equalign}

\noindent Equation (\ref{gap3Dd1}), or equivalently Eq. (\ref{gap3Dd2}),  
leads again to the chiral restoration. See Fig. [5.4], where the
constituent quark mass (being the solution of Eq. (\ref{gap3Dd1}) for
the case $m = 0$) has been plotted as a function of $p_F$. Now the
phase transition takes place at finite baryon density.  Similarly to
the finite-temperature case, there is a critical value $p_F^{\, c}$,
such that for $p_F > p_F^{\, c}$ there is only a trivial solution to
the gap equation.

\begin{figure}[ht]
\label{q3Dmps}
\xslide{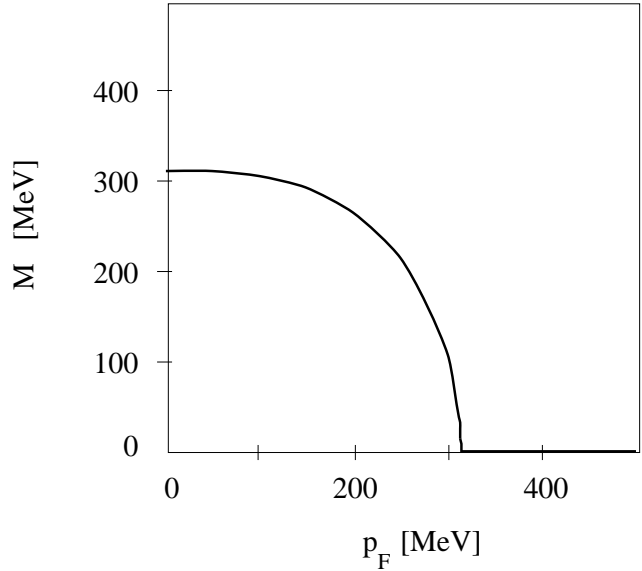}{10cm}{42}{135}{555}{700}
\caption{\small Density dependence of the constituent quark mass
$M$ (for the 3-dimensional cutoff regularization and in the chiral
limit $m=0$).  The mass is plotted as a function of the quark Fermi
momentum $p_F$. In this case, there is also a chiral restoration phase
transition happening at $p_F^{\, c}$ = 314 MeV. The numerical values
of the parameters are the same as in the previous figure.}
\end{figure}

To our knowledge, the first evaluation of the in-medium dependence of
the constituent quark mass in the NJL model was given by Hatsudo and
Kunihiro \cite{HK85}.  Afterwards, many authors repeated this kind of
calculation. The results shown in Figs. [5.3] and [5.4] are obtained
for the set of parameters equivalent to that used in \cite{ZHK94}.
The decrease of $M$ with increasing temperature, see Fig. [5.3],
implies the decrease of the quark condensate (strictly speaking of the
absolute value of $\langle {\overline q} q \rangle$).  We note that, a
similar tendency in the behaviour of $\langle {\overline q} q \rangle$
has been observed in the lattice simulations of QCD as well as in the
calculations based on the chiral perturbation theory. A more
quantitative comparison of the NJL results with other estimates will
be given in Chapter 9.

\section{Two-Flavour Symmetric Case}

We shall consider the two-flavour symmetric model in the
Hartree-Fock approximation. The appropriate form of the Schwinger-Dyson
equation, see Fig. [5.5], is now

\def\gs{G_S} \def\si{\sum_{i=0}^{3}} \def\sp{\int {d^4 p \over (2\pi)^4}}
\def\sofp{G(p) } \def\taui{\tau_i} \def\g5{\gamma_5}
\def\Tr{\hbox{tr }}

\begin{eqnarray}
\label{qse2fsy}
\Sigma & = & \gs \, i \si \sp \left\{ \taui \Tr[\taui \sofp] -
\taui \sofp \taui \vphantom{\sp} \right. \nonumber \\
& & \left. \vphantom{\sp} \mbox{} + i \taui \g5 \Tr[i \taui \g5 \sofp]
- i \taui \g5 \sofp i \taui \g5 \right\}.
\end{eqnarray}

\noindent 
Because the second term with the trace in Eq. (\ref{qse2fsy}) vanishes and
the terms without the trace cancel each other, expression (\ref{qse2fsy})
can be rewritten in a simpler form, namely

\begin{equalign}
\label{qse2s1}
\Sigma = \gs \, i \si \sp \, \taui \, \Tr \, [ \taui \sofp ].
\end{equalign}

\begin{figure}[hb]
\label{qse2}
\xslide{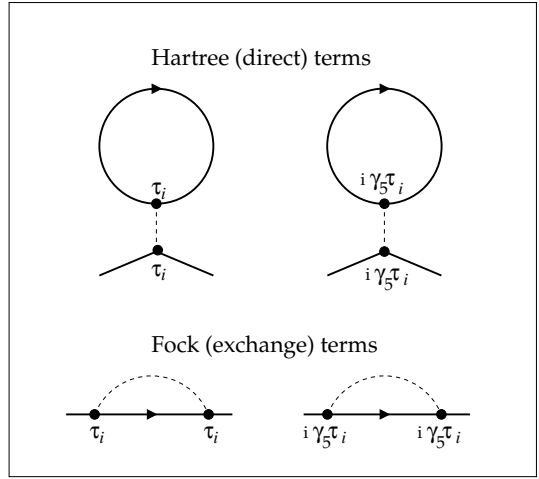}{8cm}{42}{135}{555}{700}
\caption{\small Diagrams contributing to the quark self-energy in the 
         Hartree-Fock approximation (two-flavour symmetric model).
         We note that the introduction of the dashed lines, 
         describing the point-like quark-antiquark interaction,
         allows us to distinguish graphically the Hartree and
         the Fock terms.}
\end{figure}

\newpage
We note that the quark propagator has now extra isospin indices  
(in comparison to the one-flavour model discussed before) and tr
includes now the trace over the isospin space. Consequently, in the
present case we have $2 \langle {\overline q} q \rangle = - i \,
\hbox{tr} \, G(x=0^-)$. This is due to our convention that $\langle
{\overline q} q \rangle$ describes the contribution to the quark
condensate from one flavour only, see Eq. (\ref{cond1}). Nevertheless,
one can check that the form of Eq. (\ref{gap1f}) remains unchanged in
the present case --- the replacements $\langle {\overline q} q \rangle
\rightarrow 2 \langle {\overline q} q \rangle$ and $G \rightarrow
G_S/2$ in (\ref{gap1f}) compensate each other giving $M = m - 2 G_S
\langle {\overline q} q \rangle$.

The 3-dimensional regularization scheme (introduced in the previous 
Section) has several drawbacks, e.g., it is not Lorentz invariant.
Therefore, we shall now discuss another method of regularization. As we
shall later see, it can be regarded as a version of the Pauli-Villars
subtraction scheme. At first, let us consider the function 
$I_{1, \hbox{vac}}(M^2)$ defined by Eq. (\ref{i1vac}). By making the 
Wick rotation to the variable $p_4 = -ip^0$ we can write

\begin{equalign}
\label{i1vacWR}
I_{1, \hbox{vac}}(M^2) = 8N_c \int {d^4 p_E \over (2\pi)^4} 
{1 \over p_E^2 + M^2},
\end{equalign}

\noindent where $d^4p_E = dp_4d^3p$ and $p_E^2 = p_4^2 + {\bf p}^2$.
We replace now $I_{1, \hbox{vac}}(M^2)$ by the series

\begin{equalign}
\label{i1vacRPV}
I_{1, \hbox{vac}}(M^2) \rightarrow I_{1, \hbox{vac}}^{R, PV}(M^2) = 
\sum_{i=0}^N A_i I_{1,\hbox{vac}}(\Lambda_i^2).
\end{equalign}

\noindent Here $A_0 = 1$, $\Lambda_0 = M$ and $N$ is the number of the 
so-called subtractions. The regulating masses $\Lambda_i$ and the
coefficients $A_i$ ($i>0$) should be chosen in such a way as to
provide the finite result for $I_{1, \hbox{vac}}^{R, PV}(M^2)$.  This
requirement leads to the following set of constraints

\begin{equalign}
\label{ai}
\sum_{i=0}^N A_i = 0, \,\,\,\,
\sum_{i=0}^N A_i \Lambda_i^2 = 0, \,\,\,\, ... \,\,\,\, ,
\sum_{i=0}^N A_i \Lambda_i^{2(N-1)} = 0.
\end{equalign}

\noindent Using Eq. (\ref{ai}) we find

\begin{equalign}
\label{i1vacRPV1}
I_{1, \hbox{vac}}^{R, PV}(M^2)={N_c \over 2\pi^2}\sum_{i=0}^N A_i \,
\Lambda_i^2 \, \ln \Lambda_i^2.
\end{equalign}

As it was discussed above, the temperature (matter) parts should be
regularized as well. We do it again by making the replacement

\begin{equalign}
\label{i1medRPV}
I_{1, \hbox{med}}(M^2) \rightarrow I_{1, \hbox{med}}^{R, PV}(M^2) = 
\sum_{i=0}^N A_i I_{1, \hbox{med}}(\Lambda_i^2),
\end{equalign}
where $I_{1, \hbox{med}}(\Lambda_i^2)$ is defined by Eqs. (\ref{i1tem})
and (\ref{i1mat}).

In Figs. [5.6] and [5.7] we show the in-medium dependence of $M$
obtained from the two-flavour symmetric model regularized with the
help of the Pauli-Villars method described above. Of course, these
two figures are analogs of Figs. [5.4] and [5.5] from the previous
Section. In the present case we do not limit ourselves to the limit
$m=0$. As the result we find that $M$ does not drop suddenly to zero,
but it goes smoothly to the value $M=m$ for $T \rightarrow \infty$ or
$\mu \rightarrow \infty$. Moreover, the method of subtractions
``switches off'' large momenta more gradually than the 3-dimensional
cutoff procedure. Thus, the temperature and density dependence of $M$
shown in Figs. [5.6] and [5.7] looks weaker than that shown in
Figs. [5.4] and [5.5]. Nevertheless, the qualitative behaviour of $M$
in the two versions of the model is the same. In particular, 
vanishing of the condensate at large $T$ or $\mu$ in both
cases indicates the chiral symmetry restoration phase transition.

\begin{figure}[hb]
\label{qPVtps}
\xslide{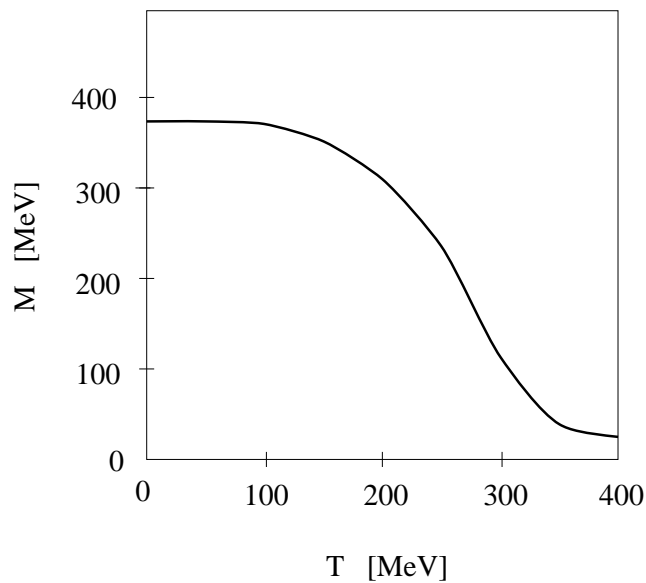}{10cm}{42}{135}{555}{700}
\caption{\small Temperature dependence of the constituent quark mass.
Pauli-Villars regularization scheme (as described in the text) was 
employed here with three regulating masses: $\Lambda_1$ = 0.68 GeV,
$\Lambda_2 = 2.1 \Lambda_1$ and $\Lambda_3 = 2.1 \Lambda_2$.
The coupling constant $G_S$ = 0.75 fm$^2$, and the current quark
mass $m$ = 8.56 MeV.}
\end{figure}

We note that at $T=0$ all values of the chemical potential in the
range $0 < \mu < M$ are equivalent and correspond to the physical
vacuum. The reason is that the chemical potential $\mu$ is the energy
needed to add a quark to the system. Clearly $\mu$ must exceed the
minimal energy, the vacuum rest mass of a quark $M$, before the
Fermi sea of quarks begins to be populated.

\begin{figure}[ht]
\label{qPVmps}
\xslide{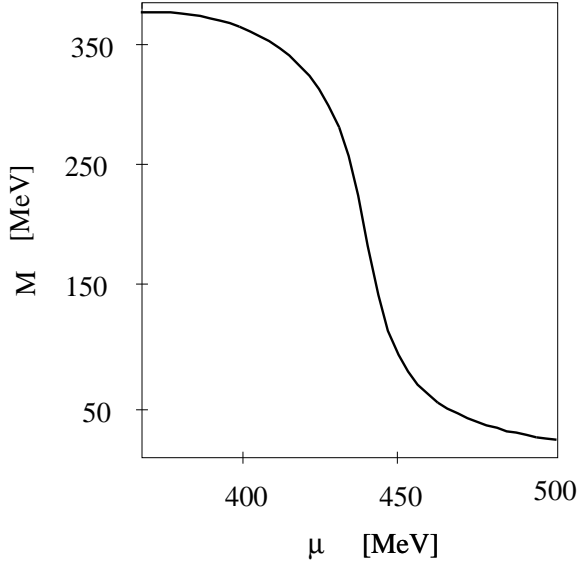}{10cm}{42}{135}{555}{700}
\caption{\small Constituent quark mass as a function of the quark
chemical potential $\mu$. Pauli-Villars regularization procedure
was used here with the same parameters as in Fig. [5.6].}
\end{figure}

\newpage
\section{Two-Flavour Standard Case}

Let us turn to the discussion of the two-flavour standard model.
Calculating the quark self-energy in the Hartree-Fock approximation
one finds

\begin{equalign}
\label{qse2fst}
\Sigma = 2 G \, i \int {d^4p \over (2\pi)^4}
 \left\{ \vphantom{{1\over 2}} \hbox{tr} [\sofp] - \sofp  
+ i \boldtau \g5 \Tr[i \boldtau \g5 \sofp]
- i \boldtau \g5 \sofp i \boldtau \g5 \right\},
\end{equalign}

\noindent 
which after calculation of the trace over spinor indices gives

\begin{equalign}
\label{qse2fst1}
M = m +  4i M  G  \left(2 N_c N_f + 1 \right)  \int {d^4p \over (2\pi)^4}
{1 \over p^2 - M^2 + i\epsilon}.
\end{equalign}

\noindent In contrast to the symmetric version of the model, in this case
the Fock terms (see Fig. [5.8]) do not cancel each other. The extra
``1'' in (\ref{qse2fst1}) comes just from the inclusion of the
exchange terms. Nevertheless, for $N_c=3$ and $N_f=2$ the contribution
from the Fock terms can be regarded as a small correction to the
result obtained only with the Hartree approximation.  Consequently,
one typically neglects the Fock terms arguing that this is allowed in
the leading order of the $1/N_c$ expansion. A more rigorous 
treatment of the model in the $1/N_c$ expansion, which generalizes the
results of this Section, will be presented in Chapter 9.

Neglecting the Fock terms, we find the gap equation in the form

\begin{equalign}
\label{gap2fst}
M = m - 4 G \langle {\overline q} q \rangle = 
m + 2 M G I_{1, \hbox{vac}}(M^2).
\end{equalign}

\noindent In this case, we can again use our definitions (\ref{cond1}) 
and (\ref{condvac}). We note that there is a factor $4G$ in the gap
equation (\ref{gap2fst}), which is in contrast to our two previous
cases where a factor $2G$ appeared.

\begin{figure}[hb]
\label{qse3}
\xslide{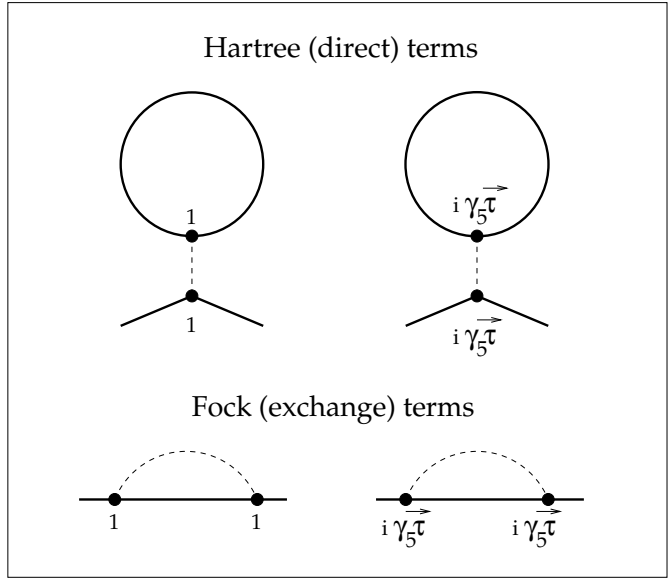}{10cm}{42}{135}{555}{700}
\caption{\small Diagrams used for the calculation of the quark 
         self-energy in the case of the standard two-flavour model.}
\end{figure}

Discussing the standard version of the model, we shall introduce
yet another regularization procedure. This is so-called Schwinger
proper-time method based on the following replacement

\begin{equalign}
\label{i1vacS1}
I_{1, \hbox{vac}}(M^2) = 8N_c \int {d^4 p_E \over (2\pi)^4} 
{1 \over p_E^2 + M^2}
\longrightarrow 8 N_c
\int {d^4 p_E \over (2\pi)^4} \int\limits_{\Lambda^{-2}}^{\infty}
ds \, \exp\left\{-s [p_E^2+M^2]\right\}. 
\end{equalign}

\noindent Introducing the exponential integral function defined as 

\begin{equalign}
\label{expint}
E_n(x) \equiv \int\limits_1^{\infty} dt \, t^{-n}
\exp(-x t),
\end{equalign}
we can write
\begin{equation}
\label{i1vacS2}
I_{1, \hbox{vac}}^{R, SPT}(M^2) = {N_c \Lambda^2 \over 2 \pi^2}
\, E_2\left[{M^2 \over \Lambda^2} \right].
\end{equation}

Using the Schwinger proper-time regularization at finite temperature,
it is not convenient to separate the vacuum and temperature parts.  We
only replace the integral over energy by the sum over the Matsubara
frequencies. According to formula (\ref{is}), we can generalize
Eq. (\ref{i1vacS2}) to the finite temperature case writing

\begin{eqnarray}
\label{I1SPT}
I_{1}^{R, SPT}(M^2,T) 
& = & 8 N_c T \sum_j \int {d^3p \over (2\pi)^3} 
\int\limits_{\Lambda^{-2}}^{\infty} ds \, 
\exp\left\{ -s \left[ (2j+1)^2 \pi^2 T^2
+ {\bf p}^2+M^2 \right] \right\} \nonumber \\
& = & N_c {T \Lambda \over \pi^{{3 \over 2}} } \sum_j
E_{3 \over 2} \left[{{M^2+(2j+1)^2 \pi^2 T^2} \over \Lambda^2} \right].
\end{eqnarray}

\bigskip
\noindent Solving the gap equation (\ref{gap2fst}) with 
$I_{1, \hbox{vac}}(M^2)$ replaced by $I_{1}^{R, SPT}(M^2,T)$ we find
again the characteristic $T$-dependence of $M$, similar to that shown
in Fig. [5.3] or [5.6].  

At this place we close our considerations of the Schwinger-Dyson
equation in the different formulations of the model. The results
derived by us in Sections 5.1 --- 5.3 will be frequently used in the
next Chapters. In particular, we shall return to the discussion of
the two-flavour standard model and the Schwinger proper-time method
in Chapter 9, where the calculations extending the HF approach will
be presented.

\chapter{\bf Mesonic Excitations}
\label{chapt:mesons}

Mesons are quark-antiquark excitations of the true ground state of the
theory. Their quantitative description requires introduction of the
meson correlation functions, as we did it in the end of Chapter 4. In
the NJL model, the overall scheme used to describe mesons
is similar to the standard HF (Hartree-Fock) + RPA (random
phase approximation) approach known from nuclear physics.  The
Hartree-Fock ground state (vacuum) contains the quark condensate. It
is determined through a nontrivial solution of the gap
equation. Mesons, being quark-antiquark excitations of the vacuum,
behave very much like particle-hole states in a strongly interacting
many-body system.

In the first Section of this Chapter we present details of the
calculation of the zeroth-order correlation functions at finite
$T$ and $\mu$. The second Section contains further discussion 
concerning the regularization of the model. In the third Section
we define the full correlation function in the RPA approximation
and discuss the difference between the dynamic and screening masses.
Finally, we show that at sufficiently large $T$ or $\mu$, the dynamic
masses of chiral partners become equal. Such degeneracy is a
characteristic feature of the chiral symmetry restoration phase 
transition.

\section{Zeroth-Order Correlation Functions}

Our investigation of the meson properties in the NJL model starts with
the calculation of the meson correlation functions. At first, using
Eqs. (\ref{mcf}) and (\ref{opo}), we construct the zeroth-order
correlation functions. They correspond to the following quark loop
diagram \footnote{\it Conventions introduced in this Chapter are
suitable for the two-flavour symmetric model defined in Sections 3.2
and 5.2.  Nevertheless, as we have seen in Chapter 5, changing $N_f$
and/or $G_S$ one can easily obtain expressions whose form is valid for
other versions of the model.}

\begin{equalign}
\label{gcf1}
{\chi}^{(0) \, ij}_{AB}(Q) = i \, \hbox{tr} \int {d^4p \over (2\pi)^4}
[\Gamma_A \tau^i G(p+Q) \Gamma_B \tau^j G(p) ].
\end{equalign}

\begin{figure}[ht]
\label{cfps}
\xslide{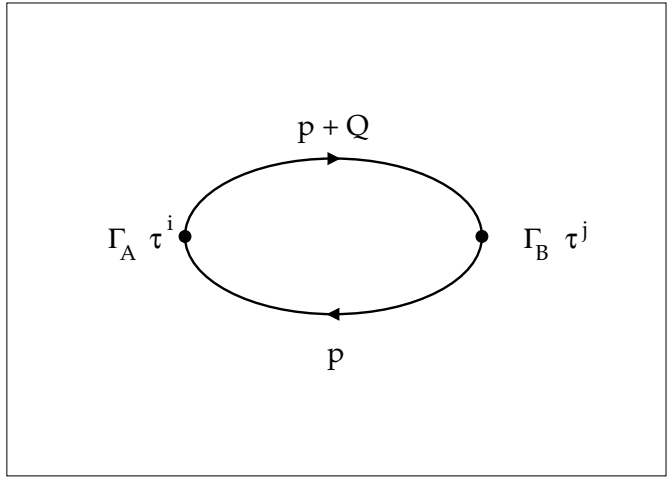}{10cm}{42}{135}{555}{700}
\caption{\small Quark loop diagram corresponding to the zeroth-order
                meson correlation function.}
\end{figure}

\noindent Here $Q^{\mu}=(\omega, {\bf q}\,)$ is the external (meson)
momentum. The indices $A$ and $B$ take on the values $P$ or $S$ and
$\Gamma_P$, $\Gamma_S$ are the Dirac matrices in the spinor space:
$\Gamma_P=i\g5$ and $\Gamma_S=1$. We note that the indices $a$ and $b$
in Eq. (\ref{opo}) correspond to two pairs of indices in Eq. (\ref{gcf1}),
namely $a = (A,i)$ and  $b = (B,j)$, see also Fig. [\ref{cfps}].  

In the isospin symmetric case the matrix ${\chi}^{(0) \, ij}_{AB}$ is 
diagonal in flavour indices and its all diagonal elements are the same.  
Consequently, we can write ${\chi}^{(0) \, ij}_{AB} = \delta^{ij} 
{\chi}^{(0)}_{AB}$, where

\begin{equalign}
\label{gcf2}
{\chi}^{(0)}_{AB}(Q) = 2 i N_c \, \hbox{Sp} \int {d^4p \over (2\pi)^4}
\left[ \Gamma_A G(p+Q) \Gamma_B G(p) \right].
\end{equalign}

\noindent Calculation of the trace over the Dirac indices, denoted
here by Sp, shows that the nondiagonal terms $(A \neq B)$ vanish.
Therefore,  we are left with two non-vanishing functions

\begin{equalign}
\label{gcfpp}
{\chi}^{(0)}_{PP}(Q) = 8 i N_c \, \int {d^4p \over (2\pi)^4}
{p^2 - M^2 + p \cdot Q \over 
[(p+Q)^2-M^2+i\epsilon \,][p^2-M^2+i\epsilon\,]}
\end{equalign}
and
\begin{equalign}
\label{gcfss}
{\chi}^{(0)}_{SS}(Q) = 8 i N_c \, \int {d^4p \over (2\pi)^4}
{p^2 + M^2 + p \cdot Q \over 
[(p+Q)^2 - M^2+i\epsilon \, ][p^2-M^2+i\epsilon \,]}.
\end{equalign}

\subsection{Vacuum Parts}

Using the imaginary-time formalism, discussed in detail in Chapter 4,
we generalize expressions (\ref{gcfpp}) and (\ref{gcfss}) to the case
of finite temperature (density). To do it, we replace the integration
over energy in Eqs. (\ref{gcfpp}) and (\ref{gcfss}) by the Matsubara
sums (\ref{is}), and convert these sums to the contour integrals using
expressions (\ref{soff2}) and (\ref{soff3}). In this case, the vacuum
parts of the correlation functions can be represented in a compact
form

\begin{equalign}
\label{ch0pv}
\chi^{(0)}_{PP, \hbox{vac}}(Q) = 
I_{1, \hbox{vac}}(M^2)-Q^2 I_{2, \hbox{vac}}(M^2,Q^2)
\end{equalign}
and
\begin{equalign}
\label{ch0sv}
\chi^{(0)}_{SS, \hbox{vac}}(Q) = I_{1, \hbox{vac}}(M^2)
-(Q^2-4M^2) I_{2, \hbox{vac}}(M^2,Q^2).
\end{equalign}

\noindent The function $I_{1, \hbox{vac}}(M^2)$, appearing in Eqs. 
(\ref{ch0pv}) and (\ref{ch0sv}), has been already defined by formula
(\ref{i1vac}).  The new function $I_{2, \hbox{vac}}(M^2,Q^2)$ is
defined by the integral

\begin{equalign}
\label{i2vac}
I_{2, \hbox{vac}}(M^2,-q^2_E) = -4N_c \int {d^4 p_E \over (2\pi)^4} {1 \over
[(p_E+q_E/2)^2+M^2][(p_E-q_E/2)^2+M^2]},
\end{equalign}

\noindent where $p_4 = -ip^0$, $q_4 = -i\omega$ and $d^4 p_E =
dp_4 d^3p$. We note that frequency $Q^0$ is purely imaginary now ($Q^0
= \omega \rightarrow iq_4$), thus the ``Euclidean'' form of expression 
(\ref{i2vac}) is natural here.

\subsection{Temperature Parts}

The form of the temperature part of $\chi^{(0)}_{PP}(Q)$ follows
from Eqs. (\ref{gcfpp}) and (\ref{soff2})

\begin{eqnarray}
\label{ch0pt}
& & \chi^{(0)}_{PP, \hbox{tem}}(Q)  =  -8 i N_c
\int\limits_{-i\infty+\epsilon}^{i\infty+\epsilon} {dp^0 \over 2\pi}
\int {d^3 p \over (2\pi)^3} 
\left[ {1 \over (p+Q)^2-M^2 } + {1 \over (p-Q)^2-M^2 } \right]
{1 \over e^{p^0/T} + 1} \nonumber \\ & &-8 i N_c \!\!\!\!
\int\limits_{-i\infty+\epsilon}^{i\infty+\epsilon} 
\!\!\! {dp^0 \over 2\pi}
\!\!\! \int {d^3 p \over (2\pi)^3}
\left[ {p \cdot Q \over [(p\!+\!Q)^2-\!M^2]  [p^2\!-\!M^2]  } 
- {p \cdot Q \over [(p\!-\!Q)^2-\!M^2] [p^2\!-\!M^2] } \right]
{1 \over e^{p^0/T} + 1}. \nonumber \\
\end{eqnarray}

\noindent Using again the fact that $Q^0$ is the bosonic Matsubara
frequency ($Q^0 = \omega \rightarrow i\omega_m = 2m\pi i T$), we make
the shifts of the integration variable $p^0$, and simplify expression
(\ref{ch0pt}) to the form \footnote{At this place we write $\omega =
i\omega_m$ rather than $\omega = iq_4$ in order to point out the
discrete character of the complex variable $\omega$. This fact allows
us to make the shifts of the argument of the exponent.}

\begin{eqnarray}
\label{ch0pt1}
& & \chi^{(0)}_{PP, \hbox{tem}}(Q)  =  - 16 i N_c
\int\limits_{-i\infty+\epsilon}^{i\infty+\epsilon} {dp^0 \over 2\pi}
\int {d^3 p \over (2\pi)^3} 
{1 \over p^2-M^2 } {1 \over e^{p^0/T} + 1} \nonumber \\ 
& & + 8 i N_c Q^2 \!\!\!
\int\limits_{-i\infty+\epsilon}^{i\infty+\epsilon} {dp^0 \over 2\pi}
\int {d^3 p \over (2\pi)^3}
{1 \over [(p+Q/2)^2-M^2]  [(p-Q/2)^2-M^2] }
{1 \over e^{(p^0+i\omega_m/2)/T} + 1}. \nonumber \\
\end{eqnarray}

\noindent The last result indicates that the analogous decomposition 
to that find in vacuum, see Eq. (\ref{ch0pv}), holds also at finite
temperature, namely

\begin{equalign}
\label{ch0pt2}
\chi^{(0)}_{PP, \hbox{tem}}(Q) = I_{1, \hbox{tem}}(M^2)-Q^2 
I_{2, \hbox{tem}}(M^2,Q),
\end{equalign}

\noindent where $I_{1, \hbox{tem}}(M^2)$ is defined by  Eq. (\ref{i1tem}) 
and the function $I_{2, \hbox{tem}}(M^2,Q)$ has the form

\begin{equalign}
\label{i2tem}
I_{2, \hbox{tem}}(M^2,Q) = - 8 i N_c \!\!\!\!
\int\limits_{-i\infty+\epsilon}^{i\infty+\epsilon} 
\!\!\! {dp^0 \over 2\pi}
\!\!\! \int {d^3 p \over (2\pi)^3}
{1 \over [(p\!+\!Q/2)^2\!-\!M^2][(p\!-\!Q/2)^2\!-\!M^2] }
{1 \over e^{(p^0+i\omega_m/2)/T}\!+\!1}.
\end{equalign}

\noindent Since $I_{2, \hbox{tem}}(M^2,Q)$ is a function of  $\omega^2$ and 
$q^2$, we shall also use the notation $I_{2, \hbox{tem}}(M^2,\omega^2,q^2)$.
In the special cases (i.e., either for $q=0$ or $\omega=0$) one finds

\begin{equalign}
\label{i2tem2}
I_{2, \hbox{tem}}(M^2,\omega^2,0) = {N_c \over \pi^2}
\int\limits_{0}^{\infty}
{dp \, p^2 \over E_p} {1 \over e^{E_p /T} + 1}
{1 \over E_p^2-{1 \over 4} \omega^2} 
\end{equalign}
and
\begin{equalign}
\label{i2tem3}
I_{2, \hbox{tem}}(M^2,0,q^2) = -{N_c \over q\pi^2} \int\limits_{0}^{\infty}
{dp \, p \over E_p} {1 \over e^{E_p /T} + 1}
\ln\left|{2p-q \over 2p+q}\right|.
\end{equalign}

\noindent We note that Eq. (\ref{i2tem2}) has been obtained for
a purely imaginary frequency $\omega = i\omega_m$. Thus, the
expression on the RHS of (\ref{i2tem2}) is a well defined
quantity. Furthermore, according to the rules discussed by us in
Chapter 4, the result (\ref{i2tem2}) can be analytically continued to
real frequencies. This can be achieved by making the substitution
$i\omega_m \rightarrow \omega \pm i \epsilon$, where $\omega$ is
$real$.

In the similar way one does the calculations in the scalar channel
finding the following decomposition

\begin{equalign}
\label{ch0st}
\chi^{(0)}_{SS, \hbox{tem}}(Q) = I_{1, \hbox{tem}}(M^2)-(Q^2-4 M^2) 
I_{2, \hbox{tem}}(M^2,Q),
\end{equalign} 

\noindent which is the analog of Eq. (\ref{ch0sv}).

\subsection{Matter Parts}

In order to calculate the matter part of the correlation function
$\chi^{(0)}_{PP}(Q)$, we use Eqs. (\ref{gcfpp}) and (\ref{soff3}),
which yield

\begin{equalign}
\label{ch0pm}
\chi^{(0)}_{PP, \hbox{mat}}(Q)  =  8 i N_c
\int\limits_{\cal C} {dp^0 \over 2\pi}
\int {d^3 p \over (2\pi)^3} \left[ {1 \over (p+Q)^2-M^2 } 
+ {p \cdot Q \over [(p\!+\!Q)^2-\!M^2]  [p^2\!-\!M^2]  } \right].
\end{equalign}

\noindent Making the shifts of the integration variable $p^0$,
one can check that the decomposition of the form (\ref{ch0pv}) or
(\ref{ch0pt2}) is valid at finite density. Indeed, after a few
manipulations we obtain

\begin{equalign}
\label{ch0pm1}
\chi^{(0)}_{PP, \hbox{mat}}(Q) = I_{1, \hbox{mat}}(M^2)-Q^2 
I_{2, \hbox{mat}}(M^2,Q),
\end{equalign}

\noindent where $I_{1, \hbox{mat}}(M^2)$ is defined by  Eq. (\ref{i1mat}) 
and the function $I_{2, \hbox{mat}}(M^2,Q)$ is given by expression

\begin{equalign}
\label{i2mat}
I_{2, \hbox{mat}}(M^2,Q) = 4 i N_c \int\limits_{\cal C} 
{dp^0 \over 2\pi} \int {d^3 p \over (2\pi)^3}
{1 \over [(p + Q/2)^2 - M^2][(p - Q/2)^2 - M^2] }.
\end{equalign}

\noindent
The calculation of $I_{2, \hbox{mat}}(M^2,\omega^2,q^2)$ proceeds in
the same way as the calculation of the function $I_{1,
\hbox{mat}}(M^2)$. Now, as remarked above, the integral over $p^0$ is
evaluated for imaginary $\omega$ and subsequently analytically
continued to real frequencies. The final results (for $\omega >$ 0)
are

\begin{eqnarray}
\label{i2mat2}
I_{2, \hbox{mat}}(M^2,\omega^2 \pm i\epsilon,0) &=& {N_c \over 2\pi^2}
\int\limits_{0}^{p_F} {dp \, p^2 \over E_p} 
{1 \over E_p^2-{1 \over 4} (\omega \pm i\epsilon)^2} \nonumber \\
&=& {N_c \over 2\pi^2}\left[\log\left(\frac{p_F+E_F}{M}\right)+
{\cal F}(M^2,\omega^2\pm
i\epsilon) \right]
\end{eqnarray}

\noindent and

\begin{eqnarray}
\label{i2mat3}
I_{2, \hbox{mat}}(M^2,0,q^2) &=& -{N_c \over 2q\pi^2} 
\int\limits_{0}^{p_F} {dp \, p \over E_p} 
\ln\left|{2p-q \over 2p+q}\right|  \nonumber \\
&=& -{N_c \over 2q\pi^2}{\cal G}(M^2,q^2),
\end{eqnarray}

\noindent where $E_F = \sqrt{p_F^2+M^2}$,

\begin{eqnarray}
\label{calf}
\!\!\!\!\!\!{\cal F}(M^2,\omega^2\pm i\epsilon)
&=&-\frac{1}{\omega}\left[\sqrt{4 M^2 - \omega^2} \,\,\arctan
\left(\frac{p_F\omega}{E_F \sqrt{4 M^2 -
\omega^2}}\right)\right] \,\,\,\, (\omega < 2 M) \nonumber \\
\!\!\!\!\!\!&=&-\frac{1}{\omega}\sqrt{\omega^2-4 M^2}\,\,
\ln\left[\frac{E_F\sqrt{\omega^2-4 M^2}+p_F\omega}
{E_F\sqrt{\omega^2-4 M^2}-p_F\omega}\right] \nonumber \\
\!\!\!\!\!\!& & \pm \,i\, {\pi \over 2 \omega}
\theta\left(\omega-2M\right) \theta\left(2E_F-\omega\right)
 \sqrt{\omega^2-4 M^2} \,\,\,\,\,\,\,\,\,\,\,\, (\omega > 2 M) 
\nonumber \\
\!\!\!\!\!\!&&
\end{eqnarray}

\noindent and

\begin{eqnarray}
\label{calg}
{\cal G}(M^2,q^2)&=& \left(\frac{\sqrt{q^2 +
4 M^2}}{2}-E_F\right)\ln\left|\frac{2 p_F+q}{2 p_F-q}\right|
\nonumber \\
& &+\frac{\sqrt{q^2+4 M^2}}{2}\,\ln\left[\frac{2 M^2 + q p_F+
E_F\sqrt{q^2+4M^2}}{2 M^2 - q p_F +E_F\sqrt{q^2+4 M^2}}\right]
\nonumber \\
& &-q\ln\left[\frac{p_F+E_F}{M}\right].
\end{eqnarray}

\noindent It becomes clear now, that for the scalar channel we find 
the analog of Eq. (\ref{ch0st}), namely  

\begin{equalign}
\label{ch0sm1}
\chi^{(0)}_{SS, \hbox{mat}}(Q) = I_{1, \hbox{mat}}(M^2)-(Q^2-4 M^2) 
I_{2, \hbox{mat}}(M^2,Q).
\end{equalign} 

\section{Regularization of the Zeroth-Order Correlation Functions}

In the last Section, we have reduced the problem of the evaluation of
zeroth order correlation functions to the calculation of the functions
$I_1 \equiv I_{1, \hbox{vac}} + I_{1, \hbox{med}}$ and $I_2 \equiv
I_{2, \hbox{vac}} + I_{2,
\hbox{med}}$. Nevertheless, one can easily notice that the function
$I_{2,\hbox{vac}}(M^2,-q_E^2)$ is defined by the divergent integral
(\ref{i2vac}).  Moreover, although the functions
$I_{2,\hbox{med}}(M^2,\omega^2,q^2)$ are finite, they should be
regularized in the same way as the vacuum functions. In this way, see
discussion in Section 5.1 following Eq.  (\ref{sumi1}), we ensure the
correct high temperature (density) behaviour of the complete
expression $I_2 = I_{2, \hbox{vac}} + I_{2, \hbox{med}}$. Thus, in the
next three Subsections, we shall present three different ways of the
regularization of the function $I_2(M^2,\omega^2,q^2)$. We shall use
the methods introduced in Chapter 5.

\subsection{3-Dimensional Regularization}

At first, let us discuss the simplest method, which is based on the
introduction of a 3-dimensional cutoff. Similarly as it was done
in Section 5.1, we perform integration over energy in
Eq. (\ref{i2vac}).  Setting ${\bf q}$=0 and collecting the
contributions from the poles one finds

\begin{equalign}
\label{i2vac1}
I_{2,\hbox{vac}}(M^2,-q_4^2) = - {N_c \over 2 \pi^2} 
\int {dp \,p^2 \over E_p} {1 \over E_p^2 + {1 \over 4} q_4^2 }.
\end{equalign}

\noindent Doing analytic continuation of expression (\ref{i2vac1}), 
$q_4 \rightarrow -i(\omega \pm i\epsilon)$ , and introducing the 
cutoff for p-integration we find ($\omega > 0$)

\begin{equalign}
\label{i2vacR3D}
I^{R, 3D}_{2,\hbox{vac}}(M^2,\omega^2 \pm i\epsilon) = 
- {N_c \over 2 \pi^2} \int\limits_0^{\Lambda} {dp \,p^2 \over E_p}
{1 \over E_p^2 - {1 \over 4}(\omega \pm i\epsilon) ^2 }.
\end{equalign}
This integral can be calculated in the same way as
expression (\ref{i2mat2}) --- there is only an overall difference in
sign between these two integrals and $p_F$ in (\ref{i2mat2})
corresponds to $\Lambda$ in (\ref{i2vacR3D}).

The final formula for the regularized function $I_2 (M^2,\omega^2 \pm 
i\epsilon,0)$ is 
\begin{equalign}
\label{i2vacR3D1}
I^{R, 3D}_2(M^2,\omega^2 \pm i\epsilon,0) = 
- {N_c \over 2 \pi^2} \int\limits_0^{\Lambda} {dp \,p^2 \over E_p}
\left[1 - {2 \over e^{E_p/T} + 1} \right]
{1 \over E_p^2 - {1 \over 4}(\omega \pm i\epsilon) ^2 }
\end{equalign}
for finite temperature ($\mu=0$), and
\begin{equalign}
\label{i2vacR3D2}
I^{R, 3D}_2(M^2,\omega^2 \pm i\epsilon,0) = 
- {N_c \over 2 \pi^2} \int\limits_{p_F}^{\Lambda} 
{dp \,p^2 \over E_p}
{1 \over E_p^2 - {1 \over 4}(\omega \pm i\epsilon) ^2 }
\end{equalign}
for finite density ($T=0$). We note that integral (\ref{i2vacR3D2})
also can be done analytically, similarly as Eqs. (\ref{i2mat2}) or
(\ref{i2vacR3D}).

Doing calculations for $q_4 = 0$ and ${\bf q} \not = 0$ one finds
the analogs of (\ref{i2vacR3D1}) and (\ref{i2vacR3D2}), namely

\begin{equalign}
\label{i2vacR3D3}
I^{R, 3D}_2(M^2,0,q^2) = 
{N_c \over 2 \pi^2 q} \int\limits_0^{\Lambda} {dp \,p \over E_p}
\left[1 - {2 \over e^{E_p/T} + 1} \right]
\ln\left|{2p - q \over 2p + q}\right|
\end{equalign}
and
\begin{equalign}
\label{i2vacR3D4}
I^{R, 3D}_2(M^2,0,q^2) = 
{N_c \over 2 \pi^2 q} \int\limits_{p_F}^{\Lambda} 
{dp \,p \over E_p}
\ln\left|{2p - q \over 2p + q}\right|.
\end{equalign} 

Although the 3-dimensional cutoff regularization is frequently used,
for many purposes it is not satisfactory since it explicitly breaks 
the Lorentz invariance. We have collected the results based on this
regularization in order to facilitate the comparison of different
formulations of the model. Nevertheless, the major part of our further 
calculations will employ the covariant Pauli-Villars method.

\subsection{Pauli-Villars Subtraction Scheme}

In this Subsection we shall calculate the function $I_2(M^2,Q) =
I_{2,\hbox{vac}}(M^2,Q^2) + I_{2,\hbox{med}}(M^2,Q)$ using the Pauli-Villars
regularization scheme. This calculation proceeds in the similar way as
the evaluation of the function $I_1^{R,PV} (M^2)$ in Section
5.2. The regularized form of (\ref{i2vac}) is obtained by making the
substitution

\begin{equalign}
\label{i2vrPV}
I_{2, \hbox{vac}}(M^2,-q_E^2) 
\rightarrow I_{2, \hbox{vac}}^{R, PV}(M^2,-q_E^2) = \sum_{i=0}^N
A_i I_{2, \hbox{vac}}(\Lambda_i^2,-q_E^2),
\end{equalign}

\noindent where the coefficients $A_i$ are defined by formula (\ref{ai}).
To calculate $I_{2, \hbox{vac}}^{R,PV}(M^2,-q_E^2)$ one has to use first the
identity

\begin{equalign}
\label{ab}
{1 \over AB} = 2 \int\limits_{-1}^{1} du { 1 \over [A(1+u)+B(1-u)]^2},
\end{equalign}

\noindent which allows us to simplify the form of the denominator in
the integral on the RHS of Eq. (\ref{i2vac}).  Afterwards exchanging
the order of integration over $p_E$ and $u$ and introducing the new
variable $p_E^{\prime}= p_E^{} + u q_E^{}/2$ one obtains the
expressions which can be easily integrated. The final result is

\def\x{{q_E \over 2 \Lambda_i}}
\begin{equalign}
\label{i2vrPV1}
I_{2, \hbox{vac}}^{R, PV}(M^2,-q_E^2)={N_c \over 2 \pi^2}\sum_{i=0}^N A_i
\left[ {2\Lambda_i \over q_E} \sqrt{1+\left(\x\right)^2}
\ln \left(\sqrt{1+\left(\x\right)^2}+\x \right)
+\ln \Lambda_i \right].
\end{equalign}

\noindent The function $I^{R, PV}_{2, \hbox{vac}}(M^2,\omega^2)$ can be 
obtained by performing the analytic continuation of the expression
given on the RHS of Eq. (\ref{i2vrPV1}).  The substitution $q
\rightarrow -i(\omega \pm i\epsilon)$ leads to the 
following formula $(\omega > 0)$

\def\y{\left({\omega \over 2 \Lambda_i }\right)}
\begin{eqnarray}
\label{i2vrPV2}
&&I_{2, \hbox{vac}}^{R, PV}(M^2,\omega^2 \pm i \epsilon) =  
{N_c \over 2 \pi^2}
\sum_{i=0}^N A_i \left\{ \Theta(2\Lambda_i - \omega) \left[
{2\Lambda_i \over \omega} \sqrt{1\!-\!\y^2} \hbox{arcsin}\y \right.
\right.
\nonumber \\
&& + \left. \left. \vphantom{\sqrt{\y}} \ln\Lambda_i \right]
 +\Theta(\omega\!-\!2\Lambda_i) \left[{2\Lambda_i \over \omega}
\sqrt{\y^2\!\!\!-\!\!1} \left( \hbox{arcosh}\y \mp {i\pi \over 2}
\right) + \ln\Lambda_i \right] \right\}.
\end{eqnarray}

\noindent The functions $I_{2, \hbox{med}}(M^2,Q)$ should be calculated 
in the same way as the function $I_{1, \hbox{med}}(M^2)$, namely we make 
the replacement

\begin{equalign}
\label{i2medPV}
I_{2, \hbox{med}}(M^2,Q) \rightarrow I_{2, \hbox{med}}^{R, PV}(M^2,Q) = 
\sum_{i=0}^N A_i I_{2, \hbox{med}}(\Lambda_i^2,Q),
\end{equalign}
where the function $I_{2, \hbox{med}}(\Lambda_i^2,Q)$ is defined by Eqs.
(\ref{i2tem}) or (\ref{i2mat}).

\subsection{Schwinger Proper-Time Regularization}

Using the Schwinger proper-time method of regularization, we rewrite
Eq. (\ref{i2vac}) in the following form

\begin{eqnarray}
\label{f0r}
I_{2, \hbox{vac}}^{R, SPT}(M^2,-q_E^2) 
&=& -4 N_c \int {d^4p_E \over (2\pi)^4} \int\limits_{\Lambda^{-2}}^{\infty}
ds \, s \int\limits_0^1 du \exp\left\{-s [p_E^2 + M^2 + u(1-u) q_E^2] \right\} 
 \nonumber \\
&=& - {N_c \over 4 \pi^2} \int\limits_0^1 du \, E_1\left[{M^2 \over \Lambda^2}
+ u (1-u) {q_E^2 \over \Lambda^2} \right].
\end{eqnarray}

\noindent Replacing the integral over energy, $p_4$, by the sum over
Matsubara frequencies we can generalize Eq. (\ref{f0r}) to the case
of finite temperatures

\begin{eqnarray}
\label{ftr}
\!\!\!\!\!\!\!\!\!\!\!\!\!\!\!& &I_2^{R, SPT}(M^2,n,q,T) = 
-4 N_c T \sum_j \int {d^3p \over (2\pi)^3} 
\int_{\Lambda^{-2}}^{\infty} ds \, s \int\limits_0^1 du \times 
\nonumber \\ 
\!\!\!\!\!\!\!\!\!\!\!\!& & \exp\left\{ -s \left[ 
M^2 + u(1-u)((E^b_n)^2 + {\bf q}^2) + \left[ {\bf p}
- {\bf q} (1-u) \right]^2 + \left[E^f_j - E^b_n (1-u) \right]^2
\right] \right\}  \nonumber \\ 
\!\!\!\!\!\!\!\!\!\!\!\!& &=-{N_c T \over 2 \pi^{{3 \over 2}} \Lambda} \sum_j
\int\limits_0^1 du \, E_{1\over 2}\left[{M^2 \over \Lambda^2} +
u(1-u) {(E^b_n)^2 + {\bf q}^2 \over \Lambda^2} \right. 
  +  \left. {[E^f_j-E^b_n(1-u)]^2 \over \Lambda^2} \right] \;.
\end{eqnarray}
where $E^b_n = 2\pi n T$ and $E^f_j = (2j+1)\pi T$.

\section{Meson Masses}
\bigskip
\subsection{Definitions}
\bigskip

In the random phase approximation (RPA) the full correlation function
has the form
\begin{equalign}
\label{cf1}
{\chi}_{AA}(Q) =
{ {\chi}^{(0)}_{AA}(Q) \over 1 - G_S {\chi}^{(0)}_{AA}(Q) }.
\end{equalign}
\noindent This is a nonperturbative expression, which follows from
the infinite summation of a certain class of diagrams, see Fig. [6.2].

\begin{figure}[ht]
\label{rpaps}
\xslide{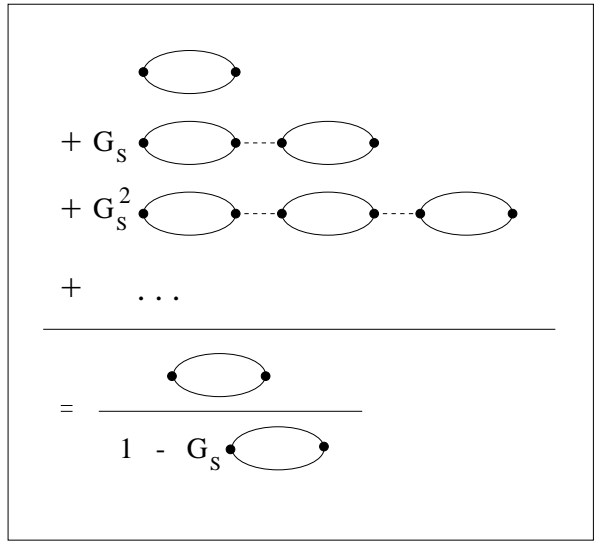}{8cm}{8}{130}{585}{712}
\caption{\small The class of diagrams used to evaluate the full
                correlation function in the RPA approximation.}
\end{figure}

The correlation function $\chi_{AA}(Q)$ depends on $Q$ through the
variables $\omega^2$ and $q^2$. Therefore, in the following we shall
use the notation $\chi_{AA}(\omega^2,q^2)$ rather than $\chi_{AA}(Q)$.
The {\it dynamic mass} is defined by the position of the nearest
to zero pole of the function $\chi_{AA}(\omega^2,0)$. In the
NJL model it can be easily found by solving the equation

\begin{equalign}
\label{dm}
1 - G_S {\chi}^{(0)}_{AA}(m_{dyn}^2,0) = 0.
\end{equalign}

\noindent This comes directly from Eq. (\ref{cf1}). On the other
hand, the {\it screening mass} is defined by the asymptotic behaviour
of the static correlation function in space, namely

\begin{equalign}
\label{sm}
m_{scr} = - \lim_{r \rightarrow \infty}
{ d \ln {\chi}_{AA}(r) \over dr},
\end{equalign}

\noindent where

\begin{equalign}
\label{ft}
\chi_{AA}(r) = \int {d^3 q \over (2\pi)^3} \,\, \chi_{AA}(0,q^2) \,\,
e^{i {\bf q} \cdot {\bf r}} =
{1 \over 4 \pi^2 i r} \int\limits_{-\infty}^{+\infty} dq \, q
\,\, \chi_{AA}(0,q^2)\,\, e^{iqr}.
\end{equalign}

\noindent To get the last term in Eq. (\ref{ft}) we first performed
the integration over the angles, and then extended the region of
integration over $q = |\bf q \,|$ to the interval from minus to plus
infinity.

Let us now consider an arbitrary correlation function $\chi_{AA}
(\omega^2,q^2)$, not necessarily obtained from the NJL model. One can
ask the question when the dynamic mass is equal to the screening
mass. At $T=0$ the system is explicitly Lorentz invariant and
$\chi_{AA}(\omega^2,0)=\chi_{AA}(0,-\omega^2)$. Therefore, if the
function $\chi_{AA}(\omega^2,0)$ has a pole for $\omega=m_{dyn}$,
the function $\chi_{AA}(0,q^2)$ has a pole for $q=im_{dyn}$.  The
latter gives the contribution to the integral (\ref{ft}), which has
the form $\sim \exp(-m_{dyn}r)$. For very large $r$ this contribution
is the dominant one, since $im_{dyn}$ is the nearest pole.
Consequently, we find that $m_{dyn}=m_{scr}$. Of course one has to be
careful because the analytic structure of the correlation function can
be complicated and there might be some other contributions to the
integral (\ref{ft}). Nevertheless at $T=0$ we expect that
$m_{dyn}=m_{scr}$. A different situation takes place when $T>0$.  In
this case it is difficult to find a simple relation between the
functions $\chi_{AA}(\omega^2,0)$ and $\chi_{AA}(0,q^2)$. The Lorentz
invariance is implicitly broken by the existence of the preferable
reference frame connected with the heat bath. The fact that $\chi_{AA}
(\omega^2,0)$ has a pole for $\omega=m_{dyn}$ does not imply that
$\chi_{AA}(0,q^2)$ has a pole for $q=im_{dyn}$. Furthermore,
at $T > 0$ the contribution to Eq. (\ref{ft}) from the cuts of
the correlation function may be important and  it can substantially change
the space asymptotics of the function $\chi_{AA}(r)$.  In consequence,
at $T > 0$, because of these at least two reasons, the dynamic mass
and the screening one are different. Of course, the same arguments
hold for the case of finite density.

\newpage
\subsection{In-Medium Meson Dynamic Masses}

The dynamic masses of the pion and the sigma are obtained from
Eq. (\ref{dm}), which has the form

\begin{eqnarray}
\label{pidm}
& & 1 - G_S \left[ I_1(M^2) -  m^2_{dyn,\pi} 
I_2(M^2,m^2_{dyn,\pi},0) \right]  \nonumber \\
& & = {m \over M} + m^2_{dyn,\pi} \, G_S 
\, I_2(M^2,m^2_{dyn,\pi},0) = 0
\end{eqnarray}

\noindent for the pseudoscalar (pion) channel, and

\begin{eqnarray}
\label{sidm}
& & 1 - G_S \left[ I_1(M^2) - (m^2_{dyn,\sigma}-4M^2)
I_2(M^2,m^2_{dyn,\sigma},0) \right]  \nonumber \\
& & = {m \over M} + (m^2_{dyn,\sigma}-4M^2) \,
G_S \, I_2(M^2,m^2_{dyn,\sigma},0) = 0
\end{eqnarray}

\noindent for the scalar (sigma) channel. In Eqs. (\ref{pidm}) and
(\ref{sidm}), we have assumed that the parameters are such that the 
gap equation has a non-trivial solution, i.e., $M = m + G_S M I_1(M^2)$. 

{\it Formula (\ref{pidm}) is the cornerstone of the NJL model. It
explicitly shows, that the spontaneous symmetry breaking leads to the
appearance of the Goldstone bosons}. If the current quark mass is
zero, $m=0$, the pion dynamic mass turns out to be zero. Moreover, at
the Hartree level the mass of the sigma is simply $2M$. We also note
that the structure of Eqs.  (\ref{pidm}) and (\ref{sidm}) is
independent of the regularization scheme (as long as our
regularization procedure modifies the functions $I_1$ and $I_2$ only.)

\begin{figure}[ht]
\label{dmtps}
\xslide{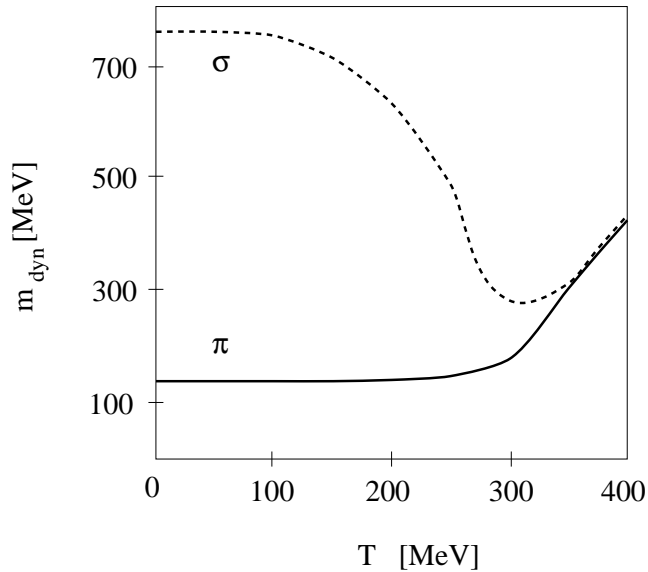}{10cm}{42}{135}{555}{700}
\caption{\small Temperature dependence of the dynamic masses of
                pion (solid line) and sigma (dashed line).}
\end{figure}

We note that Eqs. (\ref{pidm}) and (\ref{sidm}) are correct not only
in vacuum but also at finite temperature or density. As
follows from Eqs. (\ref{i2vrPV2}), (\ref{i2tem2}) and (\ref{i2mat2}),
the correlation functions have cuts for arguments larger than $2M$. At
high temperatures (densities) the meson poles merge with the
quark-antiquark cuts, which means that in this model the mesons can
decay into $q\bar{q}$ pairs. Thus, at these temperatures (densities)
there are no isolated poles which can be identified with the mass of
the pion or the sigma. We circumvent this difficulty by defining the
mass as the zero of the real part of Eq. (\ref{dm}), i.e.,

\begin{equation}
\label{dmm}
1 - G_S \, \hbox{Re} \, {\chi}^{(0)}_{AA}(m_{dyn}^2,0) = 0.
\end{equation}

\noindent We note that by using Eq. (\ref{dmm}) we implicitly neglect
the $q\bar{q}$ widths of the mesons \footnote{In a general situation,
the mass and the width of a meson are determined by the shape of its
spectral function. In our case the latter has the form $ ( 2 \, \hbox{Im}
\, {\chi}^{(0)}) / (1 - G_S\,{\chi}^{(0)}) = (2\,G_S\,\hbox{Im} {\chi}^{(0)})
/ (\,[1 - G_S\,\hbox{Re} {\chi}^{(0)} ]^2 + [\,G_S\,\hbox{Im} {\chi}^{(0)}
]^2)$.  If the imaginary part of ${\chi}^{(0)}$ (describing the width
of a meson) is small, the spectral function is strongly peaked for the
four-momenta satisfying the condition (\ref{dmm}).}. This seems
reasonable, since these widths are non-zero only because the NJL model
lacks confinement.

Our result concerning the temperature dependence of the dynamic masses
of mesons is shown in Fig. [6.3]. The calculations are based on the
two-flavour symmetric model. We use now the same computational scheme
as that used to calculate the temperature dependence of $M$ in Section
5.2, i.e., we make three subtractions with the regulating masses:
$\Lambda_1 = 680$ MeV, $\Lambda_2 = 2.1 \Lambda_1$ and $\Lambda_3 = 2.1 
\Lambda_2$.  The coupling constant $G_S = 0.75 \,\hbox{fm}^2$ and the 
current quark mass $m = 8.56$ MeV. Using these values of the
parameters we find that at zero temperature the constituent quark mass
$M = M_0 = 376$ MeV, the pion mass $m_{dyn,\pi} = 138$ MeV and the
sigma mass $m_{dyn,\sigma} = 760$ MeV. The pion and sigma masses are
connected by relation 
\begin{equalign}
\label{psm}
m^2_{dyn,\sigma} \approx 4M^2 + m^2_{dyn,\pi}.
\end{equalign}
In Fig. [6.3] one can see that the pion mass remains constant as long
as the temperature is smaller than 250 MeV.  Afterwards it increases
suddenly. The temperature dependence of the sigma mass is rather
complicated. In the interval $0<T<250$ MeV it behaves in the similar
way as the constituent quark mass, see Fig. [5.6]. Moreover, in this
region relation (\ref{psm}) is still fulfilled.  When the temperature
reaches 300 MeV, the sigma mass stops decreasing, remains for a
while constant and later increases.  At very high temperature the
sigma mass and the pion mass are with a good approximation the
same. {\it This fact signifies the chiral restoration.}

In Fig. [6.4] we show the density dependence of the dynamic masses of
mesons. We use here the same values of the parameters as in the finite
temperature calculations discussed above.  Thus, our results for $\mu
= 0$ and $T = 0$ coincide with each other.  One can see that with the
increasing value of the chemical potential the dynamic masses of pion
and sigma become degenerate. This fact indicates restoration of the
chiral symmetry at high density. The overall $\mu$-dependence shown in
Fig. [6.4] is similar to the $T$-dependence shown in Fig. [6.3].

The first calculation of the in-medium dynamic masses of the pion and
the sigma in the NJL model was done by Hatsuda and Kunihiro \cite{HK85}.
Their calculation employed the 3-dimensional regularization scheme.
The results presented in Figs. [6.3] and [6.4] follow from the calculation
based on the Pauli-Villars method \cite{FF94a}. They qualitatively agree
with those obtained in  \cite{HK85}. The advantage of using the
Pauli-Villars method in our case is that it will allow us to compare
directly the temperature dependence of the dynamic and screening masses.
As follows from the discussion in the next Chapters, the 3-dimensional 
regularization scheme introduces unphysical contributions to the Fourier
transform defining the static correlation function. Thus, this way
of regularization spoils the expected relations between the screening 
and dynamic masses.

\begin{figure}[hb]
\label{dmmps}
\xslide{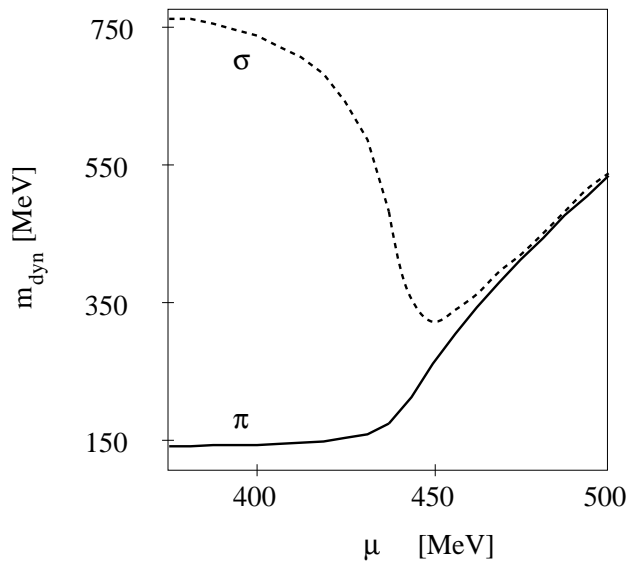}{10cm}{42}{135}{555}{700}
\caption{\small Pion (solid line) and sigma (dashed line) dynamic masses
         plotted as functions of the quark chemical potential $\mu$.}
\end{figure}

\chapter{\bf Static Meson Fields at Finite Temperature}

In this Chapter we study the damping of static meson correlation
functions at finite temperature. Our interest in these investigations
has been triggered by the results of the lattice simulations of
QCD. They indicate that high-temperature hadronic correlation
functions describe a system of (almost) non-interacting quarks.  In
the first Section of this Chapter we do the analytic calculations
based on the perturbative QCD. This allows us to understand the high
temperature behaviour of the correlation functions.  In the second
Section we come back to the study of the NJL model.  We find that 
the screening masses obtained within the model have the desired
temperature dependence. In particular, for stable mesons the screening
masses at $T=0$ are equal to the dynamic ones, whereas for extremely
high temperatures they approach the values expected from the
perturbative QCD calculation.

\section{ Screening of Meson Fields in Hot QCD}

Lattice ``measurements'' of hadronic correlation functions in QCD show
exponential behaviour at large distances \cite{DTK87,GLRST87,GRH88,
BGIKLPS91}. At high temperatures, $T>T_c$, the hadronic screening
masses of chiral partners are equal, but non-vanishing. This is
consistent with the restoration of chiral symmetry.  Eletskii and
Ioffe \cite{EI88} pointed out that the meson screening masses
approached the value $2 \pi T$, whereas those of baryons were close to
$3 \pi T$ and argued that such a behaviour is typical for a gas of
non-interacting massless quarks.  Their reasoning was the following:
In the considered case the space-time correlation function is the
product of two (mesons) or three (baryons) quark propagators.  The
space asymptotics of each propagator is determined by the lowest lying
pole in the complex $q = |\bf q \, |$ plane. In the imaginary-time
formalism this pole corresponds to the lowest fermionic Matsubara
frequency $\pi T$ and, therefore, the screening masses are multiples
of $\pi T$. Shuryak has presented alternative arguments, based on the 
large-distance behaviour of the free quark propagator at finite
temperatures, which lead to the same result \cite{Shuryak93}.

We should emphasize that the screening masses obtained from the lattice
calculations are not exactly the multiples of $\pi T$. For baryons
and (axial)vector mesons the relation of Eletskii and Ioffe is indeed
very well fulfilled. However, for (pseudo)scalar mesons one can see
a deviation from $2 \pi T$. This discrepancy indicates that there is
a non-negligible interaction in this channel, present even at very 
high temperature. Nevertheless, in the limit $T \rightarrow \infty$
we expect that in all meson channels the screening masses would go to
$2 \pi T$. 

The arguments of Eletskii and Ioffe allow us to understand intuitively
how the screening masses turn out to be $2\pi T$ and $3 \pi T$.  We
are of the opinion, however, that it is instructive to obtain this
result from a more rigorous calculation.  Actually, the aim of this
Section is to show that starting from the regularized momentum
representation of the static meson correlation function we can
analytically derive its full spatial dependence \cite{FF94b}.  Knowing
the asymptotic behaviour of the correlation function, we can find the
values of the screening masses which are in agreement with the result
of Eletskii and Ioffe.

The spatial dependence of the correlation function will be calculated
by us in two ways, namely

\begin{equalign}
\label{plane}
{\tilde \chi}(z) \equiv \int_{-\infty}^{\infty}
{dq_z \over 2\pi} \, {\tilde \chi}(0,q^2_z) \, e^{i q_z z}
\end{equalign}
or
\begin{equalign}
\label{point}
{\tilde \chi}(r) \equiv
\int {d^3 q \over (2\pi)^3} \, {\tilde \chi}(0,q^2) \, 
e^{i {\bf q} \cdot {\bf r}} =
{1 \over 4\pi^2 ir} \int\limits_{-\infty}^{\infty}\!\! dq \,q \,
{\tilde \chi}(0,q^2) \, e^{iqr}.
\end{equalign}

\noindent  The quantity ${\tilde \chi}(0,q^2)$ denotes the static meson 
correlation function obtained in the framework of the perturbative
QCD. We shall restrict our considerations to the special case of the
pseudoscalar channel. Eq. (\ref{plane}) describes the situation in
which the correlation function is used to find the response of the
system to the {\it static plane-like perturbation}.  Here $z$ is the
distance from the plane in the transverse direction and $q_z$ is
identified with the momentum in this direction. Such a case is
considered in the lattice simulations.  Eq. (\ref{point}), equivalent
to the definition (\ref{ft}), describes the situation when the
perturbation is the {\it static point}.  We note that the functions
${\tilde \chi}(z)$ and ${\tilde \chi}(r)$ depend differently on their
arguments, however, for simplicity of notation we use for them the
same symbol.

In the perturbative QCD, the leading contribution to the meson
correlation function is given by the quark loop diagram, see
Fig. [6.1] (the diagrams with gluon lines contribute only in higher
order terms). Thus, we can use our results from the previous Chapters.
In particular, if we employ the Pauli-Villars regularization
procedure, we can use expressions (\ref{ch0pv}) and (\ref{ch0pt2})
with the functions $I_{1,\hbox{vac}}(M^2)$, $I_{1,\hbox{tem}}(M^2)$,
$I_{2,\hbox{vac}} (M^2,-q^2)$ and $I_{2,\hbox{tem}}(M^2,0,q^2)$ given
by expressions (\ref{i1vacRPV1}), (\ref{i1tem}), (\ref{i2vrPV1}) and
(\ref{i2tem3}), respectively.

In practice, we do not use Eqs. (\ref{plane}) and (\ref{point})
directly in the form as they have been written down.  To evaluate the
Fourier transform, in both cases, we analytically continue ${\tilde
\chi}(0,q^2)$ on the whole complex $q$ plane. Later we deform the
contour of the integration, and represent the result as the integral
around the singularities appearing in the upper half-plane.  The
Fourier integral (\ref{plane}) or (\ref{point}) becomes in this case
the sum of two terms (the first one represents the integral over the
vacuum part and the second one is the integral over the temperature
part).

\newpage
\subsection{Vacuum Part}

The vacuum part of the correlation functions ${\tilde
\chi}_{\hbox{vac}}(0,q^2)$ is given by expression

\def\x{{q \over 2 \Lambda_i}}
\begin{equalign}
\label{vpq1}
{\tilde \chi}_{\hbox{vac}}(0,q^2) = {N_c \over 2 \pi^2}\sum_{i=0}^N A_i
\left\{ \Lambda_i^2 \ln \Lambda_i^2 + q^2
\left[ {2\Lambda_i \over q} \sqrt{1\!+\!\left(\x\right)^2}
\ln \left(\sqrt{1\!+\!\left(\x\right)^2}+\x \right)
\!+\!\ln \Lambda_i \right] \right\},
\end{equalign}

\noindent which follows from Eqs. (\ref{ch0pv}), (\ref{i1vacRPV1}) and
(\ref{i2vrPV1}). The analytic continuation of (\ref{vpq1}) has cuts
for purely imaginary $q=ik$, starting at $k=\pm 2M$, see Fig. [7.1].
Consequently, the Fourier integral can be replaced by the integral
around the cut stretching along the imaginary axis. In this way we
obtain

\begin{equalign}
\label{vpz1}
{\tilde \chi}_{\hbox{vac}}(z) =  {i \over 2\pi}
\int\limits_{2M}^{\infty}\!\! dk  
\left[{\tilde \chi}_{\hbox{vac}}(0,-k^2+i\epsilon)
-{\tilde \chi}_{\hbox{vac}}(0,-k^2-i\epsilon) \right]e^{-kz}.
\end{equalign}

\begin{figure}[hb]
\label{qcd1ps}
\xslide{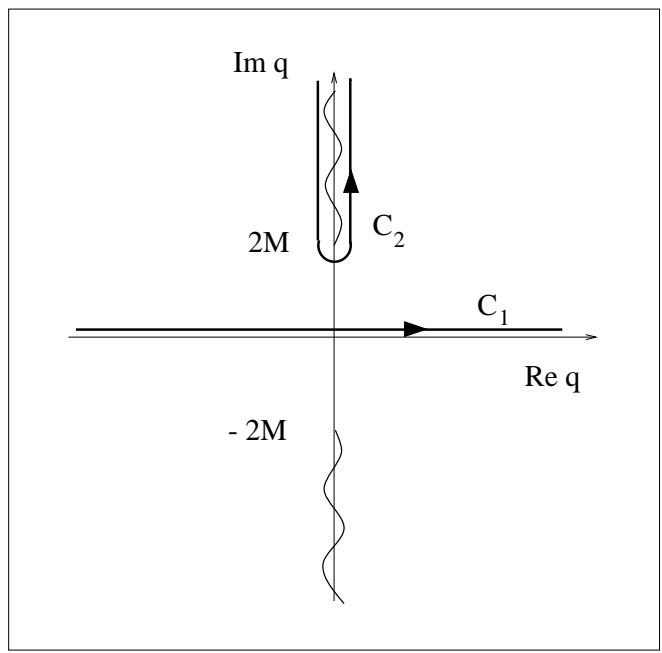}{10cm}{42}{135}{555}{700}
\caption{\small The cuts of the vacuum part of the correlation function.
    They start at $\hbox{Im}\, q = \pm 2 M$ and go along the imaginary
    axis to $\pm \infty$. The integrations along the curves ${\cal C}_1$
    and ${\cal C}_2$ are equivalent.}
\end{figure}

\noindent The real parts of 
${\tilde \chi}_{\hbox{vac}}(0,-k^2 \pm i\epsilon)$ are the same on
both sides of the cut, and the imaginary parts have opposite
signs. Thus, the expression in the square brackets on the RHS of
(\ref{vpz1}) equals $2i \,\hbox{Im} \, {\tilde
\chi}_{\hbox{vac}}(0,-k^2 +i\epsilon)$. As the final result one gets

\begin{equalign}
\label{vpz2}
{\tilde \chi}_{\hbox{vac}}(z) = {N_c \over 4\pi^2} \sum_{i=0}^N \, A_i \,
\int\limits_{2\Lambda_i}^{\infty} \!\!dk \, k
\sqrt{k^2-4\Lambda_i^2} e^{-kz}
 =  {N_c \over \pi^2 z} \sum_{i=0}^N A_i \, \Lambda_i^2 \,
K_2(2\Lambda_i z),
\end{equalign}

\noindent where $K_2$ is the modified Bessel function \cite{AS72}.  
The vacuum part of the correlation function ${\tilde \chi}_{\hbox{vac}}
(0,q^2)$ is divergent and must be regularized. Nevertheless, the
Fourier transform remains finite for $z > 0$ also when one sends the
cutoff masses to infinity. For $\Lambda_i \rightarrow \infty \,\,
(i>0)$, only the first term in the above series is non-negligible and
we find

\begin{equalign}
\label{vpz3}
{\tilde \chi}_{\hbox{vac}}(z) = {N_c \over \pi^2 z} M^2 K_2(2Mz).
\end{equalign}

\noindent Similar calculations starting from Eq. (\ref{point})
lead to the result

\begin{equalign}
\label{vpr1}
{\tilde \chi}_{\hbox{vac}}(r) = {N_c \over 2 \pi^3 r^3} M^2 
\left[ 3K_2(2Mr) +
2MrK_1(2Mr) \right].
\end{equalign}

\subsection{Temperature Part}

The temperature part of the correlation function is given by expression

\begin{equalign}
\label{mpq2}
{\tilde \chi}_{\hbox{tem}}(0,q^2)  = -{N_c \over \pi^2} 
\int\limits_{0}^{\infty} {dp
\, p^2 \over E_p} {1 \over e^{E_p /T} + 1} \left[4 + {q\over
p}\ln\left|{2p-q \over 2p+q}\right|\right],
\end{equalign}
which follows from Eqs. (\ref{ch0pt2}), (\ref{i1tem}) and (\ref{i2tem3}).
In the following we shall make the substitution

\begin{equalign}
\label{log}
\ln\left|{2p-q \over 2p+q}\right| =
{1 \over 2}\ln{ (2p-q)^2 + \epsilon^2 \over
(2p+q)^2+ \epsilon^2 }.
\end{equalign}

\noindent 
In this way ${\tilde \chi}_{\hbox{tem}}(0,q^2)$ becomes an analytic
function of $q$. Of course, at the final stage of the calculations we
have to consider the limiting case $\epsilon \rightarrow 0$.

The spatial dependence of the temperature part of the correlation
function is obtained from Eqs. (\ref{plane}) and (\ref{point}).
Exchanging the order of the integration over $q$ and $p$ we get

\begin{equalign}
\label{mpz1}
{\tilde \chi}_{\hbox{tem}}(z) = - {N_c \over 2 \pi^3} 
\int\limits_{0}^{\infty} {dp \, p^2 \over E_p} {F(p,z) \over 
e^{E_p /T} + 1},
\end{equalign}
where the function $F(p,z)$ is defined through the integral

\begin{eqnarray}
\label{fp1}
F(p,z) & \equiv & \int_{-\infty}^{\infty} dq
\left[4 + {q\over 2p}\ln{ (2p-q)^2 + \epsilon^2 \over
(2p+q)^2+ \epsilon^2 }\right]e^{iqz} \nonumber \\
& = & {\pi \over p z^2}
\left[ e^{-2ipz} \left(i - 2pz \right)
     - e^{2ipz}  \left(i + 2pz \right)
\right].
\end{eqnarray}

\noindent The calculation of $F(p,z)$ proceeds as follows. First we
analytically continue the integrand on the upper half-plane of complex
$q$.  Such a function has two cuts starting at $q=\pm 2p + i\epsilon$
and going upward, parallel the imaginary axis (see Fig. [7.2]).
Consequently, the initial integral can be replaced by the two
integrals around the cuts.  The values of the integrated function on
both sides of the cuts are different only because the logarithm has a
different phase there.  However the difference of the phase, the
quantity which enters our integrals, is constant. This simplifies our
expressions and leads, in the limit $\epsilon \rightarrow 0$, to the
result shown in the second line of Eq. (\ref{fp1}).

\begin{figure}[hb]
\label{qcd3ps}
\xslide{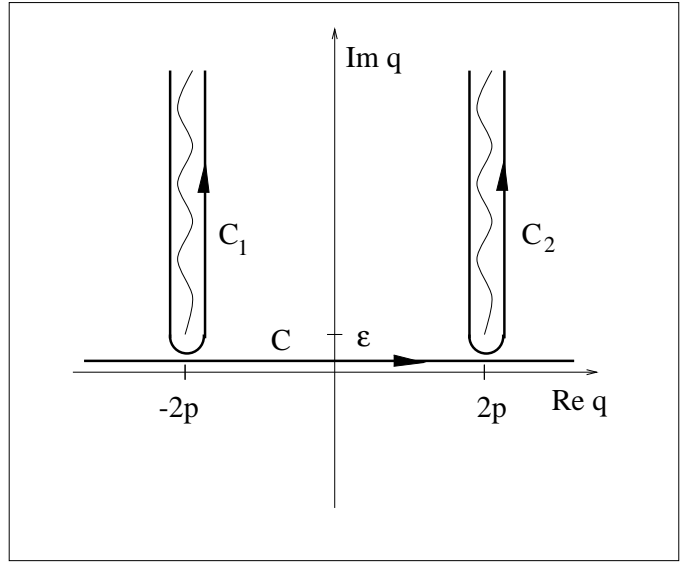}{10cm}{42}{135}{555}{700}
\caption{\small The analytic structure of the function being the 
                integrand in Eq. (\ref{fp1}). The integration along the
                real axis (contour $\cal C$) can be replaced by the two
                integrals around the cuts (contours ${\cal C}_1$ and
                ${\cal C}_2$).}
\end{figure}

Substitution of Eq. (\ref{fp1}) into Eq. (\ref{mpz1}) leads to the
following result

\begin{equalign}
\label{mpz2}
{\tilde \chi}_{\hbox{tem}}(z) = {N_c \over 2 \pi^2 } \left( {1 \over z^2}
- {1 \over z} {\partial \over
\partial z} \right) {\partial \over \partial z}
\int\limits_{0}^{\infty} {dp \over E_p}
{1  \over e^{E_p/T} + 1} \cos(2pz).
\end{equalign}

\noindent Now, using the series representation of the
Fermi-Dirac distribution

\begin{equalign}
\label{fd}
{1 \over e^x + 1} = {1 \over 2} - \sum_{l=-\infty}^{+\infty}
{x \over (2l+1)^2 \pi^2 + x^2},
\end{equalign}

\noindent we can write the temperature part in the form

\begin{equalign}
\label{mpz3}
{\tilde \chi}_{\hbox{tem}}(z) = {N_c \over 2\pi^2} \left( {1 \over z^2} -
{1 \over z} {\partial \over \partial z} \right) \left[ 
{\tilde G}^{\hbox{tem}}_1(z) + {\tilde G}^{\hbox{tem}}_2(z) \right]
\end{equalign}

\noindent where

\begin{equalign}
\label{tg1}
{\tilde G}^{\hbox{tem}}_1(z) \equiv {\partial \over \partial z} 
\int_0^{\infty} {dp \over
2 E_p} \cos(2pz) = {1 \over 2} {\partial \over \partial z}
K_0(2Mz) = -M K_1(2Mz),
\end{equalign}

\noindent and

\begin{eqnarray}
\label{tg2}
{\tilde G}^{\hbox{tem}}_2(z) & \equiv & - T {\partial \over \partial z}
\sum_{l=-\infty}^{+\infty} \int_0^{\infty} {dp \, \cos(2pz) \over
(2l+1)^2 \pi^2 T^2 + E_p^2}
=-i T \sum_{l=-\infty}^{+\infty} \int_{-\infty}^{\infty}
{dp \, p \, e^{2ipz} \over (2l+1)^2 \pi^2 T^2 + E_p^2}
\nonumber \\
& = & \pi T \sum_{l=-\infty}^{+\infty} \exp(-2z\sqrt{ (2l+1)^2 \pi^2
T^2 + M^2}).
\end{eqnarray}

In the 3-dimensional case one can derive an analogous expression

\begin{equalign}
\label{mpr1}
{\tilde \chi}_{\hbox{tem}}(r) = {N_c \over 4\pi^3 r} \left( {2 \over r^3} -
{2 \over r^2} {\partial \over \partial r} +
{1 \over r} {\partial^2 \over \partial r^2}
\right) \left[ {\tilde G}^{\hbox{tem}}_1(r)
+ {\tilde G}^{\hbox{tem}}_2(r) \right].
\end{equalign}

\subsection{Asymptotics}

Using Eqs. (\ref{vpz3}) and (\ref{mpz3}), and also employing the
recursion relations for the Bessel functions \cite{GR80} we can check
that in the full correlation function the vacuum part and the first
piece of the temperature part (coming from the differentiation of the
function ${\tilde G}^{\hbox{tem}}_1$) cancel each other.  \footnote{Such
cancellation takes place only because the vacuum part was calculated
in the limit $\Lambda_i \rightarrow \infty \, (i>0)$. If we kept
$\Lambda_i$'s finite, we would have to regularize the temperature part
as well. Otherwise, we could not get the correct result for the
screening masses.} We are left only with the term coming from the
differentiation of the function ${\tilde G}^{\hbox{tem}}_2$. Moreover, for
large distances, it is enough to consider only the leading terms in
the series which defines ${\tilde G}^{\hbox{tem}}_2$ ($l=0$ and
$l=-1$). Consequently, we find for large $z$ and finite $M$ that

\begin{equalign}
\label{cfz}
{\tilde \chi}(z) \sim {2 N_c T \over \pi z} \sqrt{\pi^2 T^2 + M^2}
\exp(-2z\sqrt{\pi^2 T^2 + M^2}).
\end{equalign}

\noindent From Eq. (\ref{cfz}) we read off the screening mass
$m_{scr} = 2\sqrt{\pi^2 T^2 + M^2}$. The same screening mass is
obtained if we use Eqs. (\ref{vpr1}) and (\ref{mpr1}).

Let us also discuss in more detail the case of massless quarks. The
expression for ${\tilde \chi}_{\hbox{tem}}(z)$ or ${\tilde
\chi}_{\hbox{tem}}(r)$ can be then given in a compact form because the
series appearing in the definition of ${\tilde G}^{\hbox{tem}}_2$
becomes a geometric one and can be easily summed up.  Using Eq.
(\ref{mpz3}) we obtain

\begin{equalign}
\label{mpz4}
{\tilde \chi}_{\hbox{tem}}(z) \stackrel{M \rightarrow 0}{\longrightarrow}
- {N_c \over 2 \pi^2 z^3}
+ {N_c T \over 2\pi z^2 \hbox{sinh}(2\pi Tz) }
\left[1 + 2 \pi T z \hbox{coth}(2 \pi T z) \right],
\end{equalign}

\noindent and similarly from Eq. (\ref{mpr1}) we get

\begin{eqnarray}
\label{mpr2}
{\tilde \chi}_{\hbox{tem}}(r) & \stackrel{M \rightarrow 0}{\longrightarrow} &
- {3 N_c \over 4 \pi^3 r^5}
+ {N_c T \over 2\pi^2 r^4 \hbox{sinh}(2\pi Tr) }
\left[
\vphantom{{1 \over 2^2}}
1 + 2\pi Tr \, \hbox{coth}(2\pi Tr)
\right. \nonumber \\
& & \left.
+ 2\pi^2 T^2 r^2 \left( {2 \over \hbox{sinh}^2 (2\pi Tr)}
+1\right)
\right].
\end{eqnarray}

\noindent Eqs. (\ref{vpz3}) and (\ref{vpr1}) in the limiting case
$M=0$ have the form

\begin{equalign}
\label{vpz4}
{\tilde \chi}_{\hbox{vac}}(z) \stackrel{M \rightarrow 0}{\longrightarrow}
{N_c \over 2 \pi^2 z^3 }
\end{equalign}
and
\begin{equalign}
\label{vpr2}
{\tilde \chi}_{\hbox{vac}}(r) \stackrel{M \rightarrow 0}{\longrightarrow}
{3 N_c \over 4 \pi^3 r^5}.
\end{equalign}

\noindent In this case we can easily see that the temperature parts
of the correlation functions vanish if $T=0$. On the other hand, in the
expression for the complete correlation function, for each finite
value of $T$, the temperature independent piece of the temperature
part cancels exactly the vacuum part.

\subsection{Remarks on the Temporal Function}

Finally, let us  briefly discuss the screening in the temporal
direction. In this case we need to evaluate the correlation
function in the imaginary time direction

\begin{equalign}
\label{temp1}
{\tilde \chi}^{(4)}(\tau) = T \sum_n e^{-i \omega_n \tau}
{\tilde \chi}^R(i \omega_n,0),
\end{equalign}
where 
\begin{equalign}
\label{temp2}
{\tilde \chi}^R(\omega,0) = {8 N_c \over \pi^2} 
\int_0^{\infty} dp \, p^2 \, \hbox{tanh}
\left({E_p \over 2T} \right) {E_p \over 4 E_p^2 - \omega^2}
+ (\hbox{reg.})
\end{equalign}
is the regularized correlation function (``reg.'' denotes the 
regularization terms).

Using standard techniques we find
\begin{equalign}
\label{temp3}
{\tilde \chi}^{(4)}(\tau) = {N_c \over \pi^2}
\int_0^{\infty} dp \, p^2 \,
{ \hbox{cosh} \left[2 E_p (\tau - \beta/2) \right] \over
 \hbox{cosh}^2 (E_p \beta /2) } + (\hbox{reg.}),
\end{equalign}
where $\beta = 1/T$. Since the integral converges for $0 <
\tau < \beta$ we can again drop  the regulating terms by letting
$\Lambda_i \rightarrow \infty$. For massless quarks, the integral
can be done analytically
\begin{equalign}
\label{temp4}
{\tilde \chi}^{(4)}(\tau) = {N_c T^3 \over 2 \pi^2}
{\partial^2 \over \partial (\tau/\beta)^2 }
\left(1 - {2 \tau \over \beta} \right) 
{\pi \over \sin(2 \pi \tau / \beta) }.
\end{equalign}
The real time correlation function is obtained by analytically
continuing to real times; for large $t$ we find ${\tilde
\chi}^{(4)}(t) \sim \exp(-2 \pi T t)$. Thus, for massless
non-interacting quarks, the mesonic screening mass in the time
direction is the same as in the spatial direction, $m_{scr} = 2 \pi
T$. Consequently, we expect the temporal correlation function
in mesonic channels to behave qualitatively in the following 
manner: For a uniform source at zero temperature, where mesonic
modes exist, ${\tilde \chi}^{(4)}(t)$ oscillates in time with
frequency $m$, the meson mass. On the other hand, at high
temperatures, where the system consists of weakly interacting
almost massless quarks, the correlation function is screened,
with the screening mass $2 \pi T$.

\section{Screening in the NJL Model}

In this Section we analyze the spatial dependence of static meson
correlation functions calculated within the NJL model.  We use again
the two-flavour symmetric model defined in Sections 3.2 and 5.2. Thus,
our considerations are restricted to the scalar and pseudoscalar
channels. We find that the screening masses (obtained from the
asymptotic behaviour of the correlation functions) differ from the
dynamic masses (defined by a pole of the meson propagator in the energy
plane, see discussions in Section 6.3). In the high temperature limit,
the screening masses approach $2 \pi T$, which corresponds to a gas of
non-interacting quarks. Nevertheless, the interaction effects
remain well beyond the chiral transition temperature. The overall
temperature dependence of the screening masses is in the agreement
with lattice results.

\subsection{Analytic Structure in Complex Momentum Space}

Similarly to the QCD case considered in the previous Section, the
Fourier transform (\ref{ft}) can be rewritten as a sum of a few
contributions, which appear due to the singularities of the correlation
function in the complex $q$-plane. Thus, at first we are going to
study the analytic structure of the correlation functions.

The results of Section 6.1 indicate that $\chi^{\,(0)}_{AA,
\hbox{vac}}(0,q^2)$ has cuts (in the following we refer to these as the {\it
vacuum cuts}) along the imaginary axis starting at $q=\pm 2iM$ and
going to $\pm i \infty$. In order to find the analytic structure of
the function $\chi^{\,(0)}_{AA, \hbox{tem}} (0,q^2)$ we study the function 
$I_{2, \hbox{tem}} (M^2,0,q^2)$ in the whole complex $q=|\bf q|$ plane. 
To this end we represent $I_{2, \hbox{tem}} (M^2,0,q^2)$ as a sum of two 
functions, namely

\begin{equation}
\label{i2m4}
I_{2, \hbox{tem}}(M^2,0,q^2) = I_{2, \hbox{tem}}^{\,(+)}(q) + 
I_{2, \hbox{tem}}^{\,(-)}(q),
\end{equation}
where
\begin{equation}
\label{i2mpm}
I_{2, \hbox{tem}}^{\,(\pm)}(z) = -{N_c \over 2z\pi^2}
\int\limits_{0}^{\infty}
{dp \, p \over E_p} {1 \over e^{E_p/T} + 1}
\ln{\,\,\,2p-z \pm i\epsilon \over -2p-z \pm i \epsilon}.
\end{equation}

\noindent At the end of the calculations we let the infinitesimal
$\epsilon$ go to zero.  The functions $I_{2, \hbox{tem}}^{\,(\pm)}(z)$ have
logarithmic cuts parallel to the real axis and stretching from minus
to plus infinity.  At the respective cut the imaginary part of the
function $I_{2,\hbox{tem}}^{\,(\pm)}(z)$ is discontinuous. On the physical
Riemann sheet, this amounts to a change in sign:
\begin{equation}
\label{iplus2}
\hbox{Im}\,I_{2, \hbox{tem}}^{\,(+)}(q_R+i\delta_\pm) = \mp {N_c T
\over 2 q_R \pi} \ln \left[ 1 + \exp\left(
- {\sqrt{M^2 + q^2_R/4} \over T} \right) \right].
\end{equation}
Here $q_R$ is real, $\delta_+ =2 \epsilon$ and $\delta_-=0$. Thus, for
the upper sign one is above the cut of $I_{2, \hbox{tem}}^{\,(+)}$ and for
the lower sign below. Obviously the imaginary part of
$I_{2,\hbox{tem}}^{\,(-)}$ above and below its cut is equal to that of $I_{2,
\hbox{tem}}^{\,(+)}$ above and below its cut. This can be summarized by the
following equations

\begin{eqnarray}
\label{iplus1}
\hbox{Im}\,I_{2, \hbox{tem}}^{\,(-)}(z=q_R - i\delta_\mp) &=&
\hbox{Im}\,I_{2, \hbox{tem}}^{\,(+)}(z=q_R + i\delta_\pm), \nonumber \\
\hbox{Im}\,I_{2, \hbox{tem}}^{\,(-)}(z=q_R - i\delta_\pm) &=&
- \hbox{Im}\,I_{2, \hbox{tem}}^{\,(+)}(z=q_R + i\delta_\pm).
\end{eqnarray}
The cuts are arranged in such a way that for $z$ on the real axis, the
imaginary part of $I_{2, \hbox{tem}}^{\,(+)}(z)$ cancels that of
$I_{2,\hbox{tem}}^{\,(-)}(z)$ and $I_{2, \hbox{tem}}(M^2,0,q^2)$ is
real for real $q$ as it should be. Concluding, we state that the
function $\chi^{\,(0)}_{AA, \hbox{tem}}(0,q^2)$ has cuts (in the
following we refer to these as the {\it temperature cuts}) which are
parallel to the real axis \footnote{We note that the Pauli-Villars
regularization of the function $I_{2, \hbox{tem}}(M^2,0,q^2)$ does not
change its analytic properties. On the other hand, the 3-dimensional
cutoff procedure makes the temperature cuts finite, which leads to
unphysical oscillations rather than to screening. This fact is the
main reason why we have decided to use the Pauli-Villars method in the
present investigations.}.  Thus, the full correlation function
$\chi_{AA}(0,q^2)$ defined by Eq.  (\ref{cf1}) has at least these two
cuts. Moreover, it can have poles for imaginary arguments in between
the cuts.  The general structure of the singularities of the
correlation function is shown in Fig. [7.3].

\begin{figure}[hb]
\label{anstrtps}
\xslide{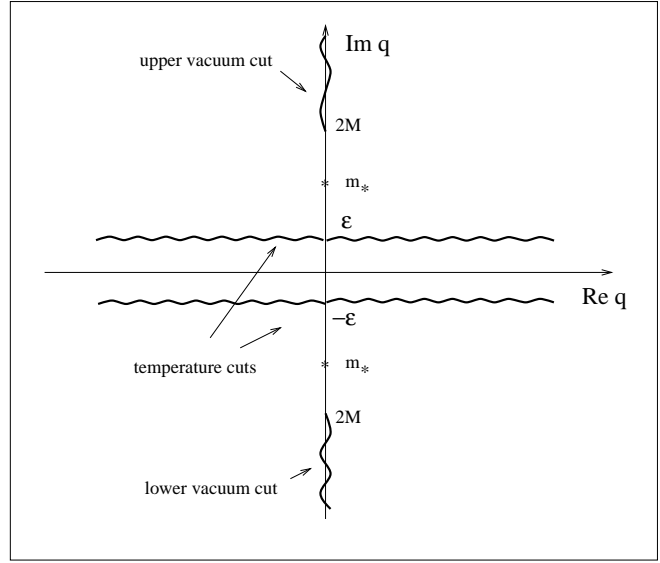}{8cm}{12}{161}{581}{681}
\caption{\small A general analytic structure of the static correlation
                function in the complex three-momentum space
                $q = | \bf q|$ (finite temperature).}
\end{figure}

The position of the pole is given by the equation whose form is
analogous to that of Eq.  (\ref{dm}), namely,

\begin{equalign}
\label{pcontr}
1 - G_S  \chi_{AA}^{(0)}(0,-m^2_{*}) = 0.
\end{equalign}

\noindent The pole can appear only the interval $0 < m_{\ast} < 2M$ because 
for $m_{\ast} = 2M$ the cut starts. At zero temperature, due to the
Lorentz invariance, we have $m_{\ast}=m_{dyn}$. On the other hand, for
$T>0$ one finds that $m_{\ast} \not = m_{dyn}$. Formally, the last
property follows from the difference between the functions
$I_{2,\hbox{tem}}(M^2,k^2,0)$ and $I_{2,\hbox{tem}}(M^2,0,-k^2)$. The
latter is the analytic continuation of $I_{2,\hbox{tem}}(M^2,0,q^2)$
and has the form

\begin{equalign}
\label{i2m5}
I_{2, \hbox{mat}}(M^2,0,-k^2) = -{N_c \over k\pi^2} \int\limits_{0}^{\infty}
{dp \, p \over E_p} {1 \over e^{E_p /T} + 1}
\left[\pi - 2 \,\hbox{arctg} \left({k \over 2p} \right) \right].
\end{equalign}

\noindent The contribution to the Fourier transform from the pole
has, of course, the form $\sim \exp(-m_{\ast} r)$.  

The integral around the vacuum cut can be written as follows

\begin{equation}
\label{vaccut}
\chi^{VC}_{AA}(r) = -{1 \over 4 \pi^2 i r} \int_{2M}^{\infty}
dk \, k \left[\chi_{AA}(0,-k^2+i\epsilon)
-\chi_{AA}(0,-k^2-i\epsilon) \right]
e^{-kr}.
\end{equation}

\noindent For large values of $r$, the exponential factor in
(\ref{vaccut}) cuts down the integrand very quickly as a function of
$k$. Thus, the remaining factors, which are slowly varying, can be
approximated by their value at $k \approx 2M$.  {\it Consequently, in
the limit $r \rightarrow \infty$ the contribution of the vacuum cut is 
also of the exponential form, $\sim \exp(-2Mr)$}.  The contribution of the 
temperature cut is

\begin{equation}
\label{thermcut}
\chi^{TC}_{AA}(r) = {1 \over 4 \pi^2 i r} 
\int\limits_{-\infty}^{+\infty} dq \, q \,
\left[\chi_{AA}(0,[q+i\delta_-]^2)-\chi_{AA}(0,[q+i\delta_+]^2)
\right]e^{iqr},
\end{equation}

\noindent where $\delta_\pm$ is defined below Eq.~(\ref{iplus2}). 
The imaginary shift of the argument, $i\delta_\pm$, denotes that the
function should be calculated just above/below the temperature cut.
We shall now discuss separately three different cases. The first two
concern our analytic results for $T=0$ and $T \rightarrow \infty$.
The third one concerns intermediate temperatures for which we do 
numerical calculations.

\subsection{Results for $T=0$} 
 
At zero temperature the temperature cut vanishes so we are left only
with one or two contributions. In the case of the pseudoscalar channel
we have an isolated pole and a cut. This pole, because of the Lorentz
invariance, coincides with that determining the dynamic mass. For
large values of $r$ the contribution from the pole, $\sim
\exp(-m_{dyn, \pi}r)$, is larger than that from the vacuum cut, $\sim
\exp(-2Mr)$. Thus, the pion screening mass at $T=0$ is equal to
the dynamic one: $m_{scr, \pi} = m_{dyn, \pi} =$ 138 MeV.  In the case
of the scalar channel we do not have an isolated pole and the only
contribution to the Fourier transform comes from the vacuum cut. In
consequence the sigma screening mass equals $2M$ and it is smaller
than the dynamic one: $m_{scr, \sigma}=$ 752 MeV and $m_{dyn, \sigma}
=$ 760 MeV. 

We note that the result $m_{scr, \sigma} = 2 M$ is of course an artifact
of the NJL model, which lacks confinement and of the approximation,
which lacks the coupling of the sigma to the two-pion continuum.
With this effect included, the screening mass in the sigma channel
would equal $2 m_{dyn, \pi}$.

\subsection{The Case of Extremely High Temperature}
 
In the limit $T \rightarrow \infty$ the zeroth-order correlation
functions vanish. It means that at sufficiently high temperature they
are very small and, at the same time, they are a good approximation
for the full correlation functions. In this case the expression $G_S
\, \chi^{(0)}_{AA}(0,q^2)$, appearing in the definition (\ref{cf1}), can
be neglected when compared to unity. Using this fact we define the
high temperature correlation function in space by equation
\footnote{The situation described in this Subsection resembles the case
of the perturbative QCD calculation discussed thoroughly in Section
7.1. The basic difference is that now the cutoffs are finite and
fixed by fitting physical observables at $T=0$.}

\begin{equalign}
\label{htcf}
\chi^{(0)}_{AA}(r) = {1 \over 4 \pi^2 i r} \int\limits_{-\infty}^{\infty}
dq \, q \, \chi^{(0)}_{AA}(0,q^2) e^{iqr}.
\end{equalign}

\noindent Here the integration of the vacuum part $\chi^{(0)}_{AA,
\hbox{vac}}(0,q^2)$ and the temperature part 
$\chi^{(0)}_{AA, \hbox{mat}}(0,q^2)$ defines the functions
$\chi^{(0)}_{AA, \hbox{vac}}(r)$ and $\chi^{(0)}_{AA,
\hbox{mat}}(r)$, respectively. Of course we have
 
\begin{equalign}
\label{dec}
\chi^{(0)}_{AA}(r) = \chi^{(0)}_{AA, \hbox{vac}}(r) + 
\chi^{(0)}_{AA, \hbox{tem}}(r).
\end{equalign}

The Fourier transform (\ref{htcf}) can be calculated
analytically. Using the same method as that developed by us in Section
7.1, we find that

\begin{equalign}
\label{htcfvac}
\chi^{(0)}_{PP, \hbox{vac}}(r) = {N_c \over 2 \pi^3 r^3} \sum_{i=0}^N A_i
\Lambda_i^2 \left[3K_2(2\Lambda_i r)+2\Lambda_i r K_1(2\Lambda_i r)\right].
\end{equalign}
On the other hand the temperature piece has the form
\begin{equalign}
\label{htcfmat}
\chi^{(0)}_{PP, \hbox{mat}}(r) = {N_c \over 4 \pi^3 r}
\left({2 \over r^3}-{2\over r^2}{\partial \over \partial r} +
{1 \over r}{\partial^2 \over \partial r^2} \right) 
\left[G^{\hbox{tem}}_1(r) + G^{\hbox{tem}}_2(r) \right],
\end{equalign}
where
\begin{equalign}
\label{g1}
G^{\hbox{tem}}_1(r) = - \sum_{i=0}^N A_i \,\Lambda_i K_1(2\Lambda_i r)
\end{equalign}
and
\begin{equalign}
\label{g2}
G^{\hbox{tem}}_2(r) = \pi T \sum_{i=0}^N A_i \sum_{l=-\infty}^{+\infty}
\exp(-2r\sqrt{(2l+1)^2 \pi^2 T^2 + \Lambda_i^2}).
\end{equalign}
 
Using the properties of the modified Bessel functions \cite{GR80} we
can check that in expression (\ref{dec}) the vacuum part
(\ref{htcfvac}) is exactly canceled by the first term of the
temperature part (\ref{htcfmat}), i.e., by the term coming from
differentiation of the function $G^{\hbox{tem}}_1(r)$.  We want to
emphasize that this cancellation is independent of the number of
subtractions $N$ and of the values of the regulating masses
$\Lambda_i$.
 
The function $G^{\hbox{tem}}_2(r)$ has a simple asymptotic behaviour. For $r
\rightarrow \infty$ it is enough to consider only the leading terms in
the series, i.e., these corresponding to $l=0,-1$ and $i=0$. The
terms for $i>0$ are suppressed since $\Lambda_i > \Lambda_0 = M$ 
(additionally we can neglect $M$ with respect to $T$). Finally, we obtain 
the following asymptotic expressions
 
\begin{equalign}
\label{g2as}
G^{\hbox{tem}}_2(r) \sim 2\pi T e^{-2\pi Tr}
\end{equalign}
and
\begin{equalign}
\label{htcfas}
\chi^{(0)}_{PP}(r) \sim {2 T^3 N_c \over r^2} e^{-2\pi Tr}.
\end{equalign}
From Eq. (\ref{htcfas}) we read off the screening mass in
the pseudoscalar channel to be $2\pi T$.  The calculations in the
scalar channel look similarly. We can even immediately guess that the
result will be the same: At very high temperature the constituent
quark mass drops down and is equal to the current one. The latter
is very small and the difference between the pseudoscalar and
scalar channel practically disappears.

Our present considerations showed again how important is the
regularization of the temperature part. It leads to cancellations
between the vacuum and temperature parts and, finally, to the asymptotic
behaviour of the form $\sim \exp(-2\pi Tr)$. If the temperature part
were not regularized, the asymptotic behaviour of the correlation
function would be governed by the lowest regulating mass, i.e., by
$\Lambda_1$. Such a case could not be accepted from the physical point
of view.
 
\begin{figure}[hb]
\label{astps}
\xslide{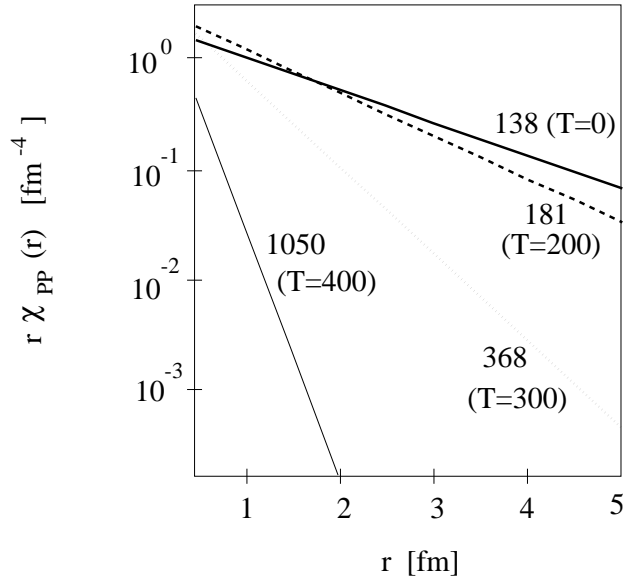}{10cm}{42}{135}{555}{700}
\caption{\small The spatial dependence of the static correlation
                function in the pion channel. The numbers at the lines
                are the values of the screening masses 
                at the corresponding temperatures (given in MeV).}
\end{figure}

\subsection{Intermediate Temperatures}
 
For the temperature range $0 \leq T \leq 400$ MeV we did numerical
calculations and read off the values of the screening masses from the
logarithmic plots representing the correlation functions in space.  We
did the calculations in two ways checking whether the results were the
same. The first method was to calculate the Fourier transform directly
from Eq. (\ref{ft}). The second method was to calculate the
contributions from the singularities separately and later to sum them
up.  In the latter case we used Eqs. (\ref{vaccut}) and
(\ref{thermcut}). Moreover, we numerically calculated the residue
corresponding to the pole (\ref{pcontr}). Both methods encounter
numerical difficulties. The direct calculation of the Fourier
transform requires an integration of a slowly converging and
oscillating function. On the other hand, the contributions from the
cuts show large cancellations. Therefore, each contribution should be
evaluated with a very high accuracy.  Let us also note here that the
parameters used to calculate the screening masses were the same as
those used in the case of the dynamic masses, see Subsection 6.3.2.

We had to restrict ourselves to rather small values of $r$, since the
correlation functions decrease very rapidly and for large $r$ their
numerical evaluation is difficult. In fact, they decrease
exponentially which suggests that the screening masses can be read off
already for small $r$'s. This exponential behaviour is shown in Figs.
[7.4] and [7.5], where we plotted the functions $r \,\chi_{PP}(r)$ and
$r \,\chi_{SS}(r)$. 

\begin{figure}[hb]
\label{si_scrps}
\xslide{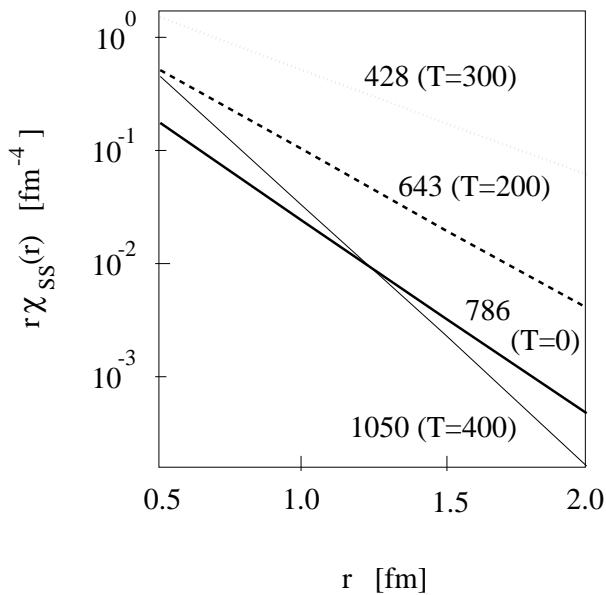}{10cm}{42}{135}{555}{700}
\caption{\small Same as Fig. [7.4] for the sigma channel.}
\end{figure}

At zero temperature our numerical procedure gives the results which
are in good agreement with our estimates based on the analytic
considerations, see Subsection 7.2.2. The numerically calculated pion
screening mass agrees exactly with the analytic result, whereas in the
sigma channel we find a small discrepancy. This difference is caused
by the fact that we do not consider asymptotic distances in the
numerical calculations.

The temperature dependence of the screening and dynamic masses is
shown in Fig. [7.6]. For $0 \leq T \leq 200$ MeV we observe that the
screening masses are close to the dynamic ones. Such situation is
expected, since for small temperatures the Lorentz invariance is only
slightly broken.  Interesting things can be observed for larger
temperatures.  {\it Both screening and dynamic masses exhibit the
restoration of chiral symmetry although they substantially differ from
each other}.  In the full interval, $0 \leq T \leq 400$ MeV, the
qualitative behaviour of the screening masses and of the dynamic ones
is similar and resembles the results of the lattice simulations of QCD.

\begin{figure}[ht]
\label{dyn+scrps}
\xslide{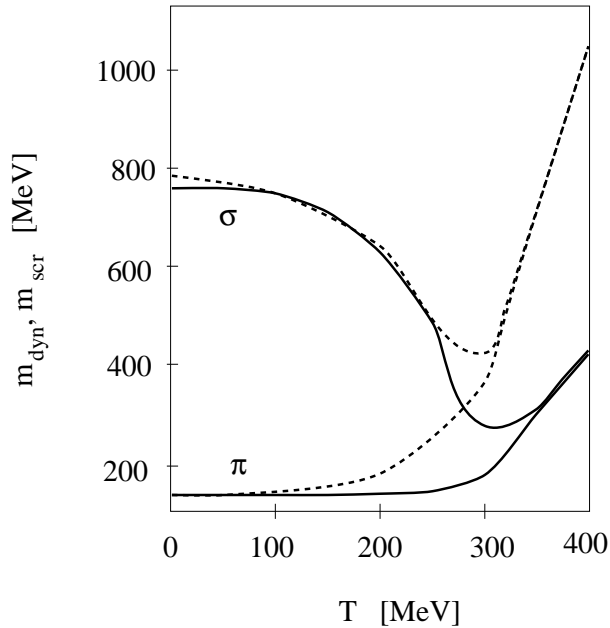}{10cm}{42}{135}{555}{700}
\caption{\small The temperature dependence of the screening (dashed
                lines) and the dynamic (solid lines) masses.}
\end{figure}

Our observation that the dynamic masses are different from the 
screening masses contradicts the results of Ref. \cite{TK91}
where it was argued that they are the same even at finite $T$.
In Ref. \cite{TK91} the chiral limit, $m=0$, is considered. Afterwards,
the gap equation is used to find that the two masses are equal.
The point is, however, that in Ref. \cite{TK91} the non-trivial
solution to the gap equation is used. It exists only if
$T<T_c$ (in the chiral limit $T_c$ is well defined) and for
$T>T_c$ we have to take into account just the trivial solution
$M=m=0$. The trivial solution does not allow us to simplify the
denominator of the correlation function, what is essential
for the proof of the equality of two masses in \cite{TK91}.
Consequently, the arguments of Ref. \cite{TK91} do not apply at $T>T_c$
and in this case we do not expect that $m_{scr}=m_{dyn}$.
The problem remains what happens at $T<T_c$. The structure of the
poles suggests that $m_{scr}=m_{dyn}$, but there exists a temperature
cut whose contribution might be not negligible. This situation 
deserves a separate study.

Another interesting point is to check whether the screening masses
approach the Eletskii-Ioffe limit, i. e., if they are equal to $2 \pi
T$ for large $T$. Our numerical results for the pseudoscalar channel
are shown in Fig. [7.7].  One can see that the ratio $m_{scr, \pi}/T$
grows with the temperature for $T > T_c$ but even at $T=450$ MeV it is
still smaller than $2 \pi $. (In the scalar channel the high
temperature behaviour is the same as in the pseudoscalar one.) As we
have already mentioned this type of behaviour is observed in the
lattice simulations.  In the (pseudo)scalar channel the screening
masses are smaller than $2 \pi T$, which is in contrast to the
(axial)vector channel where they are equal to $2 \pi T$ already for
the temperatures slightly exceeding $T_c$.  Our result suggests that
the NJL model can explain the existence of the non-negligible
(residual) interaction in the (pseudo)scalar channel. Nevertheless,
for drawing more definite conclusions the calculations in the
(axial)vector channel are required. If the difference between the
behaviour of the screening masses in the (pseudo)scalar channel and
the (axial)vector channel were found, this would act in favor
of the model.

\begin{figure}[ht]
\label{EIps}
\xslide{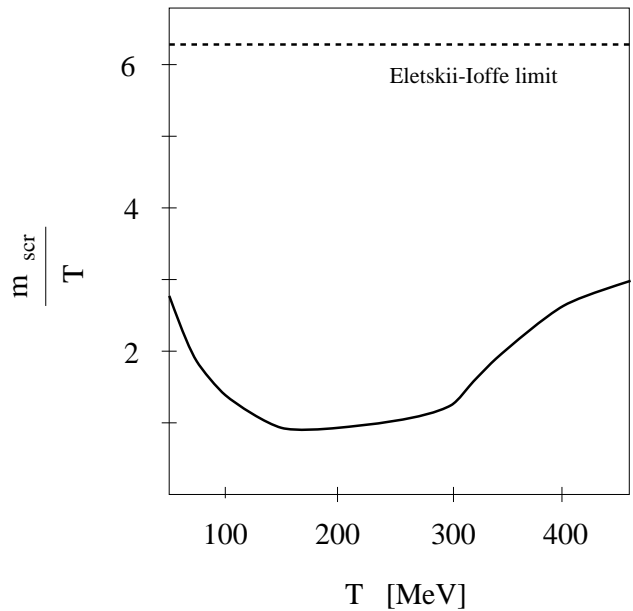}{10cm}{42}{135}{555}{700}
\caption{\small The ratio $m_{scr}/T$ plotted as a function
                of the temperature. The dashed line corresponds
                to the Eletskii-Ioffe limit: $m_{scr} = 2\pi T$.}
\end{figure}

\chapter{Static Meson Fields at Finite Baryon Density}
\label{chapt:SMF_FBD}

In this Chapter we investigate the in-medium static meson correlation
functions at finite baryon density. In contrast to the behaviour known
from the high temperature calculations, studied by us in the previous
Chapter, we find that the correlation functions at finite baryon
density are not screened but oscillate in space \cite{FF97}. This
behaviour is analogous to the Friedel oscillations in a degenerate
electron gas \cite{FW71}; in both cases the oscillations are caused by
the existence of the sharp Fermi surface.

In spite of the fact that it lacks confinement, the NJL model
successfully describes many aspects of hadron structure
\cite{pions,nucleons}.  However, a realistic description of nuclear
matter at low densities as a collection of composite nucleons would be
very complicated and outside the scope of our approach. Hence, we work
again in the HF + RPA scheme, where the correlations that bind
quarks in nucleons are neglected and consequently nuclear matter is
described as a Fermi gas of quarks. Although our model for nuclear
matter at low densities is not realistic, we believe that our results
on the correlation functions at finite baryon density are generic and
should be qualitatively the same in more realistic models. This
conjecture is supported by studies of the effective nucleon-nucleon
interaction in nuclear matter, which also indicates the presence of
the Friedel oscillations \cite{DAPS89,GDAP94,DAGP94}.

Already at moderate densities we find new singularities in the
correlation function, which probably are artifacts of the NJL model in
the HF + RPA approximation. This indicates that this approach to the
correlation functions breaks down already at moderate densities.
Hence, to complete the picture we compute the meson correlation
function at higher densities in perturbative QCD. Keeping only the
leading term, we find oscillations with a period $\delta r =
\pi/p_F$, where $p_F$ is the Fermi momentum of the quark sea.

\section{Low Density Theorem}

Before we analyze the properties of the correlation functions, let us
discuss shortly the behaviour of the quark condensate at finite
baryon density. In this way we shall supplement our discussion
from Section 5.2. The calculation of the decrease of the condensate
with increasing baryon density in the framework of the NJL model is
instructive, since there are model independent estimates of this
quantity, which impose constraints on the in-medium behaviour of the
condensate.

At $T=\mu=0$ the NJL model yields the Gell-Mann--Oakes--Renner (GOR) 
relation
\begin{equalign}
\label{GOR}
F^2_{\pi} m^2_{dyn,\pi} = - {1 \over 2} (m_u + m_d)
\langle {\overline u} u + {\overline d} d\rangle =
- 2 m \langle {\overline q} q\rangle ,
\end{equalign}
where $F_{\pi}$ is the pion decay constant. This relation is
independent of the regularization scheme.  One can use it
to find the value of $F_{\pi}$ provided the values of $m, m_{\pi}$ and
$\langle{\overline q} q\rangle$ are known. For the two-flavour
symmetric model, with the values of the parameters the same as in
Section 5.2, one finds $F_{\pi}$ = 94 MeV.

\begin{figure}[hb]
\label{rconps}
\xslide{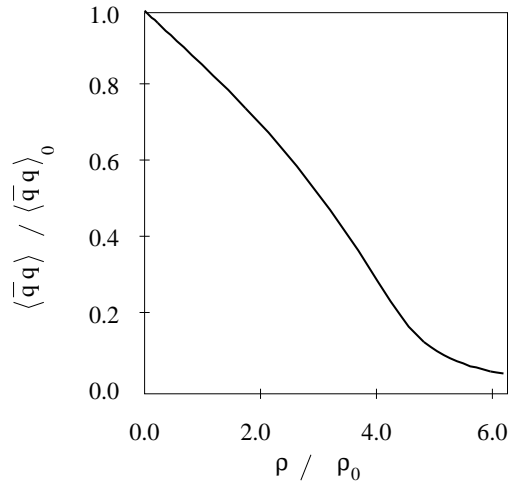}{9cm}{31}{121}{581}{720}
\caption{\small Ratio of the in-medium quark condensate to the vacuum
value, plotted as a function of baryon density.}
\end{figure}

\noindent Using the Hellmann-Feynman theorem and the GOR relation, one 
finds \cite{DL,CFG} to leading order in the density

\begin{equation}
\label{HF}
{ \langle {\overline q} q \rangle  \over
  \langle {\overline q} q \rangle_0 } =
1 -{ \Sigma_{\pi N} \over m^2_{\pi} F^2_{\pi} } \, \rho,
\end{equation}

\noindent where $\Sigma_{\pi N}$ is the pion nucleon sigma commutator,
$\langle {\overline q} q \rangle_0$ is the vacuum value of the quark
condensate, and $\rho$ is the baryon density. The baryon density
is for two flavours given by $\rho = 2p_F^3/3\pi^2$, where 
$p_F$, as usual,  is the quark Fermi momentum.
We compute the pion nucleon sigma term by using
the relation \cite{CFG}

\begin{equation}
\label{sigma}
{1 \over 3} \Sigma_{\pi N} =  \Sigma_{\pi q} = m {dM_0 \over dm},
\end{equation}

\noindent where we assume that the $\pi N$ sigma term is simply the
sum of the $\pi q$ sigma terms, like in the naive quark model.  For
our set of the parameters one gets $\Sigma_{\pi N} = $ 18 MeV,
which, using Eq. (\ref{HF}), implies that the condensate is reduced by
15 \% at the saturation density of nuclear matter, $\rho_0 = $ 0.17
fm$^{-3}$.  In Fig. [8.1] we show the numerical results for the quark
condensate (normalized to its vacuum value) as a function of the
baryon density (normalized to the saturation density) 
\footnote{We note that Fig. [8.1] corresponds exactly to Fig. [5.7]
showing dependence of $M$ on $\mu$.}. One can see that our
calculation agrees with the low-density theorem (\ref{HF}), when
we use the sigma commutator obtained within the model. Since the sigma
term is much smaller than the empirical value of 45 MeV, the density
dependence of the quark condensate is too weak. Nevertheless, the fact
that Eq. (\ref{HF}) is satisfied, shows that the calculation is
consistent.

\section{Friedel Oscillations in the NJL Model}

Let us come back to the discussion of the correlation functions.  Our
analysis of the static meson correlation functions at finite baryon
density is based on the two-flavour symmetric model. Moreover, we use
again the same values of the parameters as in Sections 5.2, 6.3 and
7.2. Thus our results for zero baryon density agree with the results
of Section 7.2 for the case $T = 0$. Our starting point are Eqs.
(\ref{ft}) and (\ref{cf1}), where the zeroth-order correlation
functions are given by formulae (\ref{ch0pv}), (\ref{ch0sv}),
(\ref{ch0pm1}) and (\ref{ch0sm1}). Thus, the full correlation function
(\ref{cf1}) can be expressed in terms of the functions $I_{1,
\hbox{vac}}(M^2)$, $I_{1, \hbox{mat}}(M^2)$, $I_{2, \hbox{vac}}(-q^2)$ 
and $I_{2,
\hbox{mat}}(0,q^2)$. We note that these four functions have been already
calculated. In the case of the Pauli-Villars regularization scheme,
which is adopted here, we can use Eqs. (\ref{i1vacRPV1}),
(\ref{i1medRPV}), (\ref{i2vrPV1}) and (\ref{i2medPV}), respectively.

\begin{figure}[hb]
\label{anstmps}
\xslide{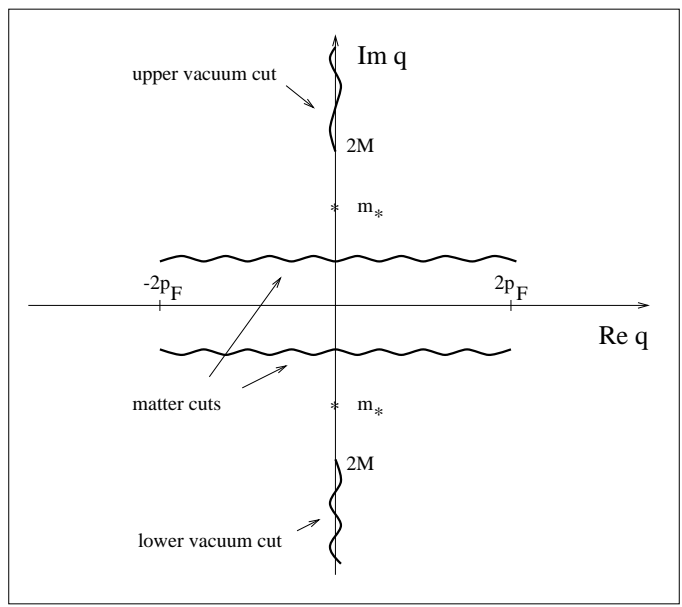}{8.5cm}{9}{151}{582}{690}
\caption{\small A general analytic structure of the static correlation
           function in the complex three-momentum space $q =
            |\bf q|$ (finite baryon density).}
\end{figure}

The Fourier transform (\ref{ft}), defining the static correlation
function in space, can be again written as a sum of a few
contributions connected with the appearance of the singularities in
the complex momentum plane. At finite baryon density, the structure of
the singularities is very similar to that at finite temperature.
It is shown in Fig. [8.2].

The basic difference between the finite temperature and finite
density cases is that the matter cut has a finite range in $q$
(the temperature cut, shown in Fig.  [7.3], is infinite).
To understand this behaviour, we can analyze the function
$I_{2, \hbox{mat}}(M^2,0,q^2)$ in the same way as the function 
$I_{2, \hbox{tem}}(M^2,0,q^2)$ was analyzed in Subsection 7.2.1.
First, we write

\begin{equation}
\label{i2mat4}
I_{2, \hbox{mat}}(M^2,0,q^2) = I_{2, \hbox{mat}}^{\,(+)}(q) + 
I_{2, \hbox{mat}}^{\,(-)}(q),
\end{equation}
where
\begin{equation}
\label{i2matpm}
I_{2, \hbox{mat}}^{\,(\pm)}(z) = -{N_c \over 4z\pi^2}
\int\limits_{0}^{p_F}
{dp \, p \over E_p} 
\ln{\,\,\,2p-z \pm i\epsilon \over -2p-z \pm i \epsilon}.
\end{equation}

\noindent The form of expression on the RHS of Eq. (\ref{i2matpm})
shows that the functions $I_{2, \hbox{mat}}^{\,(\pm)}(z)$ have
logarithmic cuts parallel to the real axis and stretching from $-2p_F
\pm i\epsilon $ to $+2p_F \pm i \epsilon $. The finite range of the
cuts is caused by the finite range of momenta available in the
integral (\ref{i2matpm}).

Let us now discuss the results of the finite density calculations.  
In the case of the pseudoscalar channel, we did the numerical
calculations in the energy range $M_0$ = 376 MeV $\leq \mu \leq 415$
MeV. In Fig. [8.3] we show the corresponding correlation for $\mu = 376,
400$ and 410 MeV.  In contrast to the exponential decay found in
vacuum (solid line), the correlation function at finite density
(dashed and dotted lines) oscillates, with a power-law decay of the
amplitude. Its period decreases with the increasing value of the
chemical potential.  From the physical point of view, such
oscillations are caused by the existence of a sharp Fermi surface.

\begin{figure}[ht]
\label{resps}
\xslide{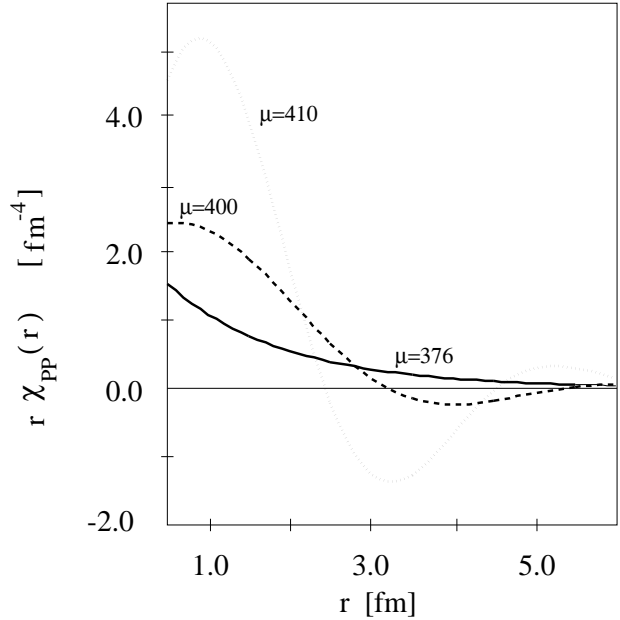}{11cm}{31}{121}{581}{720}
\caption{\small Correlation function in the pseudoscalar channel
plotted for three different values of the chemical potential: $\mu$ =
376 MeV (solid line), $\mu$ = 400 MeV (dashed line), and $\mu$ = 410
MeV (dotted line). }
\end{figure}

The oscillatory behaviour of the correlation functions at finite
density is well known for the non-relativistic degenerate electron
gas, where the phenomenon is called the Friedel oscillations
\cite{FW71}. These long range oscillations lead to many interesting
phenomena like, e.g, broadening of the nuclear magnetic resonance
lines. A characteristic feature of the Friedel oscillations (at very
large distances) is their period $\delta r = \pi/p_F$. In our case the
situation is similar; for very large values of $r$, the contribution
from the matter cut dominates over the other contributions.  Since the cut
extends over a finite range in $q$ ($|q| < 2p_F$) the correlation
function oscillates with the period $\pi / p_F$ (at $r \rightarrow
\infty$).  The numerical study of the correlation
functions at very large distances is difficult because the
amplitudes decrease with $r$. Consequently, we restrict the numerical
calculations to the interval 0.5 fm $< r <$ 6 fm. 

So far we have discussed the results for $\mu < $ 415 MeV. For larger
values of the chemical potential (corresponding roughly to $\rho >
{1\over 2} \rho_0$) the pseudoscalar correlation function acquires
additional singularity on the real axis for $q\approx 2 p_F$. In the
scalar channel, a similar singularity appears at even smaller
densities ($\rho > {1\over 3} \rho_0$). These singularities indicate
instabilities of the ground state, which lead the system to states of
lower energy.  However, most likely the singularities are artifacts
due to regularization procedure and do not correspond to physical
instabilities. In order to check this point we have redone the
calculation using a different regularization scheme with a
3-dimensional cutoff. Using the parameters obtained in
Ref. \cite{HKZV94}, we find no singularities on the real axis. This shows
that these singularities are unphysical, since their presence depends
on details in the formulation of the model. Thus, our regularization
scheme can be used only at low densities.

On the other hand, the 3-dimensional cutoff regularization is
unsatisfactory because it explicitly breaks Lorentz invariance. This
obscures the relation between the correlation functions in time-like
and space-like regions. Moreover, in the 3-dimensional cutoff
scheme, one does not recover the well known screening of the
correlation function at finite temperatures. Instead, the
correlation function oscillates, much like the Friedel oscillations at
$T=0$, with a period $\pi/\Lambda$, where $\Lambda$ is the momentum
cutoff. We stress that this behaviour is an artifact, due to the
finite range of momenta ($|q| < \Lambda$) available in the Fourier
transform. 

Consequently, we stick to the covariant Pauli-Villars method, in spite
of its shortcomings. However, since the results are inconclusive at
densities $\rho \sim \rho_0$ and higher, we restrict the
calculations to small densities $\rho < {1 \over 3} \rho_0$. The
physics at very large densities will be addressed in the following
Section.

\section{Oscillations in Perturbative QCD}

In a straightforward way, the formalism developed so far can be
used to study finite-density correlations functions in perturbative
QCD. Similarly to the high temperature case discussed in Section
7.1, the leading term in the meson correlation function is the 
lowest order quark loop diagram (as in Fig. [6.1]). Thus, for the
pseudoscalar channel we can write

\begin{equalign}
\label{dec1}
{\tilde \chi}(r) = {\tilde \chi}_{\hbox{vac}}(r) 
+ {\tilde \chi}_{\hbox{mat}}(r),
\end{equalign}

\noindent where ${\tilde \chi}_{\hbox{vac}}(r)$ is given by Eq. 
(\ref{vpr1}), and the matter piece has the form

\begin{equalign}
\label{mpr11}
{\tilde \chi}_{\hbox{mat}}(r) = {N_c \over 4\pi^3 r} \left( {2 \over r^3} -
{2 \over r^2} {\partial \over \partial r} +
{1 \over r} {\partial^2 \over \partial r^2}
\right) \left[ {\tilde G}^{\hbox{mat}}_1(r)
+ {\tilde G}^{\hbox{mat}}_2(r) \right].
\end{equalign}

\noindent Eq. (\ref{mpr11}) is the analog of Eq. (\ref{mpr1}) for
the case of finite density. One can check that ${\tilde
G}^{\hbox{mat}}_1(r) = {\tilde G}^{\hbox{tem}}_1(r)$. 
Similar cancellations to those described in Section 7.1 take
place. Thus, finally only the function ${\tilde G}^{\hbox{mat}}_2(r)$
gives a non-vanishing contribution to the correlation function.
A simple calculation yields

\begin{equalign}
\label{g2mat}
{\tilde G}^{\hbox{mat}}_2(r) = 
- {\partial \over \partial r} 
\int_{p_F}^{\infty} {dp \over 2 \sqrt{p^2+M^2}} \cos(2pr).
\end{equalign}

\noindent In the limit $M \rightarrow 0$ we find
${\tilde G}^{\hbox{mat}}_2(r) = \cos(2p_F r)/2r$, which implies that
the correlation function ${\tilde \chi}(r)$ is not screened but
oscillates in space with a period $\delta r = \pi /p_F$. For
massless quarks the behaviour in the scalar channel is identical.

\chapter{Beyond the Hartree-Fock Approximation}
\label{chapt:GBMF}

Calculations based on the NJL model show that the value of the
quark condensate is strongly modified by the in-medium effects.  It
eventually goes to zero for $T, \mu \rightarrow \infty$.  The
interesting question is whether the changes of the condensate (as described
by the model) exhibit some universal features.  At low density, we
have seen (Section 8.1) that the NJL calculations agree with the model
independent estimates.  On the other hand, it turns out that the
change of the condensate at small temperature is not compatible with
the general requirements of the chiral perturbation theory: In the
standard approach to the model (HF + RPA), the condensate remains
completely flat at small $T$, whereas the chiral perturbation theory
indicates that the absolute value of the condensate should decrease
linearly with $T^2$.

As we shall see in this Chapter, the discrepancy between the
NJL model and the chiral perturbation theory is an effect caused by
the HF approximation. The standard approach is not sufficient to
describe properly the thermodynamic properties of hadronic matter at
small temperatures, which should be dominated by the lightest hadrons,
i.e., pions. In order to describe properly the low temperature change
of the condensate in the NJL model, one has to go beyond the HF
approximation, including the so-called meson loops. In this Chapter,
we outline the characteristic features of this approach.

\section{Effective Action}

Our present considerations will be based on the two-flavour standard
model with scalar-isoscalar and pseudoscalar-isovector interactions.
The corresponding Lagrangian is given by Eq.  (\ref{l2}). In this
Section, we shall apply the formalism of the effective action
\cite{ItzZub} to this Lagrangian. Details of this procedure have been
given in \cite{NBCRG96}. In the following, we shall describe only the
basic ingredients of this method.

At first we introduce the meson mean fields $\Phi = (\Phi_0,
\mbox{\boldmath $\Phi$})$, where $\Phi_0$ and $\mbox{\boldmath $\Phi$}$ 
correspond to the pion and the sigma, respectively. At the 
quark-loop level the effective action is

\begin{equation}
{I}(\Phi) = \int d^4x \left ( {1 \over 2}{a^2} \Phi^2 
 - a^2 m \Phi_0 + {1 \over 2}{a^2} m^2 \right )
 - {1 \over 2}{\rm Tr}\,\ln (D^{\dagger }D),  
\label{eq:Seffq}
\end{equation}

\noindent where $1/a^2 = 2\, G$ and $D$ is the Dirac operator, i.e., $D =
\partial_\tau - {\rm i}{\mbox{\boldmath $\alpha$} \cdot
\mbox{\boldmath $\nabla$} } + \beta \Phi_0 + {\rm i} \beta \gamma_5
{\mbox{\boldmath $\tau$}} \cdot {\mbox{\boldmath $\Phi$}}$. For the
moment, we consider the case $T=0$ and work in Euclidean space-time
($\tau$, ${\mbox{\boldmath $x$}}$).  The generalization of our
considerations to the case $T>0$ will be described in the next
Section. In Eq.~(\ref{eq:Seffq}) we have replaced the usual ${\rm
Tr}\,\ln D$ term with ${1\over 2}{\rm Tr}\,\ln (D^{\dagger }D)$, which
is exact for SU(2).

Meson loops bring an additional term to the effective action 
\cite{ItzZub,NBCRG96}

\begin{equation}
{\Gamma}(\Phi) =  {I}(\Phi) + {1\over 2} {\rm Tr}\,\ln ({K}^{-1}), 
 \label{eq:Seffm}
\end{equation}

\noindent where $K$ is the inverse {\em meson propagator} matrix 
defined as

\begin{equalign}
K^{-1}_{ab}(x,y) = \frac{\delta^2 I \left( \Phi \right)} {\delta
\Phi_a(x) \delta \Phi_b(y)}.
\end{equalign}

\noindent In Eqs. (\ref{eq:Seffq}) and (\ref{eq:Seffm}) 
${\rm Tr}$ denotes the full trace, including functional space,
isospin, and in addition colour and spinor trace for quarks.  In the
$N_c$-counting scheme, the quark loop term ${I}(\Phi)$ is the leading
contribution of order ${\cal O}(N_c)$, and the meson loop term
$\frac{1}{2}{\rm Tr}\,\ln ({K}^{-1})$ is of order ${\cal O}(1)$.  Thus the
one-meson-loop contributions give the first correction to the
leading-$N_c$ results.

Using standard methods, Green's functions can be obtained from 
Eq.~(\ref{eq:Seffm}) via differentiation with respect to mean 
meson fields. Of particular importance is the one-point function, 
which gives the expectation value of the sigma field.  The condition

\begin{eqnarray}
\label{GAP} 
\frac{\delta \Gamma(\Phi)} {\delta \Phi_0(x)}_{\mid \Phi_0(x)=S} 
= && a^2 (S-m) - {1 \over 2} {\rm Tr}
\left( (D^{\dagger} D)^{-1} \frac{\delta (D^{\dagger }D)}{\delta \Phi_0(x)}
\right )_{\Phi_0(x)=S}  \nonumber \\ 
&& + {1 \over 2} {\rm Tr} \left ( K \frac{\delta K^{-1}}
{\delta \Phi_0(x)} \right)_{\Phi_0(x)=S} = 0
\end{eqnarray}

\noindent yields the equation for the vacuum expectation value of $\Phi_0$, 
which we denote by $S$. As shown in Ref.~\cite{NBCRG96}, introducing

\begin{eqnarray}
\label{propK}
K_\sigma(S,q_E^2) &=& \left[-I_{2, \hbox{vac}}(S^2,-q_E^2) \, 
(q_E^2+4S^2) + 
{a^2 m \over S} \right]^{-1} \; , \nonumber \\
K_\pi(S,q_E^2) &=& \left[-I_{2, \hbox{vac}}(S^2,-q_E^2)\,
q_E^2+{a^2 m \over S}\right]^{-1},
\end{eqnarray}

\noindent and retaining terms up to order ${\cal O}(N_c^0)$, 
Eq.~(\ref{GAP}) can be written in the form

\begin{eqnarray}
\label{gap0}
& & a^2 \left(S - m \right) - S \, I_{1, \hbox{vac}}(S^2)\nonumber \\
& & -  \frac{S}{16 \pi^4} \int d^4 q_E 
   \left\{ \left [2 \, I_{2, \hbox{vac}}(S^2,0)
 + \frac{d}{dS^2} \left( I_{2, \hbox{vac}}(S^2,-q_E^2) [q_E^2 + 
    4 S^2] \right) \right]
   K_\sigma(S,q_E^2) \right. \nonumber \\
& & + \left. 3 \left [2 \, I_{2, \hbox{vac}}(S^2,0)
 + \frac{d}{dS^2} I_{2, \hbox{vac}}(S^2,-q_E^2) \, q_E^2 \right]  
 K_\pi(S,q_E^2) \right\} = 0.
\end{eqnarray}

\noindent The functions $I_{1, \hbox{vac}}(S^2)$ and 
$I_{2, \hbox{vac}}(S^2,-q_E^2)$ have been defined by expressions
(\ref{i1vacWR}) and (\ref{i2vac}). One can easily check that the
leading-$N_c$ term in (\ref{gap0}) coincides with the gap equation
(\ref{gap2fst}).
 
It is important to observe that the one-meson-loop gap equation
(\ref{gap0}) requires introduction of an extra regulator for meson
momenta.  The regularization of the quark loop only is not sufficient
to make meson loops finite, since at large $q_E^2$ one has $I_{2,
\hbox{vac}}(S^2,-q_E^2) \sim 1/q_E^2$. Consequently, the structure of 
divergences is the same as, e.g., in the linear $\sigma$ model, in
particular Eq.~(\ref{gap0}) is quadratically divergent without a
cutoff in the $q_E$ integration.  The need for an additional cutoff 
is not surprising, since a non-renormalizable theory requires new
cutoffs at next levels of a perturbative expansions.  Therefore we
have to regularize the divergent integral over $d^4q_E$.  In
Ref. \cite{NBCRG96} this was achieved by the substitution 

\begin{equation}
\label{4dc}
\int d^4q_E \longrightarrow
\pi^2 \int\limits_0^{\Lambda_b^2} dq_E^2 \, q_E^2, 
\end{equation}

\noindent where $\Lambda_b$ was the four-dimensional Euclidean meson 
momentum cutoff. At the same time, the quark loops were regularized
with the Schwinger proper-time method.  In the present study we employ
the 3-dimensional cutoff procedure for the meson loops, i.e., we make 
the replacement

\begin{equation}
\label{3dc}
\int d^4q_E  \longrightarrow 4 \pi \int dq_4 
\int\limits_0^{\Lambda_b} dq \, q^2,
\end{equation}

\noindent where $q_E=(q_4,{\bf q})$ and $q = |{\bf q}|$. The form
(\ref{3dc}) is convenient for the implementation of the boundary
conditions satisfied by temperature Green's functions. Similarly to
Ref. \cite{NBCRG96}, we regularize the quark loops (i.e., the 
functions $I_{1, \hbox{vac}}(S^2)$ and $I_{2, \hbox{vac}}(S^2,-q_E^2)$)
using the Schwinger proper-time method.

If chiral symmetry is broken, then Eq. (\ref{gap0}) has a nontrivial
solution for $S$.  The quark condensate and $S$ are related by the
formula
\begin{equation}
\label{qq}
\langle \overline{q}q \rangle = - a^2 (S - m) \;,
\end{equation}
which follows immediately from the fact that $\langle \overline{q}q
\rangle = \delta \Gamma(\Phi)/\delta m$ and Eq.~(\ref{eq:Seffm}).

\newpage 
\section{Finite Temperature}

For calculations at $T > 0$ we shall again adopt the 
imaginary time formalism. In this situation, the finite-temperature analog
of Eq.~(\ref{gap0}) can be written in the following form 

\begin{eqnarray}
\label{gapT}
& & a^2 \left(S - {m} \right) - S \, I_1^{R, \,SPT}(S^2,T) - 
 { S T \over 2 \pi^2} \sum_n \int\limits_0^{\Lambda_b} dq \, q^2 \times 
\nonumber \\
& & \left\{ \left [2 I_2^{R, \,SPT}(S^2,0,0,T) 
+ \frac{d}{dS^2} \left (I_2^{R, \,SPT}(S^2,n,q,T)
\left[(E^b_n)^2 + {\bf q}^2 + 4S^2\right] \right ) \right]
 K_\sigma(S,n,q,T) \right. \nonumber \\
& & + \left. 3 \left [2 I_2^{R, \,SPT}(S^2,0,0,T) 
+ \frac{d}{dS^2} I_2^{R, \,SPT}(S^2,n,q,T) 
\left[(E^b_n)^2 + {\bf q}^2\right] \right]  
 K_\pi(S,n,q,T) \right\} = 0 \;.\nonumber \\
\end{eqnarray}

\noindent where $K_\sigma(S,n,q,T)$ and $K_\pi(S,n,q,T)$ 
are the generalized expression for the inverse meson 
propagators  

\begin{eqnarray}
\label{propT}
 K_\sigma(S,n,q,T) & = & \left\{
-I_2^{R, \,SPT}(S^2,n,q,T)\left[(E^b_n)^2 + {\bf q}^2+4S^2\right] + 
  a^2 m/S \right\}^{-1}, \nonumber \\ 
 K_\pi(S,n,q,T) & = & \left\{ 
-I_2^{R, \,SPT}(S^2,n,q,T)\left[(E^b_n)^2 + {\bf q}^2\right] + a^2 m/S  
\right\}^{-1}.
\end{eqnarray}

\noindent In Eqs. (\ref{gapT}) and (\ref{propT}) $E^b_n = 2\pi n T$, 
and the functions $I_1^{R, \,SPT}(S^2,T)$ and $I_2^{R,
\,SPT}(S^2,n,q,T)$ are given by expressions (\ref{I1SPT}) and
(\ref{ftr}) --- since in this Chapter the Schwinger proper-time
method is used to make the quark loops finite, we make the reference to 
the expressions giving the regularized form of $I_{1}(S^2,T)$
and $I_{2}(S^2,n,q,T)$.

\section{Low-Temperature Expansion in the Chiral Limit}
\label{sec:lowT}

Before presenting our numerical results for $\langle {\overline q} q
\rangle_T$ let us consider the low-temperature expansion. As shown by
Gasser and Leutwyler \cite{GL}, {\em in the exact chiral limit}
the low-temperature expansion of the quark condensate has the form

\begin{equation}
\label{eq:gl}
\langle \overline{q} q \rangle_T = \langle \overline{q} q \rangle_0 
\left ( 1 - \frac{T^2}{8 F_\pi^2} - \frac{T^4}{384 F_\pi^4} + ... \right ) .
\end{equation}
First, let us do the $N_c$ counting in this formula. Since $F_\pi \sim
{\cal O}(\sqrt{N_c})$, subsequent terms in the expansion are
suppressed by $1/N_c$. Since our one-meson-loop calculation accounts
for first sub-leading effects in the $1/N_c$ expansion, we can hope
for reproducing only the $T^2$ term in Eq.~(\ref{eq:gl}). Further
terms would require more loops.

Using techniques described in Chapter 4, the sum over the bosonic
Matsubara frequencies in Eq.~(\ref{gapT}) can be converted to a
contour integral in the complex energy plane. By deforming this
contour we collect all contributions from the singularities of the
integrand, weighted with the thermal Bose distribution. At low
temperatures, the dominant contribution comes from the lowest lying
pion pole, and other singularities are negligible.  Thus, the third
term in (\ref{gapT}) becomes

\begin{equation}
\label{ae1}
{3T \over \pi^2} \sum_n \int\limits_0^{\Lambda_b} dq \, q^2
{1 \over (E^b_n)^2 + q^2} = {3 \over 2\pi^2 }
\int\limits_0^{\Lambda_b} dq \, q 
\left[ 1 + {2 \over e^{q/T} - 1} \right].
\end{equation}

\noindent In Eq. (\ref{gapT}) we have approximated the function 
$I_2(S^2,n,q,T)$, appearing in the pion propagator, by its value at $n=q=0$. 
For sufficiently large cutoff $\Lambda_b$, the integral over the thermal 
distribution function in (\ref{ae1}) can be expressed by the Riemann
zeta function $\zeta(2) = \pi^2/6$. Thus, the final result for
(\ref{ae1}) is $ 3\Lambda^2_b/4 \pi^2 + T^2/2$.
Inserting the above result into the gap equation (\ref{gapT}) we find,
with $m=0$, the following equality:
\begin{equation}
\label{ae2}
h(S,T) \equiv a^2 - I_1(S^2,T) + {3\Lambda_b^2 \over
4 \pi^2} + {1\over 2} T^2 = 0.
\end{equation}
Eq. (\ref{ae2}) defines implicitly the function $S(T)$,
which satisfies the equation
\begin{equation}
\label{ae3}
{dS \over dT^2} = - 
{ \partial h(S,T) / \partial T^2 \over \partial h(S,T)   /  \partial S}
= \left[ 2
{\partial I_1(S^2,T) \over \partial S} \right]^{-1}.
\end{equation}
Here we have neglected the term $\partial I_1(S^2,T) /
\partial T^2$, since it is exponentially suppressed by the factor
$\exp(-S/T)$. Furthermore, the RHS of (\ref{ae3}) can be rewritten using 
the relations \cite{NBCRG96} $\partial I_1(S^2,0)/ \partial 
S = 4 S I_2(S^2,0,0,0)$ and \mbox{$-I_2(S^2,0,0,0) = 
\overline{F}_{\pi}^2/S^2$}, where $\overline{F}_{\pi}$ is the leading-$N_c$ 
piece of the pion decay constant. Collecting these equalities we 
arrive at $dS/dT^2 = -S/(8\overline{F}_{\pi}^2)$, which finally gives
\begin{equation}
\label{ae4}
S(T) = S(0) \left[ 1 - {T^2 \over 8\overline{F}_{\pi}^2} \right].
\end{equation}
Proportionality (\ref{qq}) implies that the above expression
coincides (in the large $N_c$ limit, where 
 $\overline{F}_\pi \to F_\pi$) with Eq.~(\ref{eq:gl}).
Hence our method is consistent with a basic requirement of chiral
symmetry at the one-meson-loop level.

\section{Results of Numerical Calculations}
\label{sec:res}

In the exact chiral limit the model has 3 parameters: $a$, $\Lambda$,
and $\Lambda_b$.  Since the objective of this calculation is to
explore qualitatively the role of meson loops rather than to come out
with accurate fits to many observables, we present the calculations
for the arbitrary choice $\Lambda_b/\Lambda = {1\over 2}$.  We have
also done calculations for other values of $\Lambda_b$, with similar
conclusions as long as it is not too small.  The remaining two
parameters are fixed by reproducing the physical value of
$F_\pi=93\hbox{ MeV}$ and a chosen value for $\langle \overline{q} q
\rangle_0$. For the case of $m \neq 0$ we have an extra parameter,
$m$, which is fitted by requiring that the pion has its physical mass.
We compare results with meson loops to results with the quark loop
only ($\Lambda_b = 0$).  Parameters for the two calculations are
always adjusted in such a way, that the values of $F_\pi$, $\langle
\overline{q} q \rangle_0$, and $m_\pi$ are the same.

The calculation of $F_\pi$ with meson loops, although straightforward,
is rather tedious, so we do not present it here. The method has been
presented in detail in Ref.~\cite{NBCRG96,thesis}. The only difference
in our calculation is that the three-momentum cutoff (\ref{3dc})
rather than the four-momentum cutoff of Ref.~\cite{NBCRG96} is used.

\begin{figure}[h]
\xslide{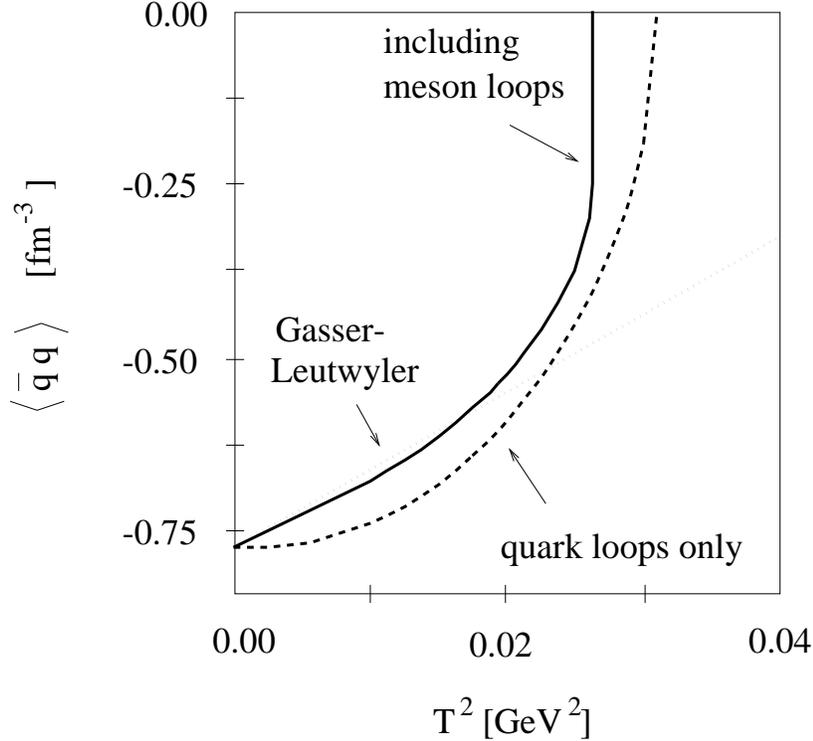}{11cm}{45}{160}{550}{680}
\caption{\small Dependence of the quark condensate on $T^2$ 
in the chiral limit $m_\pi=0$. The curves correspond to the
calculation with meson loops (solid line), with quark loops only
(dashed line), and the lowest-order chiral expansion (dotted
line). The parameters for the solid line and dashed line are adjusted
in such a way that $F_\pi = 93\hbox{ MeV}$ and $\langle \overline{q} q
\rangle_0 =-(184\hbox{ MeV})^3$. For the solid line $a=175\hbox{ MeV}$,
$\Lambda = 723\hbox{ MeV}$, and $\Lambda_b = {1\over 2} \Lambda$, whereas
for the dashed line $a=201\hbox{ MeV}$, $\Lambda = 682\hbox{ MeV}$, and
$\Lambda_b = 0$. }
\label{fig:1}
\end{figure}

Figure [9.1] shows the dependence of $\langle \overline{q} q \rangle$
on $T^2$. The solid line represents the case with meson loops. We note
that at low temperatures the curve has a finite slope, as requested by
Eq.~(\ref{ae4}). The slope is close to the leading-order
Gasser-Leutwyler result (dotted curve). As explained earlier, the
slopes would overlap in the large-$N_c$ limit.  This behaviour is
radically different from the case with quark loops only (dashed
curve). In this case at low temperatures \mbox{$\langle \overline{q} q
\rangle_T - \langle \overline{q} q \rangle_0 \sim e^{-M/T}$}, where
$M$ is the mass of the constituent quark. All derivatives of this
function vanish at $T=0$, and $\langle \overline{q} q \rangle$ is flat
at the origin.  We can also see from the figure that the fall-off of
the condensate is faster when the meson loops are included.  In fact,
for the parameters of Fig. [9.1] we have an interesting
phenomenon. At $T = 162\hbox{ MeV}$ the condensate abruptly jumps to
0. There is a first-order phase transition, with a latent heat
necessary to melt the quark condensate. Such a behaviour is not present
in the case of calculations without meson loops \cite{HK85,BMZ87a}.  
We note that with meson loops present the chiral restoration temperature 
is $162\hbox{ MeV}$, i.e., about 10\% less than $176\hbox{ MeV}$ of the 
quark-loop-only case.

Figure [9.2] shows the same study, but for the physical value of
$m_\pi$. We note that now $\langle \overline{q} q \rangle$ (solid
line) is also flat at the origin, since the pion is no more massless,
and at low $T$ we have \mbox{$\langle \overline{q} q \rangle_T -
\langle \overline{q} q \rangle_0 \sim e^{-m_\pi/T}$}.  Nevertheless,
the region of this flatness is small, and at intermediate temperatures
the curve remains close to the Gasser-Leutwyler expansion. We note
again that meson loops considerably speed up the melting of the
condensate compared to the case of quark loops only.  However, there
is no first-order phase transition such as in
Fig. [9.1]. Instead, we observe a smooth cross-over typical for
the case of $m \neq 0$.

The faster change of the quark condensate in our study is not
surprising.  It is caused by the presence of light pions which are
known to play a dominant role at low-temperatures. See, for example,
Ref. \cite{ZHK94} where the bulk thermodynamic quantities studied
within the NJL model are determined by the pion quantum numbers.  In
this case, however, the decrease of the condensate does not agree with
the requirements of the chiral perturbation theory, since the
scalar-density equation is considered at the quark-loop level only and
the slope of $\langle \overline{q} q \rangle$ vs. $T^2$ vanishes at
$T^2=0$.

\begin{figure}[hb]
\xslide{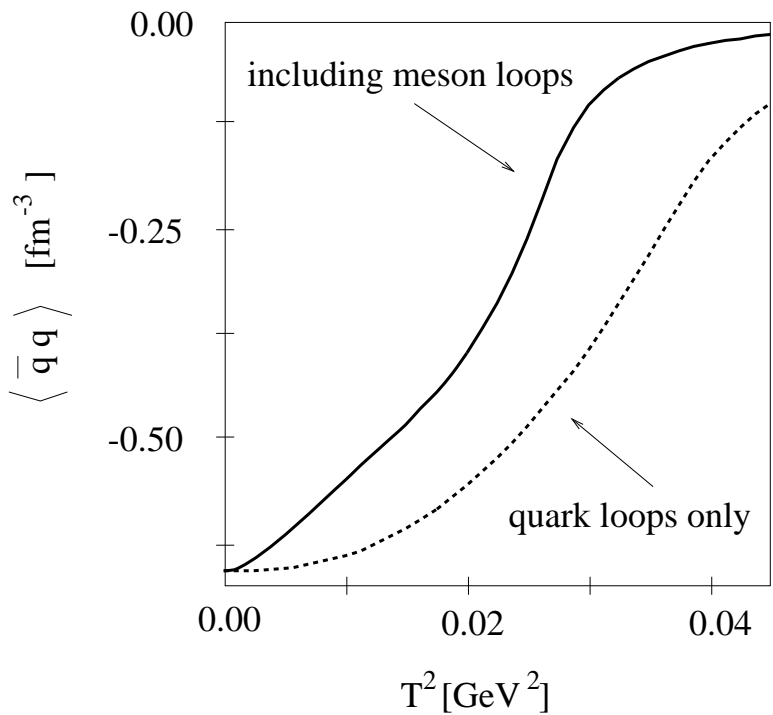}{11.cm}{45}{160}{550}{680}
\caption{Same as Fig. [9.1] for $m_\pi=139~\hbox{ MeV}$, 
 $F_\pi = 93\hbox{ MeV}$, and $\langle \overline{q} q \rangle_0 
 =-(174\hbox{ MeV})^3$.
For the solid line $a=164\hbox{ MeV}$, 
 $\Lambda = 678\hbox{ MeV}$, $\Lambda_b = {1\over 2} \Lambda$, and 
 $m = 15\hbox{ MeV}$, whereas for the
dashed line $a=175\hbox{ MeV}$, 
 $\Lambda = 645\hbox{ MeV}$, $\Lambda_b = 0$, and $m = 15\hbox{ MeV}$. }
\label{fig:2}
\end{figure}

\part{\bf NON-EQUILIBRIUM PHENOMENA}

\chapter{\bf Real-Time Formalism }
\label{chapt:real_time}

In this Chapter we give a short introduction to the real-time
formalism, presenting the most useful concepts and definitions.  For
our further considerations we do not need to know this method in more
detail, nevertheless for a better understanding it is useful to
discuss in a separate place the way in which the non-equilibrium
Green's functions are introduced. For more information about the
real-time formalism we refer the reader to \cite{D84}.

Since we are interested in the non-equilibrium evolution of a system,
we have to calculate the averages of the operators in arbitrary
states. Thus, in analogy to (\ref{rtgf}) or (\ref{vgf}), we define

\begin{equalign}
\label{negf}
i G(t_1,{\bf x}_1;t_2,{\bf x}_2) 
=  \langle  T \left[ \psi(t_1,{\bf x}_1) 
{\bar \psi}(t_2,{\bf x}_2)
\right]   \rangle,
\end{equalign}

\noindent where the symbol $\langle ... \rangle$ denotes an expectation
value with respect to the arbitrary initial state. The latter is 
specified at $t=t_0$ by the density operator ${\hat \rho}$, i.e., 
we have $\langle ... \rangle = \hbox{Tr}({\hat \rho} ...) /  
\hbox{Tr}({\hat \rho})$.

The field operators appearing in (\ref{negf}) are in the Heisenberg
picture. In the situation when the perturbation expansion is 
appropriate, the following representation is frequently used

\begin{equalign}
\label{uou}
O_{{}_H}(t) = U(t_0,t)\, O_{{}_I}(t)\, U(t,t_0),
\end{equalign}

\noindent where $O_{{}_H}(t)$ and $O_{{}_I}(t)$ are the operators in the 
Heisenberg and interaction pictures, respectively. The quantity
$U(t,t_0)$ is the evolution operator. For $t>t_0$ we have

\begin{eqnarray}
\label{utt0}
U(t,t_0) &=& \sum_{n=0}^{\infty} {(-i)^n \over n!} 
T \left[ \, \, \int\limits_{t_0}^t \, dt_1 \,\, ... 
\int\limits_{t_0}^t \, dt_n H^1_{{}_I}(t_1) ... H^1_{{}_I}(t_n) \right] 
\nonumber \\
&=& T \left[ \exp\left( -i \int\limits_{t_0}^t \, dt^{\prime} 
H^1_{{}_I}(t^{\prime}) \right) \right]
\end{eqnarray}
and
\begin{equalign}
\label{ut0t}
U(t_0,t) = {\tilde T} \left[ \exp\left( -i \int\limits^{t_0}_t \, 
dt^{\prime}  H^1_{{}_I}(t^{\prime}) \right) \right],
\end{equalign}

\noindent where $H^1_{{}_I}(t)$ is the interaction Hamiltonian in the 
interaction picture, and $T$ (${\tilde T}$) is the
chronological-ordering (antichronological-ordering) operator.

Let us now consider the expectation value of the operator $O_{{}_H}(t)$
in the ground state of an interacting many-body system, denoted later
by $| 0 \rangle$. Using a group property of the evolution operators
$U$ as well as the Gell-Mann theorem \cite{FW71} we find

\begin{equalign}
\label{pop}
\langle \, 0 \, | \, O_{{}_H}(t) \, | \, 0 \, \rangle  = 
{ \langle \, \Phi \, | \, T \left[ \exp \left( -i\int\limits_{-\infty}^{\infty}
\, dt^{\prime} H^1_{{}_I}(t^{\prime}) \right) O_{{}_I}(t) \right]
| \, \Phi \, \rangle  \over 
\langle \, \Phi \, | \, T \left[ \exp \left( -i \int\limits_{-\infty}^{\infty}
\, dt^{\prime} H^1_{{}_I}(t^{\prime}) \right) \right]| \, \Phi \, \rangle},
\end{equalign}

\noindent where $|\,\Phi\, \rangle$ is the ground state of the noninteracting
system. Eq. (\ref{pop}) can be generalized to the case of the Green's
functions, namely, we can write

\begin{eqnarray}
\label{pop1}
i G(t_1,{\bf x}_1;t_2,{\bf x}_2) 
& = & \langle \,0\,| \, T \left[ \psi(t_1,{\bf x}_1) 
{\bar \psi}(t_2,{\bf x}_2) \right] \,| 0 \,  \rangle \nonumber \\
&   & \nonumber \\
& = &
{ \langle \, \Phi \, | \, T \left[ \exp \left( -i\int\limits_{-\infty}^{\infty}
\, dt^{\prime} H^1_{{}_I}(t^{\prime}) \right) \psi_{{}_I}(t_1,{\bf x}_1)
{\bar \psi}_{{}_I}(t_2,{\bf x}_2) \right]| \, \Phi \, \rangle
\over 
\langle \, \Phi \, | \,T \left[ \exp \left( -i \int\limits_{-\infty}^{\infty}
\, dt^{\prime} H^1_{{}_I}(t^{\prime}) \right) \right]| \, \Phi \, \rangle}.
\end{eqnarray}
 
\noindent Application of the Wick theorem to (\ref{pop1}) leads to
the usual Feynman rules. In particular, the denominator cancels the
disconnected diagrams appearing in the numerator.

The method described above cannot be applied to the average in an
arbitrary state. The reason is that such a state, in general, cannot be
simply expressed in terms of the ground state of the noninteracting
particles. Nevertheless, substituting  (\ref{utt0}) and (\ref{ut0t}) 
into (\ref{uou}) we obtain expression

\begin{eqnarray}
\label{tot}
\langle O_{{}_H}(t) \rangle &=& \langle U(t_0,t) O_{{}_I}(t) U(t,t_0) \rangle
\nonumber \\
&=& 
\langle {\tilde T} \left[ \exp \left( -i \int\limits_{t}^{t_0}
\, dt^{\prime} H^1_{{}_I}(t^{\prime}) \right) \right]
O_{{}_I}(t)
T \left[ \exp \left( -i \int\limits_{t_0}^{t}
\, dt^{\prime} H^1_{{}_I}(t^{\prime}) \right) \right] \rangle,
\end{eqnarray}

\noindent which can be put into a form analogous to
Eq. (\ref{pop}). This can be achieved by introduction of an integral
contour ${\cal K}$ running along the time axis and the associated
operator $T_{{}_{\cal K}}$ ordering along this contour, see
Fig. [10.1].  Using this procedure Eq. (\ref{tot}) can be rewritten as
follows

\begin{equalign}
\label{tot1}
\langle O_{{}_H}(t) \rangle = \langle T_{{}_{\cal K}} \left[
\exp \left( -i \int_{{}_{\cal K}}
\, dt^{\prime} H^1_{{}_I}(t^{\prime}) \right) O_{{}_I}(t) \right] \rangle.
\end{equalign}

\newpage
\begin{figure}[ht]
\xslide{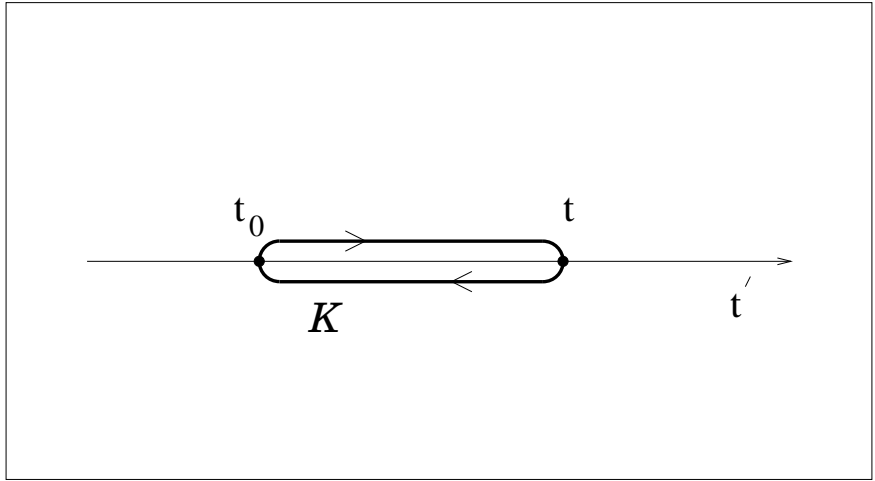}{8.cm}{24}{250}{563}{586}
\caption{\small Contour ${\cal K}$ used to calculate the operator 
                expectation value $O_{{}_H}(t)$.}
\end{figure}

\bigskip
\noindent Generalizing the last result, we introduce the Green's function
on the contour

\begin{equalign}
\label{tot2}
i G_{{}_{\cal K}}(t_1,{\bf x}_1;t_2,{\bf x}_2) 
 =  \langle  T_{{}_{\cal K}} \left[ \psi(t_1,{\bf x}_1) 
{\bar \psi}(t_2,{\bf x}_2) \right]   \rangle, 
\end{equalign}

\noindent which has the following representation

\begin{equalign}
\label{tot3}
i G_{{}_{\cal K}}(t_1,{\bf x}_1;t_2,{\bf x}_2) 
 =  \langle  T_{{}_{\cal K}} \left[ 
\exp \left( -i \int_{{}_{\cal K}}
\, dt^{\prime} H^1_{{}_I}(t^{\prime}) \right)
\psi_{{}_I}(t_1,{\bf x}_1) 
{\bar \psi}_{{}_I}(t_2,{\bf x}_2) \right]   \rangle. 
\end{equalign}

\noindent Here the contour runs from $t_0$ to the largest argument 
of the Green's function and then goes back to $t_0$. The formal
resemblance to (\ref{pop1}) is recovered since the disconnected
diagrams vanish

\begin{equalign}
\label{tot4}
1 =  \langle  T_{{}_{\cal K}} \left[ 
\exp \left( -i \int_{{}_{\cal K}}
\, dt^{\prime} H^1_{{}_I}(t^{\prime}) \right) \right]   \rangle. 
\end{equalign}

\noindent Assuming that the initial state allows for the Wick
decomposition, $G_{{}_{\cal K}}(t_1,{\bf x}_1;t_2,{\bf x}_2)$ can
be calculated using the standard technique for Feynman diagrams.

Very often it is convenient to divide the contour into the two
branches --- the upper one (stretching from smaller to larger values
of $t$) and the lower one (stretching from larger to smaller values of
$t$). According to the time ordering $T_{{}_{\cal K}}$, the time
arguments along the upper branch are regarded as earlier than those
situated along the lower branch.  Depending on the position of its
time arguments the contour Green's function fits to one of the
following categories:

\newpage
\begin{equalign}
\label{g--}
i G_{{}_{\cal K}}(t_1^-,{\bf x}_1;t_2^-,{\bf x}_2) =
i G(t_1,{\bf x}_1;t_2,{\bf x}_2),
\end{equalign}

\begin{equalign}
\label{g-+}
i G_{{}_{\cal K}}(t_1^-,{\bf x}_1;t_2^+,{\bf x}_2) =
i G^<(t_1,{\bf x}_1;t_2,{\bf x}_2) = 
- \langle {\bar \psi}(t_2,{\bf x}_2) \psi(t_1,{\bf x}_1) \rangle,
\end{equalign}

\begin{equalign}
\label{g+-}
i G_{{}_{\cal K}}(t_1^+,{\bf x}_1;t_2^-,{\bf x}_2) =
i G^>(t_1,{\bf x}_1;t_2,{\bf x}_2) = 
\langle  \psi(t_1,{\bf x}_1){\bar \psi}(t_2,{\bf x}_2) \rangle,
\end{equalign}

\begin{equalign}
\label{g++}
i G_{{}_{\cal K}}(t_1^+,{\bf x}_1;t_2^+,{\bf x}_2) =
i {\tilde G}(t_1,{\bf x}_1;t_2,{\bf x}_2).
\end{equalign}

The Green's function $G_{{}_{\cal K}}(t_1^-,{\bf x}_1;t_2^-,{\bf x}_2)$
is the usual causal (Feynman) function. On the other hand,
$G_{{}_{\cal K}}(t_1^+,{\bf x}_1;t_2^+,{\bf x}_2)$ is anti-causal.
In addition, there are two extra functions: 
$G_{{}_{\cal K}}(t_1^+,{\bf x}_1;t_2^-,{\bf x}_2)$ and
$G_{{}_{\cal K}}(t_1^+,{\bf x}_1;t_2^-,{\bf x}_2)$. Nevertheless,
one can check that only two out of these four functions are independent.  

\begin{figure}[ht]
\xslide{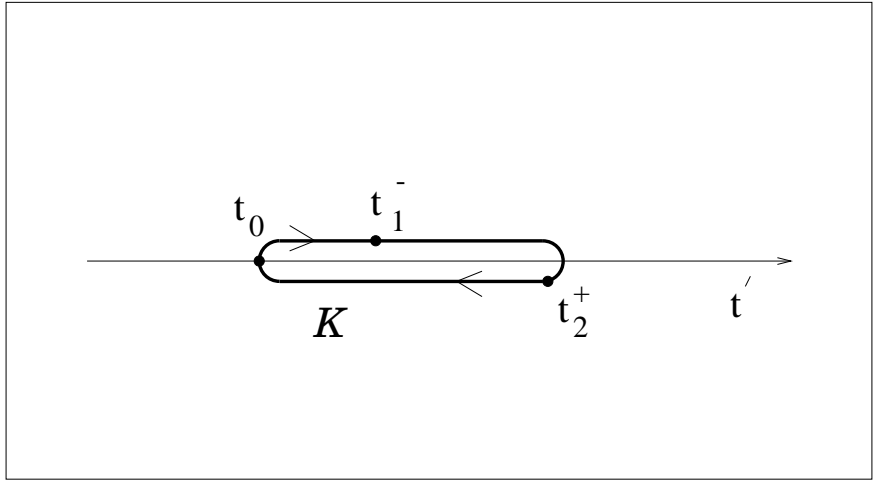}{8.cm}{24}{250}{563}{586}
\caption{\small The second time argument of the function 
                $G^<(t_1,{\bf x}_1;t_2,{\bf x}_2)$ is later
                than the first one. Thus, the time ordering along
                the contour always changes the initial ordering of the
                field operators (in addition there is a change in sign
                for fermionic operators).}
\end{figure}

The fact that at least two independent functions are required for
description of non-equilibrium situations has a simple physical
interpretation: besides the Green's function describing the spectrum
of the excitations in a medium, we need another function describing
the evolution of the medium itself. In the classical limit such
evolution is described by the kinetic equation fulfilled by the
phase-space distribution functions. In the next Chapter we shall study
how the information contained in $G^<(t_1,{\bf x}_1;t_2,{\bf x}_2)$
allows us to derive the transport equation for the quark distribution
function.

\chapter{\bf Mean-Field Transport Theory}
\label{chapt:transport}

In this Chapter, we shall study how the concept of 
chiral symmetry can be {\it explicitly} included into the quantum 
and classical transport theory based on the NJL Lagrangian. In our 
investigations, we shall restrict ourselves to the mean-field 
approximation, since the interesting phenomena (like, 
e.g., spontaneous symmetry breaking) already take place at this 
level. Our approach is based on the spinor decomposition of the 
Wigner function and follows the treatment of \cite{EETAL87} and 
\cite{VGE87}. However, in contrast to \cite{EETAL87} we do not neglect  
the spin degrees of freedom and take into account all the components
of the spinor decomposition. On the other hand, we differ from
\cite{VGE87}, where the spin dynamics is fully discussed, 
in that we study a system that is governed by a different type 
of interaction.

A transport theory for the NJL model has been initially formulated by
Zhang and Wilets \cite{ZW92}, where the closed-time-path formalism
(introduced in Chapter 10) combined with the effective-action method
has been used to derive the kinetic equations.  However,
Ref. \cite{ZW92} does not discuss how the symmetry of the underlying
theory (i.e., the chiral symmetry of the NJL Lagrangian) should
reflect itself in the transport equations. Trying to clarify this
point, we derive the general form of the transport equations (in the
mean-field approximation) which are chirally invariant. We discuss the
properties of these equations and then study their classical limit by
performing an expansion in $\hbar$ for all the functions appearing in
the spinor decomposition.  Throughout our approach, we investigate and
check the consistency of our equations under the assumptions of chiral
symmetry. In particular, this implies that the axial current should
always be conserved. For the spin evolution equation, as will be seen,
this turns out to be a non-trivial requirement.

\section{Wigner Function and its Spinor Decomposition}

In general, for a non-equilibrium transport theory, one requires two
independent Green's functions to describe the system. However, it
turns out that in the mean-field approximation (which neglects the
collisions between the particles) only one function is sufficient. We
thus choose the Green's function 

\newpage
\begin{equalign}
\label{g}
G^{<}_{\alpha \beta}(x,y) = \langle {\bar \psi}_{\beta}(y)
\psi_{\alpha}(x) \rangle
\end{equalign}

\noindent to be the fundamental quantity in our description 
(for sake of simplicity we disregard here the imaginary unit appearing
in the ``canonical'' definition (\ref{g-+})).

In the following we shall restrict our considerations to the
one-flavour model defined in Section 3.1. Moreover, we shall work
in the strict chiral limit setting $m=0$. Only in the end of this
Chapter (Section 11.7) the case $m \not = 0$ will be taken 
into consideration. One can easily notice that the Green's function
(\ref{g}) satisfies the same equation as the field $\psi(x)$ does
(compare Eq. (\ref{scmfe1}))

\begin{equalign}                                             
\label{de1}                                                   
\left[i\! \not\! \partial - \sigma(x) 
- i \gamma_5 \pi(x) \right] G^{<}(x,y) = 0.  
\end{equalign}                                               
                                 
\noindent The mean fields $\sigma(x)$ and $\pi(x)$ are determined
by the Green's function $G^{<}(x,y)$ through the relations

\begin{equalign}
\label{sp1}
\sigma(x) = -2G \,\, \hbox{Sp} \,\, G^{<}(x,x), \,\,\,\,\,
\pi(x)    &=& -2G \,\, \hbox{Sp} \, \, i\gamma_5 G^{<}(x,x), 
\end{equalign}

\noindent where the trace Sp is taken over the spinor indices.

For the description of non-uniform systems, it is convenient
to introduce Wigner transforms. This allows one to make an easier
physical interpretation of the components of the Green's function
and facilitates performing the classical limit. In the case of the 
Green's function $G^{<}(x,y)$, its Wigner transform is obtained by
introducing the center-of-mass coordinate $X = (x+y)/2$,
the relative coordinate $u=x-y$, and by taking the Fourier
transform with respect to $u$. Following the notation
of \cite{MH94}, we define

\begin{equalign}
\label{w}
W_{\alpha \beta}(X,p) = {1 \over \hbar^4}  \int \, d^4u \, 
e^{ {i \over \hbar} p \cdot u} \,
G^{<}_{\alpha \beta}\left(X+{u \over 2}, X-{u \over 2}\right).
\end{equalign}

\noindent In Eq. (\ref{w})  we have explicitly incorporated 
the Planck constant $\hbar$, since later on we wish to
investigate the classical approximation.

The Wigner transforms of the derivative of a two-point function, 
$\partial f(x,y)/ \partial x^{\mu}$, and of the product
of a one-point function with the two-point one, $f(x)g(x,y)$,
are given by the expressions 

\begin{eqnarray}
\label{wt1}
{\partial f(x,y) \over \partial x^{\mu}}  & \rightarrow &
(-i p^{\mu} + {\hbar \over 2} \partial^{\mu}) f(X,p),  \\
\label{wt2}
f(x) g(x,y) & \rightarrow &  f(X) g(X,p) -{i\hbar \over 2}
\partial_{\mu} f(X) \partial_p^{\mu} g(X,p),
\end{eqnarray}

\noindent with the notation $\partial_{\mu} = \partial / \partial 
X^{\mu}, \partial_p^{\mu} = \partial / \partial p_{\mu}$, and 
where in the second relation (\ref{wt2}), only the derivatives of
first order in $\partial_p$ and $\partial_X$ are kept.
This approximation is appropriate only for the systems which are
weakly inhomogeneous (see the discussion of this point in
\cite{MH94}).

\medskip
Using Eq. (\ref{de1}) and the properties of the Wigner transform
(\ref{wt1}) and (\ref{wt2}), we arrive at the following equation 
for the Wigner function

\begin{equalign}
\label{ke}
\left[ \left(p^{\mu} + {i \hbar \over 2} \partial^{\mu} \right) 
\gamma_{\mu}-\sigma (X) +{i \hbar \over 2} \partial_{\mu} 
\sigma(X) \partial_p^{\mu}-i \gamma_5 \pi(X) 
- {\hbar \over 2} \gamma_5 \partial_{\mu}
\pi(X) \partial_p^{\mu} \right] W(X,p) = 0.
\end{equalign}

\noindent Eq. (\ref{ke}) together with Eqs. (\ref{sp1}) and 
(\ref{w}) form the system of the coupled equations that we are 
going to study in detail in this Chapter. We note that there is 
no collision term in Eq. (\ref{ke}). This is a consequence of
our mean-field approximation. We observe that under 
the chiral transformation (\ref{chi1}), the Wigner function 
changes according to the prescription

\begin{equalign}
\label{wftr}
W \rightarrow W^{\prime} = \exp(-i\gamma_5 {\chi \over 2}) 
\,\, W \,\,\exp(-i\gamma_5 {\chi \over 2}).
\end{equalign}

\noindent Hence, one can check that Eq. (\ref{ke}) is 
invariant under transformations (\ref{chi2}), (\ref{chi3}) and 
(\ref{wftr}) and thus that the first order in the derivative
expansion of the Green's function is symmetry preserving. 

The Wigner function $W_{\alpha \beta}(X,p)$ is a 4 by 4 matrix in 
spinor indices and can be expanded in terms of 16 independent 
generators of the Clifford algebra. The conventional basis consists 
of 1, $i\gamma_5, \gamma^{\mu}, \gamma^{\mu} \gamma_5,$ and 
${1\over 2}\sigma^{\mu \nu}$ which we denote in what follows
by $\Gamma_i$. In this basis, the Wigner function has the form

\begin{equalign}
\label{sd}
W = {\cal F} + i\gamma_5 {\cal P} + \gamma^{\mu} {\cal V}_{\mu}
+ \gamma^{\mu} \gamma_5 {\cal A}_{\mu} 
+ {1\over 2} \sigma^{\mu \nu}
{\cal S}_{\mu \nu}.
\end{equalign}

\noindent The properties $\gamma^0 W^{\dagger} \gamma^0 = W$
and $\gamma^0 \Gamma_i^{\dagger} \gamma^0 = \Gamma_i$ (the first 
one results from (\ref{w}), whereas the second one is a simple 
consequence of the definition of $\Gamma_i$) indicate that the 
coefficients in the spinor decomposition ${\cal F}(X,p), 
{\cal P}(X,p), {\cal V}_{\mu}(X,p), {\cal A}_{\mu}(X,p)$ and 
${\cal S}_{\mu \nu}(X,p)$ are real functions. We note that
the spinor decomposition technique has been already used in
the formulation of the transport theory for QED \cite{VGE87,
BBGR91,ZH96} and for QHD (so-called quantum hadrodynamics)
\cite{EETAL87}.  

The vector and axial currents, see Eqs. (\ref{bcc}) and 
(\ref{acc}), are related to the functions ${\cal V}^{\mu}(X,p)$
and ${\cal A}^{\mu}(X,p)$ by the simple relations

\begin{equalign}
\label{bc}
V^{\mu}(X) = 4 \int {d^4p \over (2\pi)^4} {\cal V}^{\mu}(X,p)
\end{equalign}
and
\begin{equalign}
\label{ac}
A^{\mu}(X) = -4 \int {d^4p \over (2\pi)^4} {\cal A}^{\mu}(X,p).
\end{equalign}

\noindent Therefore ${\cal V}^{\mu}(X,p)$ and ${\cal A}^{\mu}
(X,p)$ can be interpreted as the momentum densities of these two
currents. For the physical interpretation of the other components 
we refer the reader to Refs. \cite{VGE87,BBGR91}. Under the chiral 
transformation (\ref{wftr}) the coefficients of the spinor 
decomposition (\ref{sd}) change as follows
\begin{eqnarray}
\label{chisd}
{\cal F} &\rightarrow& {\cal F}^{\prime} = {\cal F} \cos \chi
+ {\cal P} \sin \chi, \nonumber \\
{\cal P} &\rightarrow& {\cal P}^{\prime} = -{\cal F} \sin \chi
+ {\cal P} \cos \chi, \nonumber \\
{\cal V}_{\mu} &\rightarrow& {\cal V}_{\mu}^{\prime} =
{\cal V}_{\mu}, \\
{\cal A}_{\mu} &\rightarrow& {\cal A}_{\mu}^{\prime} =
{\cal A}_{\mu}, \nonumber \\
{\cal S}_{\mu \nu} &\rightarrow& {\cal S}_{\mu \nu}^{\prime}
= {\cal S}_{\mu \nu} \cos \chi + {\tilde {\cal S}}_{\mu \nu}
\sin \chi, \nonumber \\
{\tilde {\cal S}}_{\mu \nu} &\rightarrow& 
{\tilde {\cal S}}_{\mu \nu}^{\prime} =
-{\cal S}_{\mu \nu} \sin \chi +
{\tilde {\cal S}}_{\mu \nu} \cos \chi, \nonumber
\end{eqnarray}

\noindent where ${\tilde {\cal S}}^{\mu \nu}$ is the dual tensor
to ${\cal S}^{\mu \nu}$, namely

\begin{equalign}
\label{duals}
{\tilde {\cal S}}^{\mu \nu} = {1\over 2} 
\varepsilon^{\mu \nu \alpha \beta} {\cal S}_{\alpha \beta}.
\end{equalign}

\section{Kinetic Equations}

Substituting expression (\ref{sd}) into Eq. (\ref{ke}) gives the
following system of the coupled equations for the coefficients
of the decomposition (\ref{sd})

\begin{eqnarray}
\label{e1}
K^{\mu} {\cal V}_{\mu} - \sigma {\cal F} + \pi {\cal P} & = &
- {i \hbar \over 2} \left( \partial_{\nu} \sigma \partial_p^{\nu}
{\cal F} - \partial_{\nu} \pi \partial_p^{\nu} {\cal P} \right), \\
\label{e2}
-iK^{\mu}{\cal A}_{\mu} - \sigma {\cal P} -\pi {\cal F} & = &
- {i \hbar \over 2} \left( \partial_{\nu} \sigma \partial_p^{\nu}
{\cal P} + \partial_{\nu} \pi \partial_p^{\nu} {\cal F} \right), \\
\label{e3}
K_{\mu}{\cal F} + iK^{\nu}{\cal S}_{\nu \mu} 
- \sigma {\cal V}_{\mu}
+i\pi {\cal A}_{\mu} &=& -{i \hbar \over 2} \left(\partial_{\nu}
\sigma \partial_p^{\nu}{\cal V}_{\mu}
-i\partial_{\nu}\pi\partial_p^{\nu}
{\cal A}_{\mu} \right), \\
\label{e4}
iK^{\mu}{\cal P} -K_{\nu} {\tilde {\cal S}}^{\nu \mu} 
-\sigma {\cal A}^{\mu} 
+i\pi {\cal V}^{\mu} &=& -{i\hbar \over 2} \left( \partial_{\nu}
\sigma \partial_p^{\nu}{\cal A}^{\mu} -i \partial_{\nu} \pi
\partial_p^{\nu}{\cal V}^{\mu} \right), \\
\label{e5}
\!\!\!i(K^{\mu}{\cal V}^{\nu}\!\!-\!K^{\nu}{\cal V}^{\mu})
\!-\!\varepsilon^{\mu \nu \alpha \beta} K_{\alpha}{\cal A}_{\beta} 
-\! \pi {\tilde {\cal S}}^{\mu \nu}
\!+\!\sigma {\cal S}^{\mu \nu} \!\!\! &=& \!\!\!{i \hbar \over 2}
(\partial_{\gamma}\sigma \partial_p^{\gamma}{\cal S}^{\mu \nu}\!\!
- \partial_{\gamma}\pi \partial_p^{\gamma}{\tilde 
{\cal S}}^{\mu \nu}). 
\end{eqnarray}

\noindent In abbreviation we have introduced here the notation 
$K^{\mu}=p^{\mu}+{i\hbar \over 2} \partial^{\mu}$. Eqs. (\ref{e1}) -
(\ref{e5}) should be supplemented by the formulae that determine
the mean fields. They follow from (\ref{sp1}) and have the form

\begin{equalign}
\label{s}
\sigma(X) 
= -2G \int {d^4p \over (2\pi)^4} \,\, \hbox{Sp} \,\, W(X,p)
= -8G \int {d^4p \over (2\pi)^4} {\cal F}(X,p)
\end{equalign}
and
\begin{equalign}
\label{p}
\pi(X) 
= -2G \int {d^4p \over (2\pi)^4} \,\, \hbox{Sp} \,\, i\gamma_5 W(X,p)
=  8G \int {d^4p \over (2\pi)^4} {\cal P}(X,p).
\end{equalign}

\smallskip
Using the transformation properties (\ref{chi2}), (\ref{chi3}) 
and (\ref{chisd}) we can check that this system of equations is 
chirally invariant. Strictly speaking, each of Eqs. (\ref{e1}),
(\ref{e2}) and (\ref{e5}) is separately a chirally invariant
equation. On the other hand, Eqs. (\ref{e3}) and (\ref{e4}),
after a chiral transformation, form linear combinations of
each other. These combinations imply, however, that each of
their components must vanish separately. In consequence,
we can treat {\it the system of the two} equations (\ref{e3})
and (\ref{e4}) as a chirally invariant expression. 
In the analogous way, one can check that
the system of Eqs. (\ref{s}) and (\ref{p}) is chirally 
invariant. 

For further analysis, it is convenient to discuss separately
the real and imaginary parts of Eqs. (\ref{e1}) - (\ref{e5}).
The real parts give

\begin{eqnarray}
\label{r1} 
p^{\mu} {\cal V}_{\mu} - \sigma {\cal F} + \pi
{\cal P} &=& 0, \\
\label{r2}
{\hbar \over 2} \partial_{\mu}{\cal A}^{\mu} - \sigma {\cal P}
-\pi {\cal F} &=& 0, \\
\label{r3}
p_{\mu}{\cal F} - {\hbar \over 2} \partial^{\nu}{\cal S}_{\nu \mu}
-\sigma {\cal V}_{\mu} &=& -{\hbar \over 2} \partial_{\nu} \pi
\partial_p^{\nu}{\cal A}_{\mu}, \\
\label{r4}
-{\hbar \over 2} \partial^{\mu}{\cal P} - 
p_{\nu} {\tilde {\cal S}}^{\nu \mu}
-\sigma {\cal A}^{\mu} &=& -{\hbar \over 2}\partial_{\nu}
\pi \partial_p^{\nu} {\cal V}^{\mu}, \\
\label{r5}
-{\hbar \over 2} \left(\partial^{\mu} {\cal V}^{\nu} -
\partial^{\nu} {\cal V}^{\mu} \right) -
\varepsilon^{\mu \nu \alpha \beta} p_{\alpha} {\cal A}_{\beta} +
\sigma {\cal S}^{\mu \nu} - \pi 
{\tilde {\cal S}}^{\mu \nu} &=& 0.
\end{eqnarray}

\noindent One can notice that integrating of Eq. (\ref{r2}) over the 
momentum gives 

\begin{equalign}
\label{acc1}
\hbar \int {d^4p \over (2\pi)^4} 
\partial_{\mu} {\cal A}^{\mu}(X,p) = 0,
\end{equalign}

\noindent
where the definitions (\ref{s}) and (\ref{p}) have been used. 
This equation is nothing other than the statement that the
axial current should be conserved (\ref{acc}). We thus see
that after making the gradient expansion, this conservation law 
is still included in the transport equations.
The imaginary parts of Eqs. (\ref{e1}) - (\ref{e5}) yield

\begin{eqnarray}
\label{i1}
{\hbar \over 2} \partial^{\mu} {\cal V}_{\mu} &=& 
-{\hbar \over 2} \left( \partial_{\nu} \sigma
\partial_p^{\nu} {\cal F} - \partial_{\nu} \pi
\partial_p^{\nu} {\cal P} \right), \\
\label{i2}
p^{\mu} {\cal A}_{\mu} &=& {\hbar \over 2} \left(
\partial_{\nu} \sigma \partial_p^{\nu} {\cal P} +
\partial_{\nu} \pi \partial_p^{\nu} {\cal F} \right), \\
\label{i3}
{\hbar \over 2} \partial_{\mu} {\cal F} + p^{\nu}
{\cal S}_{\nu \mu} + \pi {\cal A}_{\mu} &=&
-{\hbar \over 2} \partial_{\nu} \sigma \partial_p^{\nu}
{\cal V}_{\mu}, \\
\label{i4}
p^{\mu} {\cal P} - {\hbar \over 2} 
\partial_{\nu} {\tilde {\cal S}}^{\nu \mu}
+\pi{\cal V}^{\mu} &=& -{\hbar \over 2} \partial_{\nu}
\sigma \partial_p^{\nu} {\cal A}^{\mu}, \\
\label{i5}
\left(p^{\mu} {\cal V}^{\nu} - p^{\nu} {\cal V}^{\mu} \right)
-{\hbar \over 2} \varepsilon^{\mu \nu \alpha \beta}
\partial_{\alpha} {\cal A}_{\beta} &=& {\hbar \over 2} \left(
\partial_{\gamma}\sigma \partial_p^{\gamma} {\cal S}^{\mu \nu}
- \partial_{\gamma}\pi \partial_p^{\gamma}
{\tilde {\cal S}}^{\mu \nu}
\right). 
\end{eqnarray}

Although we have neglected the higher order gradients, Eqs. 
(\ref{r1}) - (\ref{i5}) are still quantum kinetic equations.
In order to obtain the classical equations one makes an 
expansion of ${\cal F}$ in powers of $\hbar$

\begin{equalign}
\label{psh1}
{\cal F} = {\cal F}_{(0)} + \hbar {\cal F}_{(1)} + \hbar^2 
{\cal F}_{(2)} + ... \, ,
\end{equalign}

\noindent and similarly of ${\cal P}, {\cal V}^{\mu},
{\cal A}^{\mu}, {\cal S}^{\mu \nu}, \pi$ and $\sigma$.
Expansions of the form (\ref{psh1}) are inserted
into Eqs. (\ref{r1}) - (\ref{i5}) and the expressions at 
the appropriate powers of $\hbar$ are compared. The detailed
description of this procedure will be the subject of the
next Sections.

\section{Constraint Equations in the Leading Order of $\hbar$} 

Substituting expansions of the type (\ref{psh1}) into Eqs.
(\ref{r1}) - (\ref{r5}), we find to leading (zeroth) order of 
$\hbar$
\begin{eqnarray}
\label{lr1}
p^{\mu} {\cal V}_{\mu}^{(0)} 
- \sigma_{(0)} {\cal F}_{(0)}
+ \pi_{(0)} {\cal P}_{(0)} &=& 0, \\
\label{lr2}
\sigma_{(0)} {\cal P}_{(0)} 
+ \pi_{(0)} {\cal F}_{(0)} &=& 0, \\
\label{lr3}
p^{\mu} {\cal F}_{(0)} 
- \sigma_{(0)} {\cal V}^{\mu}_{(0)} &=& 0, \\
\label{lr4}
{\tilde {\cal S}}_{(0)}^{\mu \nu} p_{\nu} - 
\sigma_{(0)} {\cal A}^{\mu}_{(0)}
&=& 0, \\
\label{lr5}
-\varepsilon^{\mu \nu \alpha \beta} p_{\alpha} 
{\cal A}_{\beta}^{(0)}
+ \sigma_{(0)} {\cal S}_{(0)}^{\mu \nu} - \pi_{(0)} 
{\tilde {\cal S}}_{(0)}^{\mu \nu} &=& 0.
\end{eqnarray}

\noindent Correspondingly, the zeroth order of Eqs. (\ref{i1}) 
- (\ref{i5}) has the form

\begin{eqnarray}
\label{li2}
p^{\mu} {\cal A}_{\mu}^{(0)} &=& 0, \\
\label{li3}
p^{\nu} {\cal S}_{\nu \mu}^{(0)} 
+ \pi^{(0)} {\cal A}^{(0)}_{\mu} &=& 0, \\
\label{li4}
p^{\mu} {\cal P}^{(0)} 
+ \pi^{(0)} {\cal V}_{(0)}^{\mu} &=& 0, \\
\label{li5}
p^{\mu} {\cal V}^{\nu}_{(0)} 
- p^{\nu} {\cal V}^{\mu}_{(0)} &=& 0.
\end{eqnarray}

\noindent These are only four equations, since Eq. (\ref{i1})
is already of first order in $\hbar$. From Eqs. (\ref{lr2})
and (\ref{lr3}) one finds

\begin{equalign}
\label{scriptp}
{\cal P}_{(0)} = - \pi_{(0)} { {\cal F}_{(0)} \over
\sigma_{(0)} }
\end{equalign}
and 
\begin{equalign}
\label{scriptv}
{\cal V}_{(0)}^{\mu} =
p^{\mu} { {\cal F}_{(0)} \over \sigma_{(0)} }. 
\end{equalign}

\noindent Then Eqs. (\ref{li4}) and (\ref{li5}) are automatically 
fulfilled. Moreover, Eqs. (\ref{lr1}) - (\ref{lr3}) lead to the 
mass-shell constraint for the function ${\cal F}_{(0)}(X,p)$, namely

\begin{equalign}
\label{msf}
[p^2 - M^2(X)]{\cal F}_{(0)}(X,p) = 0, \,\,\,\,\,\, M^2(X) = 
\sigma^2_{(0)}(X) + \pi^2_{(0)}(X).
\end{equalign}

\noindent From Eq. (\ref{lr5}), we find the following
expression for the spin tensor 

\begin{equalign}
\label{st}
{\cal S}^{\mu \nu}_{(0)} = 
- {\pi_{(0)} \over M^2} \left[ p^{\mu}{\cal A}^{\nu}_{(0)} 
- p^{\nu} {\cal A}^{\mu}_{(0)} \right]
+ {\sigma_{(0)} \over M^2} \varepsilon^{\mu \nu \alpha \beta}
p_{\alpha} {\cal A}^{(0)}_{\beta},
\end{equalign}

\noindent and the dual spin tensor 

\begin{equalign}
\label{dst}
{\tilde {\cal S}}^{\mu \nu}_{(0)} = 
- {\sigma_{(0)} \over M^2}\left[ p^{\mu}{\cal A}^{\nu}_{(0)}
- p^{\nu} {\cal A}^{\mu}_{(0)} \right]
- {\pi_{(0)} \over M^2} \varepsilon^{\mu \nu \alpha \beta}
p_{\alpha} {\cal A}^{(0)}_{\beta}.
\end{equalign}

\noindent Substituting now Eq. (\ref{dst}) into Eq. (\ref{lr4}), 
and using (\ref{li2}) one obtains a mass-shell constraint
for ${\cal A}^{\mu}_{(0)}(X,p)$ also, i.e.,

\begin{equalign}
\label{msa}
[p^2 - M^2(X)] {\cal A}^{\nu}_{(0)}(X,p) = 0.
\end{equalign}

\noindent Finally, using (\ref{st}), (\ref{li2}) and (\ref{msa}) we
find that (\ref{li3}) is satisfied. In summary, Eqs. (\ref{lr1})
- (\ref{li5}) lead to the expressions for ${\cal P}_{(0)}$ and
${\cal V}^{\mu}_{(0)}$ in terms of ${\cal F}_{(0)}$, and
for ${\cal S}^{\mu \nu}_{(0)}$ and ${\tilde {\cal S}}^{\mu
\nu}_{(0)}$ in terms of ${\cal A}^{\mu}_{(0)}$. Furthermore,
they lead to the two mass-shell constraints.

The mean fields appearing in Eqs. (\ref{lr1}) - (\ref{lr5}),
(\ref{li3}) and (\ref{li4}) are required to be calculated 
in a self-consistent way from Eqs. (\ref{s}) and (\ref{p}). 
Doing so, we find                          
\begin{eqnarray}                                                    
\label{s1}                                                          
\sigma_{(0)}(X) &=&                                                 
-8G \int {d^4p \over (2\pi)^4} {\cal F}_{(0)}(X,p), \\              
\label{p1}                                                          
\pi_{(0)}(X) &=& 8G \int {d^4p \over (2\pi)^4} {\cal P}_{(0)}(X,p). 
\end{eqnarray}

\noindent Due to the relation (\ref{lr2}), one finds that these       
two equations are not independent and can be reduced to a single
equation which determines the {\it invariant} mass $M(X)$. 
To see this relation more clearly, let us define the distribution 
function $F(X,p)$ via the expression                                    
\begin{equalign}                        
\label{df}                              
F(X,p) = { {\cal F}_{(0)}(X,p) \over \sigma_{(0)}(X) } = -                   
{ {\cal P}_{(0)}(X,p) \over \pi_{(0)}(X) }                                 
\end{equalign}                            

\noindent and the angle $\Phi(X)$ via the relations

\begin{equalign}
\label{dca}
\pi_{(0)} = M(X) \sin \Phi(X), \,\,\,\,\,
\sigma_{(0)} = M(X) \cos \Phi(X).
\end{equalign}
                                                                      
\noindent $F(X,p)$ is a chirally invariant 
quantity, and knowledge of it determines the value of $M(X)$. 
In addition, it allows us to calculate the vector current density 
${\cal V}^{\mu}_{(0)}$. However, 
Eqs. (\ref{s1}) and (\ref{p1}) do not determine separately the
fields $\pi_{(0)}$ and $\sigma_{(0)}$, i.e., the angle $\Phi(X)$.
This fact is in agreement with the requirements of the chiral
symmetry of the problem: under the chiral transformations 
$\Phi(X) \rightarrow \Phi^{\prime}(X) = \Phi(X) + \chi$ and,
consequently, the absolute value of $\Phi(X)$ has no
physical significance. 

With this result, we conclude the discussion of the expressions obtained 
to zeroth order in $\hbar$. The derived equations all display the
chirally invariant form. Let us turn now to the discussion 
of the equations which follow from Eqs. (\ref{i1}) - 
(\ref{i5}), considered to first order in $\hbar$.

\section{Kinetic Equation for the Quark Distribution Functions}

Employing Eqs. (\ref{scriptp}) and (\ref{scriptv}) in Eq. 
(\ref{i1}), one immediately finds 

\begin{equalign}
\label{ke1}
p^{\mu} \partial_{\mu} F(X,p) + M(X)\partial_{\mu}M(X)
\partial_p^{\mu} F(X,p) = 0,
\end{equalign}
where we have used the definition (\ref{df}). This is again a
chirally invariant equation. Due to the mass-shell condition
(\ref{msf}), we can express $F(X,p)$ as the sum of the quark 
and antiquark distribution functions $f^{+}(X,{\bf p})$ and
$f^{-}(X,{\bf p})$ defined by 

\begin{equalign}
\label{df1}
F(X,p) = 2\pi \left\{
{\delta(p^0-E_{p}(X)) \over 2E_{p}(X)} f^{+}(X,{\bf p}) +
{\delta(p^0+E_{p}(X)) \over 2E_{p}(X)} 
\left[f^{-}(X,-{\bf p}) - 1 \right] \right\}.
\end{equalign}

\noindent Here $E_{p}(X) = \sqrt{{\bf p}^2 + M^2(X)}$, and 
$\bf p$ is a three-momentum. Substituting expression 
(\ref{df1}) into Eq. (\ref{ke1}), and integrating over $p^0$ 
gives  

\begin{equalign}
\label{ke2}
p^{\mu}\partial_{\mu} f^{\pm}(X,{\bf p}) +
M(X)\partial_{\mu}M(X)\partial_p^{\mu} f^{\pm}(X,{\bf p}) = 0.
\end{equalign}

Since the last procedure involves non-trivial cancellations
let us discuss it in more detail. Eq. (\ref{ke1}) can be shortly
written as  ${\hat L} F(X,p) = 0$, where ${\hat L}$ is a 
linear differential operator, i.e.,  ${\hat L} = p^{\mu}
\partial_{\mu} + M \partial_{\mu}M \partial_p^{\mu}$.
After some algebra, we find that ${\hat L} E_p(X) =
(p^0 M \partial_0 M) / E_p(X)$ and ${\hat L} \delta(p^0
\mp E_p(X)) = M \partial_0 M \delta^{\prime}(p^0 \mp E_p(X))
[1 \mp p^0 / E_p(X)]$, where the prime denotes the
derivative with respect to $p^0$. Since ${\hat L}$ is a
linear operator, one also gets ${\hat L} \{[\delta(p^0 
\mp E_p(X))/E_p(X)] f^{\pm}(X,\pm {\bf p}) \} =
{\hat L} \left[\delta(p^0 \mp E_p(X))/E_p(X)\right]
f^{\pm}(X,\pm {\bf p}) + [\delta(p^0 \mp E_p(X)) / E_p(X)]
{\hat L} f^{\pm}(X,\pm {\bf p})$. Of course, if ${\hat L}$
acts on $f^{\pm}(X,\pm {\bf p})$ its energy component 
$\partial_p^0$ produces zero. Now using our previous results
we can check that $\int_{\Delta_{\pm}} dp^0 {\hat L}
\left[\delta(p^0 \mp E_p(X)) / E_p(X) \right]$ $=0$,
where $\Delta_{\pm}$ is an interval containing either $E_p(X)$
or $-E_p(X)$. Consequently, integrating the whole formula
(\ref{ke1}) over $\Delta_{\pm}$, one gets the desired equation
(\ref{ke2}). We note that in the case of antiquarks still the
replacement ${\bf p} \rightarrow - {\bf p}$ should be made. 

\bigskip
Eq. (\ref{ke2}) displays a typical form for fermion
mean-field theories. In our case, however, a characteristic
feature of this equation is the appearance of the chirally 
invariant mass $M(X)$. We emphasize that the four-momentum $p^{\mu}$ 
occurring in (\ref{ke2}) is to be taken on the mass shell, i.e., 
in this case $p^0 = E_{p}(X)$. At the same time the distribution 
functions $f^{\pm}$ are functions of the space-time coordinate 
$X$ and the three-momentum $\bf p$, with the derivative with respect 
to the energy defined to be zero. 

Using Eqs. (\ref{s1}), (\ref{df}) and (\ref{df1}) we find that the 
NJL gap equation takes the form

\begin{equalign}
\label{gap}
\sigma_{(0)}(X) = \sigma_{(0)}(X) \,\, 4 G
\int {d^3p \over (2\pi)^3} {1 \over E_{p}(X)} 
\left[1 - f^{+}(X,{\bf p}) - f^{-}(X,{\bf p}) \right],
\end{equalign}

\noindent which determines the mass $M(X)$ 
appearing in $E_p(X)$ in terms of the distribution functions
$f^{\pm}(X,{\bf p})$. One can easily notice that using Eq. (\ref{p1})
instead of (\ref{s1}) we would obtain the same equation as
(\ref{gap}), with the appearance of $\pi_{(0)}$ instead of
$\sigma_{(0)}$. {\it We note that Eq. (\ref{gap}) is the 
generalization of formulae (\ref{gap3Dt}) and (\ref{gap3Dd1}) to the
case of non-equilibrium quark distribution functions.}

\newpage
Using now Eqs. (\ref{bc}) and (\ref{df1}),  we 
find the following expression for the baryon current

\begin{equalign}
\label{bc1}
V^{\mu}(X) = 2 \int {d^3p \over (2\pi)^3 }
{p^{\mu} \over E_{p}(X)} \left[f^+(X,{\bf p}) - f^-(X,{\bf p}) 
+ 1 \right],
\end{equalign}

\noindent where $p^{\mu}$ is again on the mass shell, i.e.,
$p^0 = E_{p}(X)$. The constant appearing at the end of the square
bracket can be neglected, since it does not contribute to 
${\bf V}(X)$, and for $V^0(X)$ it introduces a constant (although
infinite) charge.

We note that Eqs. (\ref{ke2}) and (\ref{gap}) form a closed system of
equations: the distribution functions $f^{\pm}(X,{\bf p})$ determine the
invariant mass $M(X)$ through Eq. (\ref{gap}), whereas the space
dependence of $M(X)$ determines in (\ref{ke2}) the time evolution of
$f^{\pm}(X,{\bf p})$. In the case $\pi_{(0)}=0$ this system of equations was
first derived by Zhang and Wilets \cite{ZW92}, whose calculations were
based on the closed-time-path formalism. Our approach represents an
alternative derivation and generalizes their results to the case
$\pi_{(0)} \not = 0$.  The numerical solutions of Eqs. (\ref{ke2}) and
(\ref{gap}), describing the expansion of quark matter possibly created
in the ultra-relativistic heavy-ion collisions, were found in
\cite{AA95}.  Another class of solutions of this system of equations
will be discussed by us in the next Chapter.

\section{Spin Evolution}

In the classical limit the spin dynamics is described by the
behaviour of the function ${\cal A}^{\mu}_{(0)}(X,p)$ 
\cite{VGE87,BBGR91}. In order to derive the kinetic equation for 
${\cal A}^{\mu}_{(0)}(X,p)$, we use Eq. (\ref{i4}) and substitute 
into it the expression for ${\cal V}^{\mu}$ obtained from 
Eq. (\ref{r3}). Such a procedure gives, to the first order 
in $\hbar$, the following equation

\begin{equalign}
\label{se0}
p^{\mu} \partial_{\nu} {\cal A}^{\nu}_{(0)} - \sigma_{(0)}
\partial_{\nu} {\tilde {\cal S}}^{\nu \mu}_{(0)} - \pi_{(0)}
\partial_{\nu} {\cal S}^{\nu \mu}_{(0)} + M \partial_{\nu} M
\partial_p^{\nu} {\cal A}^{\mu}_{(0)} = 0.
\end{equalign}

\noindent Substituting the expression for the spin tensor 
(\ref{st}) and the dual spin tensor (\ref{dst}) into Eq. (\ref{se0})
leads to the equation determining the space-time evolution of 
${\cal A}^{\mu}_{(0)}(X,p) $, namely

\begin{equalign}
\label{se}
p^{\nu}\partial_{\nu} {\cal A}^{\mu}_{(0)} 
+ M \partial_{\nu} M \partial_p^{\nu} {\cal A}^{\mu}_{(0)}
+ {\partial_{\nu} M \over M} \left[p^{\mu} {\cal A}^{\nu}_{(0)} 
- p^{\nu} {\cal A}^{\mu}_{(0)} \right]
-\varepsilon^{\mu \nu \alpha \beta} \partial_{\nu} \Phi
p_{\alpha} {\cal A}^{(0)}_{\beta} = 0.
\end{equalign} 

\noindent We note that Eq. (\ref{se}) is again a chirally
invariant equation, since both the mass $M(X)$ and 
the gradient of the angle $\Phi(X)$
are chirally invariant quantities. Due to the condition (\ref{li2}), 
only three out of four equations in (\ref{se}) are independent. 
In fact, one can check that multiplication of (\ref{se}) by 
$p^{\mu}$ gives zero. 

The mass $M(X)$ appearing in (\ref{se}) 
has to be calculated from the system of equations (\ref{ke2}) 
and (\ref{gap}). Hence, it can be treated in (\ref{se}) 
as an externally prescribed function. On the other hand, 
the gradient $\partial_{\mu}\Phi(X)$ is not known, and until we 
specify how to calculate it, Eq. (\ref{se}) cannot be used 
to determine the time evolution of ${\cal A}^{\mu}_{(0)}(X,p)$. 
Moreover, one has to check whether the solutions of (\ref{se}) 
satisfy the requirement of axial current conservation (\ref{acc}). 
The last two points will be discussed in more detail in the 
next Section, where we examine the consistency of our all 
equations up to the first order in $\hbar$. 

In analogy to the QED calculations \cite{VGE87} we introduce the 
spin up and spin down phase-space densities

\begin{equalign}
\label{fpms}
F_{\pm s}(X,p) = F(X,p)
\pm S_{\mu}(X,p) { {\cal A}^{\mu}_{(0)}(X,p) \over M(X) },
\end{equalign}

\noindent where $S_{\mu}(X,p)$ is defined by

\begin{equalign}
\label{spin}
S_{\mu}(X,p) = { {\cal A}^{(0)}_{\mu}(X,p) 
\over \left[-{\cal A}^{(0)}_{\nu}(X,p)
{\cal A}^{\nu}_{(0)}(X,p) \right]^{1\over 2} }.
\end{equalign}

\noindent Due to the condition (\ref{li2}), the four-vector 
${\cal A}^{\mu}_{(0)}(X,p)$ is space-like, ${\cal A}^{\mu}_{(0)}
(X,p){\cal A}_{\mu}^{(0)}(X,p) < 0$, therefore $S_{\mu}(X,p) 
S^{\mu}(X,p) = -1$. The quantity $S^{\mu}(X,p)$ describes the 
mean spin orientation of the classical particles placed at the 
space-time point $X$ and having the four-momentum $p$. 
On the other hand, the magnitude of ${\cal A}^{\mu}_{(0)}(X,p)$,
through Eq. (\ref{fpms}), determines how many spin up
and spin down particles are present in the system.
Using Eq. (\ref{se}), we find

\begin{equalign}
\label{se2}
p^{\nu}\partial_{\nu} S^{\mu} + M \partial_{\nu}
M \partial_p^{\nu} S^{\mu} +
{ \partial_{\nu} M \over M} p^{\mu} S^{\nu} 
- \varepsilon^{\mu \nu \alpha \beta} 
\partial_{\nu} \Phi p_{\alpha} S_{\beta} = 0 ,
\end{equalign}

\noindent and consequently

\begin{equalign}
\label{kefs}
p^{\nu}\partial_{\nu} F_{\pm s}(X,p) + M \partial_{\nu}
M \partial_p^{\nu} F_{\pm s}(X,p) = 0.
\end{equalign}

\noindent Hence, similarly as in transport equations for
QED \cite{VGE87}, we find that the spin components decouple. Since 
both functions $F(X,p)$ and ${\cal A}_{(0)}^{\mu}(X,p)$ are 
the mass-shell, we can write the analogous decomposition as in 
(\ref{df1}), namely

\begin{equalign}
\label{dfs}
F_{\pm s}(X,p) = 2\pi \left\{ 
{ \delta(p^0-E_p(X)) \over 2E_p(X) } f^+_{\pm s}(X,{\bf p}) +
{ \delta(p^0+E_p(X)) \over 2E_p(X) } 
\left[ f^-_{\mp s}(X,-{\bf p}) - 1 \right] \right\}.
\end{equalign}

\noindent Substituting Eq. (\ref{dfs}) into Eq. (\ref{kefs}), 
and integrating over $p^0$ gives

\begin{equalign}
\label{kfs1}
p^{\mu}\partial_{\mu} f^{\pm}_{\pm s}(X,{\bf p}) +
M \partial_{\mu} M \partial_p^{\mu}
f^{\pm}_{\pm s}(X,{\bf p}) = 0.
\end{equalign}

\noindent By a direct comparison of (\ref{df1}) and (\ref{dfs}),
we find $f^{\pm}(X,{\bf p}) = {1\over 2} 
\left[f^{\pm}_{+s}(X,{\bf p})+f^{\pm}_{-s}(X,{\bf p}) \right]$.
Therefore the functions $f^{\pm}(X,{\bf p})$ simply
represent the spin-averaged densities. This has been given  
in agreement with the normalization of (\ref{gap}) and (\ref{bc1}).

\section{Consistency of the Classical Transport Equations}

The kinetic equations for the quark densities (\ref{ke1}) and for
the spin (\ref{se}) follow from Eqs. (\ref{i1}) and (\ref{i4})
considered to the first order in $\hbar$. We have yet to examine
(i) whether equations (\ref{i2}), (\ref{i3}) and 
(\ref{i5}) are consistent
with these to the same order and (ii) whether the additional
information for the angle $\Phi(X)$ appearing in (\ref{se}) can
be obtained. We note that this kind of checking has been done 
in the case of QED in \cite{VGE87}. 

First, in order to illustrate that there is no contradiction 
between (\ref{i1}) and (\ref{i3}) we substitute the quantities 
${\cal A}_{\mu}$ and ${\cal S}_{\nu \mu}$ obtained from Eqs. 
(\ref{r4}) and (\ref{r5}), respectively into Eq. (\ref{i3}). 
In this way, we find the following equation              
                                                                       
\begin{eqnarray}                                                       
\label{con1}                                                           
\hbar \left\{ \partial_{\mu} {\cal F}_{(0)} - { p^{\nu} \over          
\sigma_{(0)} } \left[ \partial_{\mu} \left( p_{\nu}                    
{{\cal F}_{(0)} \over \sigma_{(0)} } \right) - \partial_{\nu} \left(   
p_{\mu} { {\cal F}_{(0)} \over \sigma_{(0)} } \right) \right]          
- {\pi_{(0)} \over \sigma_{(0)} } \partial_{\mu} {\cal P}_{(0)}        
\right. & & \nonumber \\                                               
\left.                                                                 
+ {1\over \sigma_{(0)} } \left[ \pi_{(0)} \partial_{\nu} \pi_{(0)}     
\partial_p^{\nu} {\cal V}^{(0)}_{\mu} + \sigma_{(0)}                   
\partial_{\nu} \sigma_{(0)} \partial_p^{\nu} {\cal V}^{(0)}_{\mu}      
\right] \right\} = 0. & &                                              
\end{eqnarray}                                                         
                                                                       
\noindent Since we are interested in retaining terms to the first 
order in $\hbar$ we can use the relations (\ref{lr2}), (\ref{lr3}), 
(\ref{msf}), and the kinetic equation (\ref{ke1}) to find that 
(\ref{con1}), and consequently (\ref{i3}), is fulfilled. 

In an analogous fashion, using the kinetic equation for spin (\ref{se}),
the constraints (\ref{li2}) and  (\ref{msa}),  and the definition
(\ref{dca}), we can show that Eq. (\ref{i5}) is fulfilled         
again to the first order in $\hbar$. To do this, we multiply        
both sides of (\ref{i5}) by $\varepsilon_{\alpha \beta \mu \nu}$.    
Using the formula for ${\cal V}^{\nu}$ from (\ref{r3}), we        
find that the multiplication of the left-hand-side of (\ref{i5})
gives the expression                       
                                                                     
\begin{equalign}                                                     
\label{lhs}                                                          
-{\hbar \over \sigma_{(0)} } \left\{ \varepsilon_{\alpha \beta       
\mu \nu} p^{\mu} \left[ \partial_{\gamma} {\cal S}^{\gamma           
\nu}_{(0)} - \partial_{\gamma} \pi_{(0)} \partial_p^{\gamma}         
{\cal A}^{\nu}_{(0)} \right] - \sigma_{(0)} \left(                   
\partial_{\alpha} {\cal A}^{(0)}_{\beta} -                           
\partial_{\beta}  {\cal A}^{(0)}_{\alpha} \right) \right\}.          
\end{equalign}                                                       
                                                                     
\noindent On the other hand, the multiplication of the  
RHS of (\ref{i5}) yields  
                                                                     
\begin{equalign}                                                     
\label{rhs}                                                          
\hbar \left[ \partial_{\gamma} \sigma_{(0)} \partial_p^{\gamma}      
{\tilde {\cal S}}^{(0)}_{\alpha \beta} + \partial_{\gamma}           
\pi_{(0)} \partial_p^{\gamma} {\cal S}^{(0)}_{\alpha \beta}          
\right].                                                             
\end{equalign}                                                       
                                                                     
\noindent Using now the expressions for ${\cal S}^{(0)}_{\mu \nu}$   
and ${\tilde {\cal S}}^{(0)}_{\mu \nu}$, after a rather lengthy      
calculation in which Eqs. (\ref{li2}), (\ref{msa}), (\ref{dca})      
and (\ref{se}) are used, we find that the two expressions 
above are equal to each other. Consequently, Eq. (\ref{i5}) is             
fulfilled up to the first order in $\hbar$ as well.    
                                                                     
In this way, we have shown that Eqs. (\ref{i1}), (\ref{i3})
- (\ref{i5}) are satisfied in the first order. To prove this 
fact, we have used Eqs. (\ref{r1}) - (\ref{r5}), treating them as 
if they were satisfied up to the first order. Is this procedure
admissible? The answer to this question is yes. Only ${\cal F}$           
and ${\cal A}^{\mu}$ are independent functions (with ${\cal A}^{\mu}$ 
restricted by the axial current conservation (\ref{acc1})). Eqs. 
(\ref{r2}), (\ref{r3}) and (\ref{r5}) are just the definitions           
of ${\cal P}$, ${\cal V}^{\mu}$ and ${\cal S}^{\mu \nu}$, and
therefore are in our approach always satisfied. On the other hand
Eq. (\ref{r1}) gives in the first order      

\begin{equalign}                                                         
\label{r11}                                                              
(p^2-M^2) {\cal F}_{(1)} = 2\left[\sigma_{(0)} \sigma_{(1)}              
+ \pi_{(0)} \pi_{(1)} \right] {\cal F}_{(0)} +                           
\partial_{\nu} \pi_{(0)} {\cal A}^{\nu}_{(0)}                            
\end{equalign}                                                           
and
\begin{equalign}
\label{r11bis}
(p^2-M^2) {\cal P}_{(1)} = 2\left[\sigma_{(0)} \sigma_{(1)}
+ \pi_{(0)} \pi_{(1)} \right] {\cal P}_{(0)} +
\partial_{\nu} \sigma_{(0)} {\cal A}^{\nu}_{(0)},
\end{equalign}

\noindent whereas using Eq. (\ref{r4}) one finds 

\begin{equalign}                                                         
\label{r41}                                                              
(p^2-M^2) {\cal A}^{\mu}_{(1)} = 2\left[\sigma_{(0)} \sigma_{(1)}        
+ \pi_{(0)} \pi_{(1)} \right] {\cal A}^{\mu}_{(0)}                       
-M^2 F \partial^{\mu} \Phi.                                              
\end{equalign}                                                           
                                                                         
\noindent One can notice that Eq. (\ref{r41}) as it stands
is a chirally invariant expression. On the other hand Eqs. 
(\ref{r11}) and (\ref{r11bis}) form a system of chirally 
invariant equations in the same way as equations (\ref{e3}) 
and (\ref{e4}), or equations (\ref{s}) and (\ref{p}). One can 
also check, that using formula (\ref{r2}) as the definition 
of ${\cal P}$ and Eq. (\ref{r11}), we can derive Eq. (\ref{r11bis}). 
Therefore, only one of equations (\ref{r11}) and (\ref{r11bis}) 
is really independent. We observe that all the expressions 
(\ref{r11}) - (\ref{r41}) are just the generalized mass-shell  
constraints for ${\cal F}_{(1)}$, ${\cal P}_{(1)}$ and 
${\cal A}^{\mu}_{(1)}$, and they do not influence 
the relations in the zeroth order. 

The last equation that we are still required to check is Eq. 
(\ref{i2}). Substituting the quantity ${\cal P}$, 
calculated from (\ref{r2}) into (\ref{i2}), gives 

\begin{equalign}
\label{i21}
{M^2(X) \over 2} \partial_{\mu} \Phi(X) \partial_p^{\mu}
F(X,p) = p^{\mu} {\cal A}^{(1)}_{\mu}(X,p).
\end{equalign}

\noindent We see that Eq. (\ref{i21}) defines the parallel 
part of ${\cal A}^{\mu}$ in the first order.
Multiplying (\ref{r41}) by $p^{\mu}$ and using Eqs.          
(\ref{i21}) and (\ref{msf}) we obtain zero. Hence Eqs.         
(\ref{i21}) and (\ref{r41}) are consistent.

After deriving all possible equations up to the first order in 
$\hbar$ and checking their consistency,  we are still faced with 
two problems. Firstly, the system of our equations in the leading 
order is not closed, since we do not have an equation for $\Phi(X)$.
Secondly, it is still not clear whether our kinetic equation
for spin satisfies the requirement of the axial current conservation.

One {\it a priori} possible situation is that
the condition of the axial current
conservation eliminates the freedom connected with the
choice of $\partial_{\mu}\Phi(X)$. However, by studying some
simple situations, one can convince oneself that this is
not the case. For example, assuming that $M$ = const and
that initially the spin distribution is homogeneous
${\cal A}^{\mu}_{(0)}(t=0,{\bf x},p) = B^{\mu}(p)$ we find 
$p^0 \partial_0 {\cal A}^0_{(0)} = \varepsilon^{0 i j k} \, 
\partial_i \Phi \, p_j \, B_k(p)$. In this situation, the axial
current conservation at $t=0$ requires that $\partial_i \Phi
\int d^4p \, \varepsilon^{0 i j k} \, (p_j/p^0) \, B_k(p) =0$.
We can see that this equation is not sufficient to determine
the gradient of $\Phi$ and consequently the derivative
$\partial_0 {\cal A}^0_{(0)}$ remains not well defined. 
Consequently, in order to obtain the closed system 
of equations in the classical limit we have to make some 
assumption on the form of $\partial_{\mu}\Phi(X)$. The 
simplest chirally invariant choice is

\begin{equalign}
\label{phiconst}
\partial_{\mu} \Phi(X) = 0.
\end{equalign}

In the case (\ref{phiconst}), our kinetic equation (\ref{se}) 
allows us to determine the time evolution of the function 
${\cal A}^{\mu}_{(0)}(X,p)$ from its knowledge at some initial 
time. One can notice, however, that this equation allows for 
solutions which do not satisfy the requirement of the axial 
current conservation. One of these solutions, for the
case $M$ = const, has the form ${\cal A}^{\mu}_{(0)}(X,p) =
s^{\mu}(p) s_{\nu}(p) X^{\nu} \delta^{(4)}(p-{\tilde p})$,
where $s^{\mu}(p)$ is a vector satisfying the conditions
$s^{\mu}(p) p_{\mu} = 0$ and $s^{\mu}(p) s_{\mu}(p) = -1$, 
and ${\tilde p}$ is some given
value of the four-momentum, with ${\tilde p}^{\mu}
{\tilde p}_{\mu} = M^2$. On the other hand, there also exist 
solutions of Eq. (\ref{se}) which satisfy the requirement
of the axial current conservation. These are, e.g.,
the static homogeneous solutions (for the case $M$ = const)
or the solutions which are odd functions of momentum. 

In consequence, we observe that Eq. (\ref{se}) neither
guarantees nor contradicts Eq. (\ref{acc}). Therefore
the axial current conservation should be used as an
external condition which selects the physical solutions
of (\ref{se}).

\section{Explicit Breaking of Chiral Symmetry}

In the case $m \not =0$, the chiral invariance of the
Lagrangian (\ref{l1}) is explicitly broken by the
expression $-m {\bar \psi} \psi$. In this situation,
the additional term in the Dirac equation leads to a
simple modification of our kinetic equations (\ref{e1}) -
(\ref{e5}); we have to replace the mean field $\sigma$
by the sum $\sigma + m$. Of course, in practice, 
this change affects only the real parts of our equations, 
i.e., the formulae (\ref{r1}) - (\ref{r5}), since the
imaginary parts (\ref{i1}) - (\ref{i5}) depend only
on the gradients of $\sigma$. On the other hand,
we observe that Eqs. (\ref{s1}) and (\ref{p1}) remain 
unchanged because they are just definitions of the 
mean fields.

The important fact in the non-symmetric case is that
the relations between the functions ${\cal F}_{(0)},
{\cal P}_{(0)}$ and ${\cal V}^{\mu}_{(0)}$ look 
different. In particular, instead of Eqs. (\ref{scriptp})
and (\ref{scriptv}), we now find 

\begin{equalign}
\label{newscript}
{\cal P}_{(0)} = - \pi_{(0)} { {\cal F}_{(0)} \over
\sigma_{(0)} + m }, \,\,\,\,\,
{\cal V}^{\mu}_{(0)} = p^{\mu} { {\cal F}_{(0)} \over
\sigma_{(0)} + m }.
\end{equalign}

\noindent Using the first of the relations (\ref{newscript})
in Eqs. (\ref{s1}) and (\ref{p1}), we obtain

\begin{equalign}
\label{s2}
\sigma_{(0)}(X) + 8G \int {d^4p \over (2\pi)^4} 
{\cal F}_{(0)}(X,p) = 0,
\end{equalign}

\noindent and

\begin{equalign}
\label{p2}
\pi_{(0)}(X) \left[ \sigma_{(0)}(X) + m 
+ 8G \int {d^4p \over (2\pi)^4} {\cal F}_{(0)}(X,p) \right]
= 0.
\end{equalign}

\noindent A straightforward consequence of these two
equations is the condition 

\begin{equalign}
\label{phizero}
\pi_{(0)}(X) = 0,
\end{equalign}

\noindent which implies also that ${\cal P}_{(0)}(X,p) = 0$ and 
$\Phi(X) = 0$. Therefore, we observe that if the system is not 
chirally invariant, equations (\ref{s1}) and (\ref{p1}) allow 
for the determination of the angle $\Phi(X)$.

For $m \not = 0$, the mass-shell constraints (\ref{msf})
and (\ref{msa}), as well as the kinetic equations (\ref{ke1}) 
and (\ref{se}) preserve their form, only in this case we should 
use the definitions 

\begin{equalign}
\label{newdef}
F(X,p) = { {\cal F}_{(0)} \over \sigma_{(0)} + m }, \,\,\,\,\,\,
M(X) = \sigma_{(0)} + m.
\end{equalign}

\noindent The gap equation (\ref{gap}) has now the form

\begin{equalign}
\label{newgap}
M(X) = m + 4GM \int {d^3p \over (2\pi)^3 } {1 \over
\sqrt{M^2(X)+{\bf p}^2} } \left[1 - f^{+}(X, {\bf p}) -
f^{-}(X,{\bf p}) \right].
\end{equalign}

Let us now discuss the behaviour of the axial current in the case 
when $m \not = 0$. In this situation, we no longer have an 
axial current conservation law, and using Eq. (\ref{r2}), 
we can write

\begin{equalign}
\label{pcac}
\hbar \left\{ {1\over 2} \partial_{\mu}
{\cal A}^{\mu}_{(0)} - (\sigma_{(0)} + m)
{\cal P}_{(1)} - \pi_{(1)} {\cal F}_{(0)} \right\} =0.
\end{equalign}

\noindent The integration of (\ref{pcac}) over momentum gives

\begin{equalign}
\label{pcac1}
\partial_{\mu} A^{\mu}_{(0)}(X) = - {m \over G} \pi_{(1)}(X).
\end{equalign}

\noindent In contrast to the $m = 0$, case we observe that
such an integration does not lead to any constraint for 
${\cal A}^{\mu}_{(0)}$ itself. Therefore, without any 
restrictions on $\partial_{\mu} {\cal A}^{\mu}_{(0)}$ 
we can treat Eq. (\ref{pcac}) as a defining equation for
${\cal P}_{(1)}$. One can still check that Eq. (\ref{pcac})
leads to the mass-shell constraint for ${\cal P}_{(1)}$
which agrees with (\ref{r11bis}). The latter formula,
for the case $m \not = 0$, should be rewritten in the form
$(p^2-M^2) {\cal P}_{(1)} = \partial_{\nu} M 
{\cal A}^{\nu}_{(0)}$. We note that Eq. (\ref{pcac1}) describes
the partial conservation of the axial current (PCAC) in our 
case.  

\chapter{\bf Large Time-Scale Fluctuations of the Quark Condensate}
\label{chapt:fluctuations}

In this Chapter we deal with the transport equations, again reducing
our interest only to quark dynamics in the mean (Vlasov) field. We
shall consider the systems which are not uniform but close to
equilibrium. This allows us to linearize the kinetic equations and to
use a combination of analytical and numerical methods to look for
solutions.

Our main aim is to study the properties of a hadronic system in which
the energy density is very large but which still exhibits a chirally
broken phase. One could think of a fireball formed in high-energy
heavy-ion collisions but at energies below the threshold for production
of a chirally restored phase. The temperature range of interest for
us lies between 100 and 150 MeV, and we assume that the critical 
temperature is 190 MeV. The baryon density is zero in our case,
therefore, our considerations are appropriate for the description
of the central rapidity region. We cannot study the systems at
temperatures close to $T_c$, because one of the ingredients of our
approach is that the mean field contains a large constant component
in addition to the fluctuating part. The former vanishes (becomes
negligible) at $T_c$ and this fact makes our approach inappropriate
for very high temperatures.

Because the energy density of such a fireball is large, we hope that
the description of its space-time evolution in terms of quarks is
meaningful. As we shall see later, our equations will not lead to the
separation of colour, although the NJL model does not include
confinement. In principle, one has to take into account the dynamics
of mesons as well. However, in practice it is not completely clear how
this can be done in the situation out of equilibrium and in the fully
satisfactory way, i.e., protecting all the conservation laws and
symmetries.

It is important to notice that already a quantum fluctuation of the
condensate, i.e., the sigma meson, exists. However, due to its large
mass the typical time scale over which the sigma meson wave function
changes is a fraction of fermi: $\Delta t < 1/m_{\sigma} \approx $
0.2 fm. In the approach presented here we study the large time-scale
(classical) fluctuations of the condensate and neglect such
oscillating short time-scale changes, although if the latter are
taken into account the interesting phenomena can occur from couplings
of the fluctuations on both scales. Nevertheless, the inclusion
of these short time-scale fluctuations is equivalent to taking
into account the mesonic degrees of freedom and, as it was discussed
above, represents a separate, as yet unsolved problem.

The non-uniformity of our system will lead to fluctuations of the 
mean field around its average value that is determined by the
thermal background. Since in the NJL model a change in the mean field
is proportional to a change in the condensate, the latter will also
fluctuate. As a consequence, through the use of our transport equations 
for the quark-antiquark plasma we are able to predict the space-time
development of such fluctuations.

\section{Linearized Kinetic Equations}
\label{sect:linearization}

The starting point for our investigations are Eqs. (\ref{ke2})
and (\ref{newgap}). Their straightforward generalization to the
case of arbitrary number of flavours ($N_f$) and colours ($N_c$)
has the form

\begin{equalign}
\label{lke1}
p^{\mu}\partial_{\mu} f(X,{\bf p}) +
M(X)\partial_{\mu}M(X)\partial_p^{\mu} f(X,{\bf p}) = 0
\end{equalign}
and
\begin{equalign}
\label{lgap}
1 = {m \over M(X)} + 4 N_c N_f G \int {d^3p \over (2\pi)^3 } 
{1 \over p^0} \left[1 - f(X, {\bf p}) \right].
\end{equalign}

\noindent Here $p^{\mu} = (p^0,{\bf p})$ is the four-momentum of
a quark, $X^{\mu} = (t,{\bf r})$ is its space-time position, $m$ is
the current quark mass, and $G$ is the coupling constant. The function
$f(X,{\bf p})$ is the sum of the quark and antiquark distribution
functions, i.e., $f(X,{\bf p}) = f^{+}(X,{\bf p}) + f^{-}(X,{\bf
p})$. Each of the latter two functions satisfies equation (\ref{ke2})
so we can add and subtract such two equations. Assuming that
$f^{+}(X,{\bf p}) = f^{-}(X,{\bf p})$ the equation for the difference
of two distributions is trivially fulfilled and we are left only with
Eq. (\ref{lke1}). In the case $N_f=2$ the constituent quark mass $M(X)$ 
is related to the quark condensate by the relation

\begin{equalign}
\label{mcon}
M(X) = m - 4 G \langle {\overline q} (X) q(X) \rangle,
\end{equalign}

\noindent which is a generalization of formula (\ref{gap2fst}) for
non-equilibrium situations.

In our approach quarks are always on the mass shell, i.e., $p^0 =
E_p(X) = \sqrt{{\bf p}^2 + M^2(X)}$, but their mass is a function of
space-time position. Therefore, the distribution function depends only
on space-time variables and three-momentum ${\bf p}$.  Consequently,
the derivative with respect to energy, $p^0$, in (\ref{lke1})
vanishes.  However, we have included it in our expression (\ref{lke1})
to show the relativistic invariance of the kinetic equation
explicitly. The gap equation (\ref{lgap}) is also relativistically
invariant, but the integral appearing there is divergent \footnote{The
quark distribution function vanishes fast for ${\bf p} \rightarrow
\infty$, however the integrand in Eq. (\ref{lgap}), containing the
difference $1 - f(X,{\bf p})$, does not.}. For the sake of 
simplicity and also in order to make connection to other calculations
done in the framework of the NJL model, we shall regularize (\ref{lgap})
using a 3-dimensional cutoff $\Lambda$. This procedure, of course,
introduces an explicit breaking of the Lorentz invariance. In the
following, whenever the integration is restricted to the region
${\bf p} < \Lambda$ we shall indicate this explicitly in the expressions.

It is interesting to observe that the equality of the quark and the
antiquark distribution functions (for each colour separately) has some
nice consequences for us; we can think that on the average colour is
compensated by anticolour at each point and, therefore, the discussed
states of the plasma are colourless. In other words, although we
describe the evolution of our system in terms of the motion of quarks
our equations will not lead to the separation of colour. (We note that
our assumption concerning the equality of the quark and the antiquark
distribution functions is reasonable only for the systems with net
baryon number density equal to zero.)

In the following we are going to consider the situation in which the
distribution function can be written as a sum of the isotropic
background distribution and a small perturbation, namely,

\begin{equalign}
\label{eq4}
f(t,{\bf r},{\bf p}) = f_0(p^2) + \delta f(t,{\bf r},{\bf p}).
\end{equalign}

\noindent Correspondingly, the mean field can be written as 

\begin{equalign}
\label{eq5}
M(t,{\bf r}) = M_c + \delta M(t,{\bf r}),
\end{equalign}

\noindent where $M_c$ is the constant part and $\delta M(t,{\bf r})$
is the temporally and spatially varying fluctuation. We shall assume
that the background distribution is a thermal one

\begin{equalign}
\label{eq6}
f_0(p^2) = {2 \over e^{\sqrt{p^2+M^2_c}/T} + 1}.
\end{equalign}

\noindent Here $T$ is the temperature and the factor 2 comes from summing 
up the quark and the antiquark distribution functions as was discussed
above (the chemical potential $\mu = 0$ in our case). According to
Eq. (\ref{lgap}) the fluctuating part of the condensate can be written
as $\delta \langle {\overline q} (X) q(X) \rangle = - \delta M(X)/4G$,
therefore, from now on we shall discuss simply the disturbances of the 
mean field. 

Substituting expressions (\ref{eq4}) and (\ref{eq5}) into Eqs.
(\ref{lke1}) and (\ref{lgap}) and keeping only the zeroth and first
order terms in the deviations $\delta f(t,{\bf r},{\bf p})$ and
$\delta M(t,{\bf r})$ gives us the following three equations:

\begin{equalign}
\label{eq7}
1 = {m \over M_c} + 4 N_c N_f G \int^{\Lambda} {d^3p \over (2\pi)^3}
{1 \over E_p} \left[1 - f_0(p^2) \right],
\end{equalign}

\begin{equalign}
\label{eq8}
{\partial \over \partial t} \, \delta f(t,{\bf r},{\bf p}) +
{\bf v} \cdot {\partial \over \partial {\bf r}} \,
\delta f(t,{\bf r},{\bf p}) -
{M_c \over E_p} {\partial \over \partial {\bf r}}
\delta M(t,{\bf r}) \cdot {\partial \over \partial {\bf p}}
\, f_0(p^2) = 0,
\end{equalign}

\begin{equalign}
\label{eq9}
\delta M(t,{\bf r}) \left[ {m \over M_c} + 4 N_c N_f G \int^{\Lambda}
{d^3p \over (2\pi)^3} {M_c^2 \over E_p^3} \left(1 - f_0(p^2) \right)
\right] = -4 N_c N_f G \int^{\Lambda} {d^3p \over (2\pi)^3}
{M_c \over E_p} \delta f(t,{\bf r},{\bf p}).
\end{equalign}

\noindent Here $E_p$ and ${\bf v}$ are the leading order expressions 
for the energy of a quark, $E_p = \sqrt{p^2 + M_c^2}$, and its velocity
${\bf v} = {\bf p}/E_p$.

Equation (\ref{eq7}) is the zeroth order gap equation connecting the
constant part of the mean field $M_c$ with the background distribution
$f_0(p^2)$ and has a form known from the NJL calculations at finite
temperature (see, e.g., Eq. (\ref{gap3Dt})). Solving (\ref{eq7}) we find
$M_c$ as a function of the temperature $T$. Assuming that $m=0$ and
fitting other parameters as in Ref. \cite{ZHK94}, i.e., $N_c = 3, N_f
= 2, G$ = 5.01 GeV $^{-2}$, and $\Lambda$ = 650 MeV, we find that at
zero temperature $M_c$ = 313 MeV and the corresponding value of the
quark condensate is (-250 MeV)$^3$.  We note that this set of
parameters gives the same temperature dependence of the mean field as
the parameters of the one-flavour model used in Section 5.1.

\begin{figure}[ht]
\label{bt}
\xslide{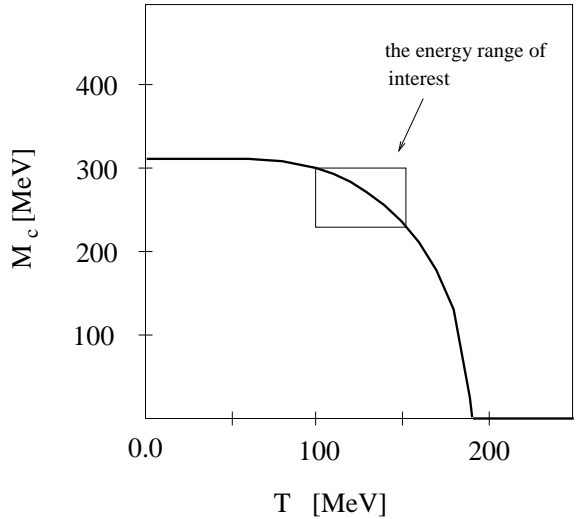}{8cm}{49}{181}{555}{671}
\caption{\small Temperature dependence of the background mean field $M_c$.
                The values of the parameters were taken in such a way 
                as to get the agreement with the temperature dependence
                shown in Fig. [5.3]. The temperature range of interest
                (together with the corresponding range of $M_c$)
                is framed by the box.}
\end{figure}

\section{Conservation Laws}
\label{sect:conservations}

Because we neglect the processes of $q {\overline q}$ annihilation
the total number of quarks and antiquarks should be conserved. We
can write this conservation law formally as

\begin{equalign}
\label{eq10}
\partial_{\mu} N^{\mu}(X) = 0,
\end{equalign}

\noindent where the number current is defined as

\begin{equalign}
\label{eq11}
N^{\mu}(X) = 2 N_c N_f \int {d^3p \over (2\pi)^3} {p^{\mu} \over p^0}
f(X,{\bf p}).
\end{equalign}

\noindent In fact, starting from (\ref{eq11}) one can derive (\ref{eq10})
using the kinetic equation (\ref{lke1}) and integrating by parts.
Similarly, we can show that the entropy of the system is conserved,
namely,

\begin{equalign}
\label{eq12}
\partial_{\mu} S^{\mu}(X) = 0,
\end{equalign}
where
\begin{equalign}
\label{eq13}
S^{\mu}(X) = - 2 N_c N_f \int {d^3p \over (2\pi)^3} {p^{\mu} \over p^0}
\left[ f(X,{\bf p}) \ln f(X,{\bf p}) + \left[1-f(X,{\bf p})\right]
\ln \left[1-f(X,{\bf p})\right] \right]
\end{equalign}
is the entropy flux. The latter is conserved because we consider a 
collisionless plasma.

Let us now define the energy-momentum tensor of quarks in the standard 
way by

\begin{equalign}
\label{eq14}
T^{\mu \nu}_{\hbox{quarks}}(X) = 2 N_c N_f \int {d^3p \over (2\pi)^3} 
{p^{\mu} p^{\nu} \over p^0} f(X,{\bf p}).
\end{equalign}
Using again the kinetic equation (\ref{lke1}) and integrating by parts
one obtains
\begin{equalign}
\label{eq15}
\partial_{\mu} T^{\mu \nu}_{\hbox{quarks}}(X) = 2 N_c N_f M(X)
\partial^{\nu} M(X) \int {d^3p \over (2\pi)^3}
{1 \over p^0} f(X,{\bf p}).
\end{equalign}

\noindent The result (\ref{eq15}) shows that the total energy and the 
total momentum of quarks alone are not conserved; this is due to the
fact that (\ref{eq14}) includes only the kinetic energy of particles.
The question arises: What tensor, describing the potential energy,
should be added to (\ref{eq14}) in order to obtain a conserved
quantity?  In the thermodynamic approach of \cite{HKZV94,ZHK94} it was
found that the expression

\begin{equalign}
\label{eq16}
\Omega_{\hbox{vac}}(X) = { \left[ M(X) - m \right]^2 \over 4 G}
- 2 N_c N_f \int^{\Lambda} {d^3p \over (2\pi)^3} p^0
\,\,\,\,\,\,\,\,\, (\,p^0 = E_p(X)\,)
\end{equalign}

\noindent represents the energy density of the vacuum (with the corresponding
vacuum pressure given by $-\Omega_{\hbox{vac}}(X)$). Because the
vacuum is isotropic we write its energy-momentum tensor as

\begin{equalign}
\label{eq17}
T^{\mu \nu}_{\hbox{vac}}(X) = g^{\mu \nu} \Omega_{\hbox{vac}}(X),
\end{equalign}
where $g^{\mu \nu}$ is the metric tensor. Using now the gap
equation (\ref{lgap}) we can show that

\begin{equalign}
\label{eq18}
\partial_{\mu} T^{\mu \nu}_{\hbox{vac}}(X) = - 2 N_c N_f M(X)
\partial^{\nu} M(X) \int^{\Lambda} {d^3p \over (2\pi)^3}
{1 \over p^0} f(X,{\bf p}).
\end{equalign}

\noindent We see that for $\Lambda \rightarrow \infty$ the summed
energy of the quarks and of the vacuum is conserved. This is an
interesting result showing that one has to include the energy of the
zero modes of quarks (the second term on the RHS of Eq. (\ref{eq16}))
to have the conservation law. Since, $\Lambda$ is finite in our
approach, it means that the energy is not exactly conserved.
However, as long as we look at the leading terms of the quantities in
our expansion defined by Eqs. (\ref{eq4}) and (\ref{eq5}) we find that
this violation is very small and, therefore, can be disregarded.

One can check that it is possible to obtain a strict conservation of
energy if the gap equation (\ref{lgap}) is regularized in a slightly
different way: The integral on the RHS of (\ref{lgap}) should be
written as a sum of two terms. The first one, independent of $f(X,{\bf
p})$, has to be regularized with the cutoff, whereas the second term
containing $f(X,{\bf p})$ is converging and should not be additionally
regularized. Such a scheme yields the exact conservation laws but at
very high temperatures it can lead to negative and, therefore,
non-acceptable solutions for the mean field --- this is the situation
discussed in Section 5.1 below Eq. (\ref{i1mat}). 

\section{Energy-Momentum Representations}
\label{sect:en_mom}

We now come back to the discussion of Eqs. (\ref{eq8}) and (\ref{eq9}).
Since these equations are linear, it is useful to expand the functions
$\delta M(t,{\bf r})$ and $\delta f(t,{\bf r},{\bf p})$ as Fourier
integrals

\begin{equalign}
\label{eq19}
\delta f(t,{\bf r},{\bf p}) = \int {d^3k \over (2\pi)^3}
\delta f(t,{\bf k},{\bf p}) e^{i {\bf k} \cdot {\bf r}},
\end{equalign}

\begin{equalign}
\label{eq20}
\delta M(t,{\bf r}) = \int {d^3k \over (2\pi)^3}
\delta M(t,{\bf k}) e^{i {\bf k} \cdot {\bf r}},
\end{equalign}
and substitute them into (\ref{eq8}) and (\ref{eq9}). In this way we
find two coupled equations for each Fourier component separately

\begin{equalign}
\label{eq21}
\left[ {\partial \over \partial t} + i {\bf k} \cdot {\bf v} \right] 
\delta f(t,{\bf k},{\bf p}) - {i M_c \over E_p} \, \delta M(t,{\bf k})
\, {\bf k} \cdot {\partial f_0 \over \partial {\bf p}} = 0,
\end{equalign}

\begin{equalign}
\label{eq22}
\delta M(t,{\bf k}) \left[{m \over M_c} + 4 N_c N_f G \int^{\Lambda}
{d^3p \over (2\pi)^3} {M^2_c \over E_p^3} (1 - f_0) \right] =
-4 N_c N_f G \int^{\Lambda} {d^3p \over (2\pi)^3} {M_c \over E_p}
\delta f(t,{\bf k},{\bf p}).
\end{equalign}

In the situation under consideration the system is perturbed at $t =
0$ and the perturbations propagate forward in time, i.e., $\delta
M(t,{\bf k})$ and $\delta f(t,{\bf k},{\bf p})$ are only different
from zero for $t > 0$. Consequently, to solve (\ref{eq21}) and
(\ref{eq22}), we use a one-sided Fourier transform in time
\footnote{In what follows we adopt the Landau method \cite{LDL65,Sogan94} 
for studying the relaxation of the initial perturbations in 
electron-ion plasmas.}. For the fluctuating part of the mean field
it is defined as 

\begin{equalign}
\label{eq23}
\delta M(\omega,{\bf k}) = \int_0^{\infty} e^{i \omega t} 
\delta M(t,{\bf k}) \, dt.
\end{equalign}

\noindent Assuming $|\delta M(t,{\bf k})| < {\cal M} \exp(\sigma t)$,
where ${\cal M}$ and $\sigma$ are some positive numbers, we find that
Eq. (\ref{eq23})  is well defined for frequencies $\omega$ having an
imaginary part $\omega_I > \sigma$. Thus, the inverse transformation
is given by

\begin{equalign}
\label{eq24}
\delta M(t,{\bf k}) = \int_{-\infty + i\sigma}^{\infty + i\sigma} 
{d\omega \over 2 \pi} e^{- i \omega t} \delta M(\omega,{\bf k}),
\end{equalign}

\noindent where the integration contour is taken to be a straight line 
parallel to the real axis. We note that the assumption discussed above
means that all singularities of $\delta M(\omega,{\bf k})$ lie below
the contour. The one-sided Fourier transform of $\delta f(t,{\bf
k},{\bf p})$ is defined in the analogous way.

Multiplying (\ref{eq21}) by $\exp(i\omega t)$ and integrating over time
from zero to infinity gives

\begin{equalign}
\label{eq25}
\delta f(\omega,{\bf k},{\bf p}) = {1 \over i ({\bf k} \cdot
{\bf v} - \omega)} \left[ g({\bf k},{\bf p}) + {i M_c \over E_p}
\, \delta M(\omega,{\bf k}) \, {\bf k} \cdot {\partial f_0 \over 
\partial {\bf p}} \right],
\end{equalign}

\noindent where $g({\bf k},{\bf p})$ represents the initial condition 
for the distribution function $\delta f(t,{\bf r},{\bf p})$, namely,

\begin{equalign}
\label{eq26}
\delta f(t=0,{\bf r},{\bf p}) = g({\bf r},{\bf p}) =
\int {d^3k \over (2\pi)^3} g({\bf k},{\bf p}) e^{i {\bf k}
\cdot {\bf r}}.
\end{equalign} 

Making the one-sided Fourier transformation of Eq. (\ref{eq22})
and substituting into its RHS Eq. (\ref{eq25}) we find the following
expressions

\begin{equalign}
\label{eq27}
\delta M(\omega,{\bf k}) \, \, \gamma(\omega,{\bf k}) = 
{\tt G}(\omega,{\bf k}),
\end{equalign}
where
\begin{equalign}
\label{eq28}
\gamma(\omega,{\bf k}) = {m \over M_c} + 4 N_c N_f \,G \, 
\int^{\Lambda} {d^3p \over (2\pi)^3 } \left[ {M_c^2 \over E_p^3}
(1 - f_0) + {M_c^2 \over E_p^2} \, {1 \over {\bf k} \cdot 
{\bf v} - \omega} \, {\bf k} \cdot {\partial f_0 \over
\partial {\bf p}} \right]
\end{equalign}
and
\begin{equalign}
\label{eq29}
{\tt G}(\omega,{\bf k}) = 4i N_c N_f \,G\,  
\int^{\Lambda} {d^3p \over (2\pi)^3 } {M_c \over E_p}
{g({\bf k},{\bf p}) \over {\bf k} \cdot {\bf v} - \omega}.
\end{equalign}

Equation (\ref{eq27}) is very much similar to that found in the
case of investigations of the oscillations of the electron-ion
plasma \cite{LP81}. Relying on this analogy, we shall call 
$\gamma(\omega,{\bf k})$ the {\it permittivity} of the quark-antiquark
plasma. Substituting Eq. (\ref{eq27}) into (\ref{eq24}) we obtain

\begin{equalign}
\label{eq30}
\delta M(t,{\bf k}) = \int_{-\infty + i\sigma}^{\infty + i\sigma}
{d\omega \over 2\pi} e^{-i \omega t} 
{ {\tt G}(\omega,{\bf k}) \over \gamma(\omega,{\bf k}) }.
\end{equalign}

\noindent The analysis of this equation will be the subject of the 
next two Sections.

\section{Dispersion Relation and Analytic Properties of the Functions 
$\gamma$ and $\tt G$}
\label{sect:gandG}

We already know that the integration in (\ref{eq30}) must be performed
above all singularities of the function ${\tt G}(\omega,{\bf k}) /
\gamma(\omega,{\bf k})$. Eq. (\ref{eq28}) defines $\gamma(\omega,{\bf k})$
as the function of the complex frequency, $\omega = \omega_R + 
i \omega_I$, which is analytic everywhere except for the interval on
the real axis where $-v^{\max} k \le \omega_R \le v^{\max} k$.
Here $v^{\max} = \Lambda/\sqrt{\Lambda^2 + M_c^2}$ is the limiting
maximal velocity of quarks, which because of the presence of the cutoff
is slightly smaller than 1. The structure of the singularities of
${\tt G}(\omega,{\bf k})$ is analogous to that of 
$\gamma(\omega,{\bf k})$; they all lie on the real axis in the interval
defined above. The additional singularities which can appear in the
integrand of (\ref{eq30}) come from the zeros of the denominator,
i.e., they are solutions to the equation

\begin{equalign}
\label{eq31}
\gamma(\omega,{\bf k}) = 0.
\end{equalign}

\noindent Equation (\ref{eq31}) is nothing else than the dispersion
relation defining modes in the plasma, which have momentum ${\bf k}$
and energy $\omega(k)$ such that $\gamma[\omega(k),k]=0$.

Let us now try to find a solution to Eq. (\ref{eq31}). We first 
perform the angle integration in (\ref{eq28}) and find the following
representation of the function $\gamma(\omega,k)$:

\begin{equalign}
\label{eq32}
\gamma(\omega,k) = {m \over M_c} + {2 N_c N_f G M_c^2 \over \pi^2}
\int_0^{\Lambda} \, dp \, {p \over E_p} \left[ {p \over E_p^2}
(1 - f_0) - {\partial f_0 \over \partial p} \, \Phi \left(
{E_p \omega \over pk} \right) \right],
\end{equalign}

\noindent where

\begin{equalign}
\label{eq33}
\Phi(z) = {1\over 2} z \ln {z + 1 \over z - 1} - 1.
\end{equalign}
One can notice that $\Phi(z)$ has a cut for real values of $z$
contained in the interval $-1 \le z_R \le 1$. This is in agreement
with an earlier remark that $\gamma(\omega,k)$ has a cut for
$-v^{\max} k \le \omega_R \le v^{\max} k$.

For the arguments slightly shifted from the real axis we find

\begin{equalign}
\label{eq34}
\Phi(z = z_R \pm i\epsilon) = {1\over 2} z_R \left[
\ln {|z_R + 1| \over |z_R - 1|} 
\mp i \pi \theta(1-|z_R|) \right] -1,
\end{equalign}

\noindent where $\theta(x)$ is the step function. One of the consequences 
of Eqs. (\ref{eq32}) and (\ref{eq34}) is that the dispersion relation
(\ref{eq31}) can have a solution for real frequencies $(\omega=\omega_R)$
only if $|\omega_R| > k v^{\max}$. Otherwise, $\gamma(\omega,k)$ has a
non-vanishing imaginary part. A closer examination of the function
(\ref{eq33}) shows, however, that for $|\omega_R| > k v^{\max}$ the real
part of $\Phi(E_p \omega_R/p k)$ is always positive and so is the real
part of $\gamma(\omega,k)$. To see the last point one can notice that
the derivative of the thermal background distribution is always
negative. Consequently, we conclude that the dispersion relation
(\ref{eq31}) has no solution for real frequencies.

For purely imaginary arguments $\Phi(z)$ has the form

\begin{equalign}
\label{eq35}
\Phi(z = i z_I) = |z_I| \left( {\pi \over 2} - \hbox{arctan} |z_I|
\right) - 1,
\end{equalign}
hence it is a negative real function. In this case (\ref{eq32}) is a
difference of two real positive functions and one can numerically
check if there exists a solution to Eq. (\ref{eq31}). In the
temperature range of interest, i.e., for 100 MeV $\le T
\le$ 150 MeV, we have found that $\gamma(\omega,k)$ is always larger
than zero. This fact again means that our dispersion relation has no
solution.

To complete our investigation we have studied numerically the
imaginary part of $\gamma(\omega,k)$ for $\omega_R, \omega_I > 0$
again in the temperature range 100 MeV $\le T \le$ 150 MeV. We have
checked that Im $\gamma(\omega,k)$ is always negative in the discussed
region and vanishes at infinity, i.e., for $|\omega| \rightarrow
\infty$. Employing the symmetries $ \hbox{Im}
\gamma(\omega_R-i\omega_I,k) = - \hbox{Im}
\gamma(\omega_R+i\omega_I,k)$ and $ \hbox{Im}
\gamma(-\omega_R+i\omega_I,k) = - \hbox{Im}
\gamma(\omega_R+i\omega_I,k)$ we find that Im$\gamma(\omega,k)$ is
different from zero also in other quadrants. This result, together
with the two previous ones, indicates that there is no solution
to Eq. (\ref{eq31}) for any finite $\omega$.

Because the only singularities of the integrand in (\ref{eq30}) are
those coming from the cut discussed earlier (placed on the real
axis) our parameter $\sigma$ can be taken to be an arbitrarily small
positive number. For $t < 0$ we can close the contour of integration
in the upper half-plane. Because the integrand in (\ref{eq30})
has no singularities in this region, we find in this case that
$\delta M(t,{\bf k})=0$, what is in agreement with the requirements
of causality. For $t>0$ we can close the contour of the integration 
in the lower half-plane and, later, we contract it to go just around
the cut. As the result of this procedure we obtain the expression

\begin{equalign}
\label{eq36}
\delta M(t,{\bf k}) = \int\limits_{-k v^{\max}}^{k v^{\max}}
{d\omega \over 2\pi} e^{-i\omega t} 
\left[ { {\tt G}(\omega + i\epsilon,{\bf k}) \over 
   \gamma(\omega + i\epsilon,{\bf k})} -
{ {\tt G}(\omega - i\epsilon,{\bf k}) \over 
   \gamma(\omega - i\epsilon,{\bf k})} \right].
\end{equalign}

For further use, it will be convenient to represent $\gamma(\omega_R
\pm i\epsilon,k)$ explicitly as the sum if its real and imaginary
parts, i.e.,

\begin{equalign}
\label{eq37}
\gamma(\omega_R \pm i\epsilon,k) =
\alpha\left({\omega_R \over k} \right) \pm i
\beta\left({\omega_R \over k} \right),
\end{equalign}
where
\begin{equalign}
\label{eq38}
\alpha(\Omega) = {m \over M_c} + {2 N_c N_f G M_c^2 \over \pi^2}
\left[\int\limits_0^{\Lambda} {dp \, p \over E_p} \left[
{p \over E_p^2} (1 - f_0) + {\partial f_0 \over \partial p} \right]
- {\Omega \over 2} \int\limits_0^{\Lambda} \,dp\,{\partial f_0 \over
\partial p} \ln {|\Omega + p/E_p| \over |\Omega - p/E_p|} \right]
\end{equalign}
and
\begin{equalign}
\label{eq39}
\beta(\Omega) = {N_c N_f G M_c^2 \over \pi} \Omega
\left[ f_0(\Lambda) - 
f_0 \left({M_c \Omega \over \sqrt{1 - \Omega^2}} \right) \right]
\theta\left(v^{\max} - |\Omega|\right).
\end{equalign}

The reader may be surprised by our result, since we have a contribution
from the cut but, on the other hand, we do not find a solution to the
dispersion relation (\ref{eq31}). In the standard case
\cite{LDL65,LP81} the situation is quite opposite: the permittivity is
an entire function of the frequency and one does find solutions to
(\ref{eq31}). The point is that for a non-relativistic Maxwellian
plasma one can analytically continue $\gamma(\omega,k)$ from the upper
half-plane into the lower half-plane. In this way, one obtains a
function $\gamma^{\prime}(\omega,k)$ such that (i) for $\omega_I > 0$
it coincides with $\gamma(\omega,k)$, (ii) for $\omega_I = 0$ it has
no cut, (iii) for $\omega_I < 0$ it is different from
$\gamma(\omega,k)$, (iv) it is an entire function of $\omega$, and
finally (v) there are solutions to the dispersion relation
$\gamma^{\prime}(\omega,k) = 0$. Because these solutions are for
complex energies in the lower half-plane it means that the modes are
damped --- this is the famous Landau damping
\cite{LDL65,Sogan94,LP81} caused by a transfer 
of the energy from the wave to the particles 
of the medium.

In our case, although the mathematics is slightly different we can expect
similar physics. {\it Our contribution from the cut is a superposition of
waves which can exhibit destructive coherence phenomena, i.e., they
can be damped as in the standard case}. The main mathematical difference
between our approach here and the typical non-relativistic one is that
we do not perform the analytic continuation of $\gamma(\omega,k)$
from the upper half-plane into the lower one because, in our case,
this does not appear to be a straightforward procedure.

\section{Results}

As the initial conditions for our equations, we shall consider two extreme
cases which lead to spectacularly quite different behaviour of the 
plasma. The first case corresponds to the perturbation which is at $t=0$
independent of the momentum ${\bf p}$, whereas the second case describes
the situation when the initial perturbation is strongly peaked at some
value of ${\bf p}$. The second case can be regarded as a more interesting
one, because any initial perturbation (including the first one) can
be treated as a superposition of such peaked elementary perturbations.
Moreover, as we shall see later, the solutions are also superpositions
of elementary solutions corresponding to such initial conditions. Let
us, however, start our considerations with the discussion of the first
case.

\subsection{Initial Perturbation Independent of Momentum}

Writing the initial perturbation of the quark distribution function
in the form

\begin{equalign}
\label{eq40}
g({\bf k},{\bf p}) = g_1({\bf k}),
\end{equalign}
and using the above expression in (\ref{eq29}) one can find that
\begin{equalign}
\label{eq41}
{\tt G}(\omega_R \pm i\epsilon,{\bf k}) = g_1({\bf k})
{M_c \over k} \left[\pm A\left({\omega_R \over k} \right)
+ i B\left({\omega_R \over k} \right) \right],
\end{equalign}
where
\begin{equalign}
\label{eq42}
A(\Omega) = {N_c N_f G \over 2 \pi} \left[ {M_c^2 \Omega^2 \over
1 - \Omega^2} - \Lambda^2 \right] \theta(v^{\max} - |\Omega|)
\end{equalign}
and
\begin{equalign}
\label{eq43}
B(\Omega) = -{N_c N_f G \over \pi^2} \int\limits_0^{\Lambda}
dp\, p \, \ln {|\Omega+p/E_p| \over |\Omega-p/E_p|}.
\end{equalign}

Equations (\ref{eq36}), (\ref{eq37}) -- (\ref{eq39}), and (\ref{eq41})
-- (\ref{eq43}) allow us to write the final expression for the 
fluctuating part of the mean field as

\begin{equalign}
\label{eq44}
\delta M(t,{\bf k}) = g_1({\bf k}) \, M_c \, F_1(kt),
\end{equalign}
where
\begin{equalign}
\label{eq45}
F_1(kt) = \int\limits_{-v^{\max}}^{v^{\max}} {d\Omega \over \pi}
{\alpha A + \beta B \over \alpha^2 + \beta^2} \cos(\Omega kt).
\end{equalign}

One can obtain the space-time dependent function $\delta M(t,{\bf r})$
by calculation of the Fourier transform (\ref{eq20}).  The time
evolution of the fluctuations is completely determined by the function
$F_1(kt)$.

In Fig. [12.2] we show the function $F_1(kt)$ calculated numerically
for two different values of the temperature: $T$ = 100 MeV and $T$ =
150 MeV, respectively. The parameters were fitted as in \cite{ZHK94},
see also the discussion following Eqs. (\ref{eq7}) -- (\ref{eq9}). For
small time arguments $F_1(kt)$ is negative as it should be --- if we
add particles to the system, $g_1({\bf k}) > 0$, the mean field
decreases. For $t=0$ the change of the mean field can be independently
calculated either from Eq. (\ref{eq44}) or directly from
Eq. (\ref{eq9}). The condition that two methods give the same result
was used by us as a check of our numerical code.

\begin{figure}[ht]
\label{oscps}
\xslide{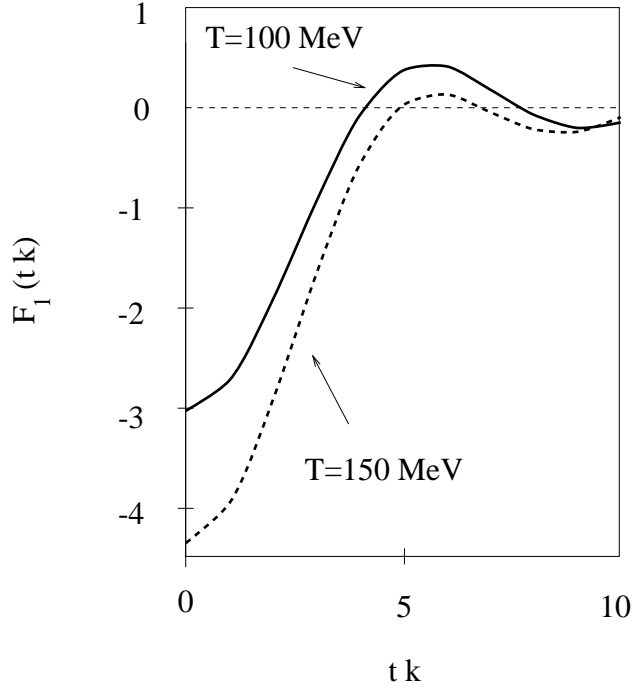}{11cm}{47}{147}{554}{732}
\caption{\small Function $F_1(kt)$ calculated for two values of the
                temperature: $T$ = 100 MeV and $T$ = 150 MeV.
                The values of other parameters are: $m$ = 0,
                $N_c$ = 3, $N_f$ = 2, $G$ = 5.01 GeV${}^{-2}$,
                and $\Lambda$ = 650 MeV.}
\end{figure}

For both values of the temperature we observe that $F_1(kt)$
oscillates in time but with the amplitude strongly decreasing. In our
opinion, we observe here a phenomenon having features in common with
the Landau damping in ordinary plasmas, i.e., the energy of the wave
is transferred to the medium without entropy production, see
Eq. (\ref{eq12}). There exist, however, differences between the
damping observed here and the standard Landau damping. First of all,
one can check that the suppression of the oscillations is much weaker
than exponential. Moreover, the damping seems to be
independent of the temperature. In fact, our calculations indicate
that it is a relativistic effect connected with the existence of the
finite maximal velocity, i.e., the velocity of light.  The latter
determines the size of the integration interval in (\ref{eq45}) --- in
practice we have always $v^{\max} \approx 1$. For the considered values
of the temperatures one can check that the function $(\alpha A + \beta
B)/(\alpha^2 + \beta^2)$ is very regular and, therefore, we expect
that for $kt = 2\pi$ the integral becomes negligible.  This result
is due to the fact that $\cos(2\pi\Omega)$ takes on equally positive
and negative values in the interval between -1 and 1. By inspection of
Fig. [12.2], one can convince oneself that the initial perturbation of
the mean field becomes strongly suppressed for $kt > 2\pi$.

It is perhaps also interesting to observe that our suppression of
oscillations has some features in common with the Friedel oscillations
analyzed in Chapter 8.  The latter are caused by the existence of the
sharp Fermi surface, in our case there exists a sharp maximal velocity
$v^{\max} \approx 1$.

\subsection{Initial Perturbation Strongly Peaked in Momentum}

Let us now turn to the discussion of the opposite case, where
the initial perturbation is strongly peaked in momentum. Employing
(\ref{eq29}) and (\ref{eq36}) and exchanging order of integration
we obtain the formula

\begin{equalign}
\label{eq46}
\delta M(t,{\bf k}) = 4 N_c N_f G \int^{\Lambda} {d^3p \over (2\pi)^3}
{1 \over E_p} \delta M(t,{\bf k};{\bf p}),
\end{equalign}

\noindent where we used the definitions

\begin{equalign}
\label{eq47}
\delta M(t,{\bf k};{\bf p}) = \int\limits_{-k v^{\max}}^{k v^{\max}}
{d\omega \over 2\pi} e^{-i\omega t}
\left[ { {\tt G}(\omega + i\epsilon,{\bf k};{\bf p}) \over 
   \gamma(\omega + i\epsilon,{\bf k})} -
{ {\tt G}(\omega - i\epsilon,{\bf k};{\bf p}) \over 
   \gamma(\omega - i\epsilon,{\bf k})} \right]
\end{equalign}
and
\begin{equalign}
\label{eq48}
{\tt G}(\omega + i\epsilon,{\bf k};{\bf p}) = g({\bf k},{\bf p})
{i M_c \over {\bf k} \cdot {\bf v} - \omega \mp i \epsilon}.
\end{equalign}

\begin{figure}[ht]
\label{vbc100}
\xslide{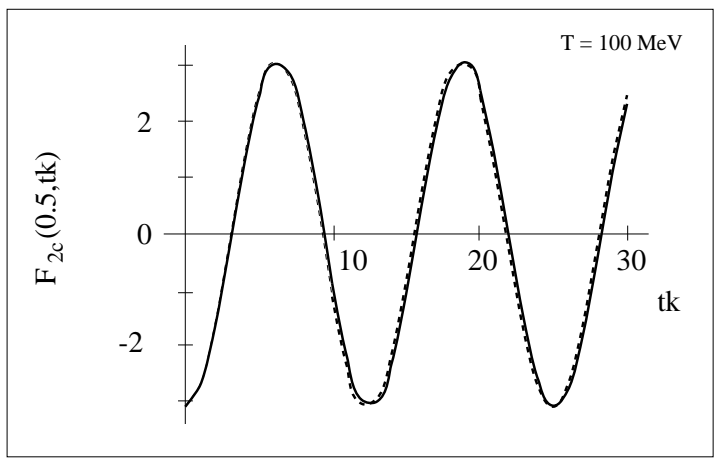}{8cm}{17}{203}{577}{660}
\caption{\small The function $F_{2c}(0.5,kt)$ plotted for $T$ = 100 
                MeV. The dashed line shows the complete result obtained
                from Eq. (\ref{eq50}), whereas the solid line is the
                approximation for $F_{2c}(0.5,kt)$ obtained by keeping
                only the pole contribution (the first term on the RHS
                of (\ref{eq50})). Other parameters as in Fig. [12.2].}
\end{figure}

Equations (\ref{eq46}) -- (\ref{eq48}) have a clear physical meaning,
they say that the fluctuation of the mean field (\ref{eq46}) can be
treated as a superposition of the modes (\ref{eq47}) which correspond
to elementary initial conditions (\ref{eq48}). Such elementary
initial conditions are realized by adding to the system, at $t=0$,
particles with well-defined momentum ${\bf p}$.

Using Eqs. (\ref{eq47}), (\ref{eq37}) -- (\ref{eq39}), and (\ref{eq48}) we
find that

\begin{equalign}
\label{eq49}
\delta M(t,{\bf k};{\bf p}) = g({\bf k},{\bf p}) M_c
\left[F_{2c}\left( {{\bf v} \cdot {\bf k} \over k}, kt \right)
-i F_{2s}\left( {{\bf v} \cdot {\bf k} \over k}, kt \right) \right],
\end{equalign}

\noindent where

\begin{equalign}
\label{eq50}
F_{2c}(x,y) = -\cos(xy) {\alpha(x) \over \alpha^2(x) + \beta^2(x) }
+ {\cal P} \int\limits_{-v^{\max}}^{v^{\max}} {d\Omega \over \pi}
\cos(\Omega y) {\beta(\Omega) \over \alpha^2(\Omega) + \beta^2(\Omega)}
{1 \over x - \Omega},
\end{equalign}
and
\begin{equalign}
\label{eq51}
F_{2s}(x,y) = -\sin(xy) {\alpha(x) \over \alpha^2(x) + \beta^2(x) }
+ {\cal P} \int\limits_{-v^{\max}}^{v^{\max}} {d\Omega \over \pi}
\sin(\Omega y) {\beta(\Omega) \over \alpha^2(\Omega) + \beta^2(\Omega)}
{1 \over x - \Omega}.
\end{equalign}

\begin{figure}[ht]
\label{vbc150}
\xslide{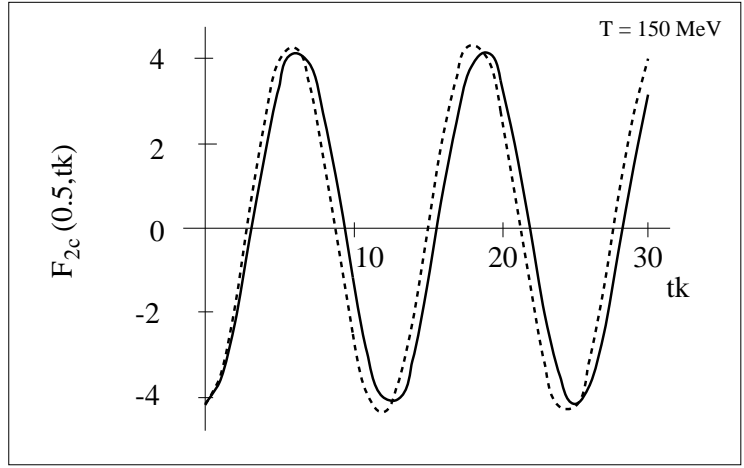}{8cm}{17}{203}{577}{660}
\caption{\small The same as Fig. [12.3] but for $T$ = 150 MeV.}
\end{figure}

In Figs. [12.3] and [12.4] we show the function $F_{2c}({\bf v} \cdot
{\bf k}/k,kt)$ for two different values of the temperature, i.e., for
$T$ = 100 MeV and $T$ = 150 MeV. All the parameters are fitted in the
same way as in the previous cases. We fixed the projection of the
velocity ${\bf v}$ on the wave vector ${\bf k}$ to be ${\bf v} \cdot
{\bf k} /k = 0.5$. The dashed curve represents in both cases the
complete result obtained from (\ref{eq50}), whereas the solid line
shows only the pole contribution (only the first term on the RHS of
Eq. (\ref{eq50})). Similarly, in Figs. [12.5] and [12.6] we show the
function $F_{2s}({\bf v} \cdot {\bf k}/k,kt)$.

One can see in Figs. [12.3] -- [12.6] that the functions
$F_{2c}({\bf v} \cdot {\bf k}/k,kt)$ and
$F_{2s}({\bf v} \cdot {\bf k}/k,kt)$ oscillate in time without any
damping. It means that our elementary fluctuations (\ref{eq49}) will
be undamped as well. This phenomenon is caused by the singularity of 
the initial quark distribution function.
 
\begin{figure}[ht]
\label{vbs100}
\xslide{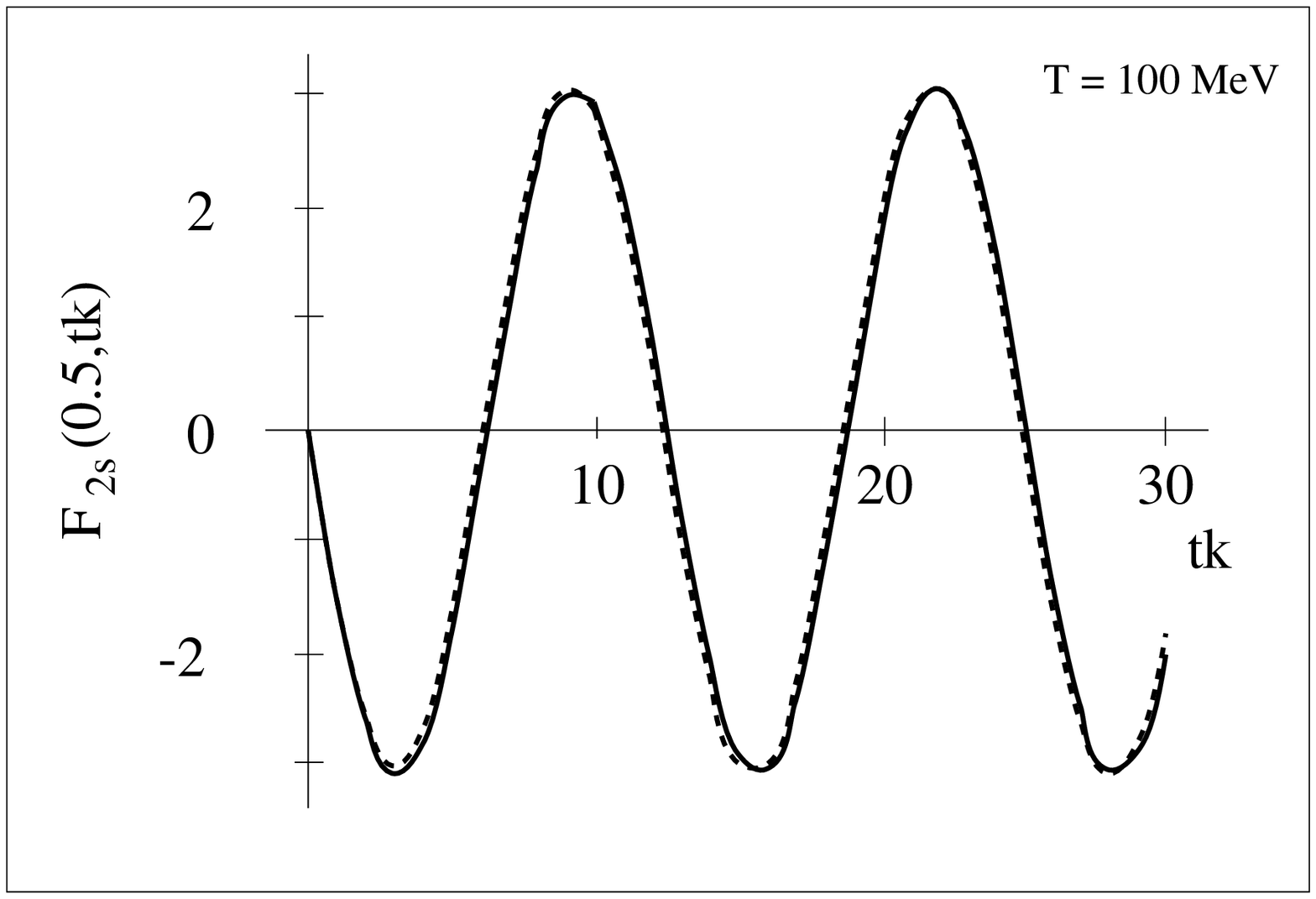}{8cm}{17}{203}{577}{660}
\caption{\small The function $F_{2s}(0.5,kt)$ plotted for $T$ = 100 
                MeV. The dashed line shows the complete result obtained
                from Eq. (\ref{eq51}), whereas the solid line is the
                approximation for $F_{2s}(0.5,kt)$ obtained by keeping
                only the pole contribution (the first term on the RHS
                of (\ref{eq51})).}
\end{figure}

We want to point out that one finds similar waves in the theory of
ordinary plasmas in the form of so-called Van Kampen modes \cite{VK55}. 
Landau's calculation \cite{LDL65} emphasized the case where the
initial perturbations were entire functions of the momentum. The
Van Kampen modes are excited by the perturbation being a $\delta$
function in momentum. Of course, such a perturbation is not an entire
function in the complex $\omega$ space. It gives rise to singularities
on the real axis and leads to undamped oscillations.

\bigskip
In Figs. [12.3] -- [12.6] one can also see that the functions
$F_{2c}({\bf v} \cdot {\bf k}/k,kt)$ and
$F_{2s}({\bf v} \cdot {\bf k}/k,kt)$ can be quite well approximated
just by the first terms on the RHS of Eqs. (\ref{eq50}) and (\ref{eq51}).
Using this fact we can find the following approximate form for the
fluctuating part of the mean field:

\begin{eqnarray}
\label{eq52}
& & \delta M(t,{\bf r}) \approx -4 N_c N_f G \int^{\Lambda} 
{d^3p \over (2\pi)^3} {M_c \over E_p} \int {d^3k \over (2\pi)^3}
{\alpha({\bf v} \cdot {\bf k}/k) \over \alpha^2({\bf v} \cdot {\bf k}/k) 
+ \beta^2({\bf v} \cdot {\bf k}/k)}\times \nonumber \\
& & \left[ {g({\bf k} \cdot {\bf p}) + g(-{\bf k} \cdot {\bf p}) \over 2}
\cos({\bf v} \cdot {\bf k} \, t - {\bf k} \cdot {\bf r}) +
{g({\bf k} \cdot {\bf p}) - g(-{\bf k} \cdot {\bf p}) \over 2i}
\sin({\bf v} \cdot {\bf k} \, t - {\bf k} \cdot {\bf r}) \right].
\nonumber \\
\nonumber \\
\end{eqnarray}

\bigskip
\noindent Equation (\ref{eq52}) shows that an arbitrary fluctuation
can be written as a sum of undamped waves. Nevertheless, such a 
superposition itself is very likely to be damped because of the destructive
coherence phenomena among the elementary excitations; this is the
reason why the fluctuations are damped in our first case.

\newpage
$\mbox{}$

\begin{figure}[t]
\label{vbs150}
\xslide{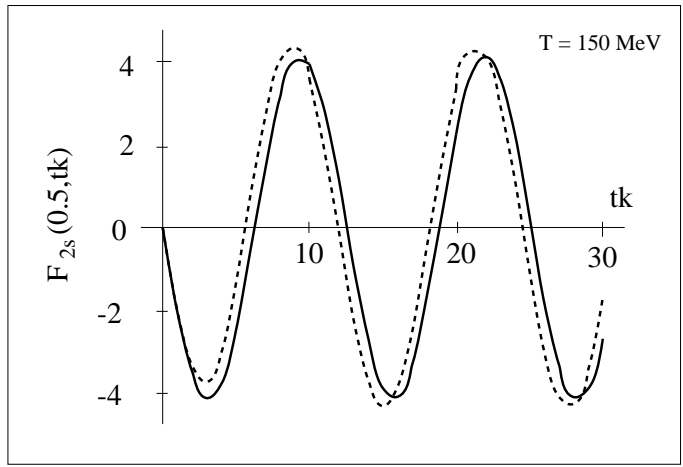}{8cm}{17}{203}{577}{660}
\caption{\small The same as Fig. [12.5] but for $T$ = 150 MeV.}
\end{figure}

\chapter{\bf Critical Scattering at the Chiral Phase Transition}
\label{chapt:critical}

The calculations based on the NJL model indicate the presence of
singularities in the elastic and hadronization cross sections
\cite{CS}. This is a manifestation of criticality at a second order
phase transition which favors the production of low-energy pions. In
this Chapter we study the time evolution of the transition from
partons to mesons and investigate to which degree the low-$p_T$
enhancement observed in ultra-relativistic heavy-ion collisions may
reflect criticality.

\section{Critical Opalescence}
\label{sect:opalescence}

\def\d3p{d^3p}
\def\tp3{(2\pi)^3}

\def\fq{f_q(p,t)}
\def\fm{f_{\pi}(p,t)}
\def\fe{f_{em}(p,t)}
\def\fii{f_i(p,t)}

\def\eq{f_{q, \,th}(p,t)}
\def\emm{f_{\pi, \,th}(p,t)}
\def\ee{f_{em, \,th}(p,t)}

\def\tth{\tau_{th}(p,t)}
\def\thd{\tau_{had}(p,t)}
\def\tdc{\tau_{dec}(p,t)}
\def\tem{\tau_{em}(p)}

\def\vq{{\bf q} \,}
\def\vp{{\bf p} \,}
\def\vr{{\bf r} \,}
\def\vrp{{\bf r}^{\prime}\,}

Critical scattering (called also a critical opalescence) is a rather
general phenomenon observed in the vicinity of a second
order phase transition \cite{STANLEY}: The cross section
$d^3\sigma/d\vq$ for the scattering of light, X-rays or neutrons
on a medium is given by the expression

\begin{equalign}
\label{cr1}
{d^3\sigma  \over d\vq} \,\, \propto  \,\,
\int d^3r \int d^3r^{\prime} e^{i\vq (\vr - \vrp)}
\langle n(\vr) n(\vrp) \rangle_T,
\end{equalign}

\noindent
where $\vq$ is the momentum transfer and $\langle n(\vr) n(\vrp)
\rangle_T$ is the density-density correlation function for
the particles in the medium at the temperature $T$. In the neighborhood
of a phase transition a long-range ordering develops and
the cross section for small values of $\vq$ has the form

\begin{equalign}
\label{cr2}
{d^3\sigma \over d\vq} \,\, \propto \,\, {1 \over k^2(T) + {\vq}^2},
\end{equalign}
where
\begin{equalign}
\label{cl}
k^2(T) \,\, \propto \,\,   \mid 1 - {T \over T_c} \mid^{\gamma}
\end{equalign}

\bigskip \noindent
is the inverse correlation length with $\gamma$ being the
critical exponent. Criticality manifests itself by the fact
that the system becomes opaque to an external observer.

An experiment, where an {\it external} test particle is scattered on
the system in its critical state, cannot be performed for a
quark-gluon plasma which only lives for $10^{-23}$~s.  Nevertheless,
also particles which are created {\it inside} the plasma can
experience critical scattering. One has to find, however, a suitable
observable, which makes the internal critical scattering visible to an
outside observer. In the following we shall argue that critical
scattering leads to a faster production of low momentum pions and thus
to an intermediate non-thermal distribution. The pions which escape to
the detector, before the thermal equilibration of the hadron gas sets
in, can be observed as an enhancement at low momenta of the pion
distribution.

A low-$p_T$ enhancement in pion and kaon spectra is indeed observed in
ultra-relativistic heavy-ion collisions.  At small values of the
transverse mass, $m_T= \sqrt{m^2+p_T^2}$, the observed distributions
of pions and kaons, $dN/dm_T$, considerably exceed the thermal
distribution \cite{QM90,QM91,QM93}. Various effects have been proposed
to explain this phenomenon: e.g., existence of the transverse flow
\cite{TA,LH}, creation of small plasma droplets \cite{VH}, decays of
resonances
\cite{GB}, formation of the pion system out of chemical equilibrium
\cite{KR,GR}, or the medium modification of the pion dispersion
relation \cite{ES}.  To which degree the critical phenomena contribute
to the observed phenomenon, is the subject of this Chapter.

\section{Kinetic Description of the Hadronization of the Quark Plasma}
\label{sect:hadronization}

The phase transition from a quark-gluon plasma to a hadron gas has two
aspects: chiral restoration and deconfinement. The NJL model is a
useful framework to study the chiral aspect. For a vanishing mass of
the current quarks, a phase transition of the second order is
predicted at a critical temperature $T_c$ above which the constituent
quarks have zero mass while below $T_c$ the pions are the Goldstone
particles with $m_{\pi}=0$.
                                                         
In explicit calculations \cite{CS}, using the NJL model for a
description of the quark-meson plasma (no gluons), it has been shown
that singularities occur in the cross sections at small center-of-mass
energies $\sqrt s$ of the colliding particles, when the phase
transition is approached.  In particular, the integrated elastic
quark-antiquark cross-section $\sigma_{q {\overline q} \rightarrow q
{\overline q}}(s,T)$ diverges like $s^{-1}$ for $T \rightarrow T_c$.
A singularity is also found for the hadronization cross-section $q
{\overline q} \rightarrow \pi \pi$. In both cases the singularity
arises because the quark condensate $\langle{\overline q}q\rangle_T$,
which is the order parameter of the chiral phase transition, goes to
zero at the critical temperature.

Although these results have been obtained for a particular model
they may be of more general relevance, provided the parton-hadron
phase transition is second order. We use the results for
the energy dependence of the cross sections in order to model
the time evolution from quark to hadron matter: The cross sections
are incorporated into the kinetic equations proposed for the
description of the transition of a spatially homogeneous 
quark plasma (described by the single particle quark distribution 
function $\fq,\,  p=|\vp|$) into a pion gas
(with a corresponding distribution function $\fm$).

In order to keep the model as transparent as possible we work
in the relaxation time approximation. However, in contrast to the 
usual approaches, the relaxation times $\tau (p,t)$ depend on momenta via 
the energy dependence of the cross sections. 
The singularities in the cross sections due to criticality of the medium 
translate themselves into singularities of the inverse 
relaxation times $1/\tau (p,t)$ in the momentum $p$.

The above ideas are translated into the following set of kinetic equations

\begin{eqnarray}    
\label{q}
{d\fq \over dt} &=& - {\fq - \eq \over \tth}
- {\fq \over \thd} + {\fm \over \tdc},
\\                 
& & \nonumber \\ 
& & \nonumber \\
\label{m}
{d\fm \over dt} &=& - {\fm - \emm \over \tth} + {\fq \over \thd}
- {\fm \over \tdc} - {\fm \over \tem} ,
\\              
& & \nonumber \\
& & \nonumber \\    
\label{e}
{d\fe  \over dt} &=& {\fm \over \tem}.
\end{eqnarray}      

\bigskip
Eq. (\ref{q}) determines the time evolution of the quark distribution
function $\fq$ (it describes both quarks and antiquarks, i.e.,
$\fq = f_{\hbox{quarks}}(p,t) + f_{\hbox{antiquarks}}(p,t)$).
The first term on the RHS of Eq. (\ref{q}) is the collision term
written in the relaxation time approximation; it is responsible for 
the thermalization of quarks since the distribution function
$\fq$ is always attracted to the thermal one $\eq$. 
The second (third) term describes the loss (gain) of quarks
due to the hadronization (deconfinement) process where we limit
ourselves to the reaction $q {\overline q} \rightarrow \pi \pi$
($q {\overline q} \leftarrow \pi \pi$).  

In Eq. (\ref{m}), for the time evolution of the distribution
functions of pions, the first three terms on the RHS describe
the thermalization of pions, appearance of pions due to
hadronization of quarks, and the two-pion reaction into
quark-antiquark pairs, respectively. The last (fourth) term 
accounts for the emission of pions from the plasma into the
detector. For $t \rightarrow \infty$ the distribution function
$\fe$ describes the observed spectrum of pions.

Eqs. (\ref{q}) - (\ref{e}) can be solved for any initial
conditions (i.e., assuming some particular form of the
distribution functions at the initial time $t=0$) provided the
thermal distribution functions $\eq$ and $\emm$ are known at all times.
Since the first terms on the RHS of Eq. (\ref{q}) and (\ref{m})
represent the collision terms (in the relaxation time approximation)
they must obey the symmetry leading to the energy conservation. 
This gives the following constraint for $\eq$ and $\emm$ at each time $t$

\begin{equalign}
\label{c}
\int {\d3p \over \tp3} \, \sqrt{p^2 + m_i^2} \,\,
{f_i(p,t) - f_{i, th}(p,t) \over \tth} &=& 0
\end{equalign}

\bigskip \noindent (here $i = q$ or $\pi$). Eq. (\ref{c}) 
determines the temperature $T_i(t)$ appearing in the thermal
distributions. For simplicity we assume the low-density
high-temperature form of these functions, namely a Boltzmann
distribution

\begin{equalign}
\label{boltz}
f_{i, th}(p,t) =  g_i
\exp\left[\,- {\sqrt{p^2+m_i^2} \over T_i(t) } \,\right],
\end{equalign}

\bigskip \noindent
where $g_i$ are the degeneracy factors:
$g_q = 24$ (quarks and antiquarks having two
different spin projections, 3 colours and 2 flavours) and
$g_{\pi} = 3$ (three different values of the isospin).
One can notice that the distributions (\ref{boltz}) correspond
to the case when the chemical potential $\mu_i$ is zero,
consequently we allow in our approach for number changing
processes like: $q {\overline q} \rightarrow 2( q {\overline q})$
or $\pi \pi \rightarrow 2(\pi \pi)$.

\section{Relaxation Times and Cross Sections}
\label{sect:cross}

The analysis of the exact collision term in the
Boltzmann kinetic equation (analogous to that in \cite{KH})
leads to an expression for the average time
for hadronization of two quarks into two pions

\begin{equalign}
\label{iht}
{1 \over \tau_{had}(p,t)} = {1 \over 2\sqrt{p^2+m_q^2}}
\int {d^3p_1 \over \tp3 \sqrt{p_1^2+m_q^2}}
f_q(p_1,t) F_{qq}(s) \sigma_{q {\overline q} \rightarrow \pi \pi}(s),
\end{equalign}

\noindent where $\sigma_{q {\overline q} \rightarrow \pi \pi}(s)$ 
denotes the total cross section
for this process. The relativistic flux factor of incoming quarks is
$F_{qq}(s) = {1 \over 2} \sqrt{s(s-4m_q^2)}$, with 
$\sqrt s$ being the center-of-mass energy
of the quarks with momenta $p$ and $p_1$, respectively.
The expression giving the deconfinement relaxation time,
$\tdc$, has the form similar to Eq. (\ref{iht}).  In the
analogous way one also defines the relaxation time for
thermalization $\tau_{th}(p,t)$ which is mainly due to
elastic processes $(qq  \rightarrow qq, q \pi \rightarrow
q \pi)$ with the respective cross sections.

In the chiral limit and at the temperature of the phase
transition, only for which the singularities in the
cross sections are observed, the masses of quarks and pions 
have to be set to zero. We shall also assume that all cross 
sections: $\sigma_{q {\overline q} \rightarrow \pi \pi}(s), 
\sigma_{qq \rightarrow qq}(s),
\sigma_{\pi\pi \rightarrow \pi\pi}(s)$ and 
$\sigma_{q\pi \rightarrow q\pi}(s)$ 
are given by the generic expression

\begin{equalign}
\label{cs}
\sigma(s) = \sigma_0 \left[1 + {s_0 \over s} \right],
\end{equalign}

\bigskip \noindent
where $\sigma_0$ represents a constant contribution to the
cross sections, and  the appearance of the singular term
$s_0/s$ accounts for the critical phenomena. 
The hadronization and deconfinement cross sections,
$\sigma_{q {\overline q} \rightarrow \pi\pi}(s)$ and
$\sigma_{\pi \pi \rightarrow q {\overline q}}(s)$, are
related to each other by the principle of detailed balance

\begin{equalign}
\label{db}
g_q^2 \sigma_{q {\overline q} \rightarrow \pi \pi}(s) =
g_{\pi}^2 \sigma_{\pi\pi \rightarrow q {\overline q}}(s).
\end{equalign}

\noindent This relation guarantees that in the absence
of emission ($\tau_{em} \rightarrow \infty$), the kinetic
equations lead to the chemical equilibrium $n_q/n_{\pi}
= g_q/g_{\pi}$.

The possibility of the emission of preequilibrium pions, 
i.e., the emission of pions at each stage of the 
hadronization process is crucial for our description.
In the model defined by the kinetic equations
(\ref{q}) - (\ref{e}) this emission is only treated in a 
very global and rather crude way. The emission
time $\tau_{em}$ is a free parameter of the theory, which
allows us to decouple a certain fraction of pions from
the interacting, non-equilibrium system. We have no good
model for the momentum dependence of $\tau_{em}$. Therefore
we test two assumptions:

\begin{equalign}
\label{ema}
{1 \over \tau^{(a)}_{em}} = 
{{\cal R} \over \tau_{th}(p = T_c, t = 0)}
\end{equalign}
and
\begin{equalign}
\label{emb}
{1 \over \tau^{(b)}_{em}} = {p \over T_c}  \,
{{\cal R} \over \tau_{th}(p = T_c, t = 0)}.
\end{equalign}

\bigskip
\noindent 
In the first case the emission time is independent of momentum, 
whereas in the second case the inverse emission time depends 
linearly on $p$ (the emission of low energetic pions is suppressed).
The magnitude of $\cal R$ indicates how much faster 
the emission is, in comparison to the rate of the 
thermalization processes inside the system.

\section{Results of Numerical Calculations}

In order to reduce the number of free parameters in our calculation we
have set all the cross sections: $\sigma_{q {\overline q} \rightarrow
\pi\pi}(s), \sigma_{qq \rightarrow qq}(s), \sigma_{q \pi \rightarrow q
\pi}(s)$ and $\sigma_{\pi\pi \rightarrow \pi\pi}(s)$ to be equal
(within a factor of 3 such a result is supported by calculations
within the NJL model) and use the form (\ref{cs}) which is appropriate
for $T = T_c$. Then, we are left with three parameters: $\sigma_0,
s_0$ and $\cal R$.  One can notice that $\sigma_0$ sets the overall
time-scale and is unimportant if we are interested in the final $(t
\rightarrow \infty$) pion distributions.

We start solving our kinetic equations by assuming that the initial
quark distribution is a thermal one and that there are no pions in the
system. After integrating Eqs. (\ref{q}) -- (\ref{e}) till the time
when all quarks are hadronized and all pions are emitted, one obtains
the distribution function $f_{em}(p,t \rightarrow
\infty)$ of the observed pions for a set of parameters 
$s_0$ and $\cal R$.

In Fig. [13.1] we show our results for the initial temperature of
quarks $T_q(t=0)= T_c$ = 140 MeV, and for $s_0 = $ 0.5 GeV${}^2$ and
${\cal R}$ = 10.  The two dashed lines represent the distribution
functions of the emitted (observed) pions. The upper one (a)
corresponds to the emission independent of momentum (according to
Eq. (\ref{ema})), and the lower one (b) corresponds to $p$-dependent
emission (according to Eq. (\ref{emb})).  One can notice that the
resulting pion distributions are non-equilibrium ones. We can also see
that they coincide for large values of momenta. The solid line
represents the thermal distribution fitted to these two curves in the
high momentum region. One can see that for small momenta, 25 MeV $< p
<$ 200 MeV, both pion distributions exceed the thermal one. In the
case (a) the pion distribution is very much peaked in the limit $p
\rightarrow 0$. This behaviour reflects directly the criticality of
the hadronization cross section.  On the other hand, in the case (b)
the pion distribution function drops to zero for $p \rightarrow 0$
since the effect of critical hadronization is, in this situation,
partially destroyed by the slow emission of low energetic pions.

In consequence, our model calculation indicates that the criticality
in the hadronization cross sections leads to the excess in the
production of low energetic pions.  The effect becomes particularly
strong if the produced pions, independently of their momenta, can
manage to decouple fast from the interacting system. The question
whether the discussed scenario is really the realistic one and
responsible for the observed low-$p_T$ enhancement can only be checked
by more detailed calculations.  They should include, in particular,
the dynamics of the space-time evolution of the system and of the
freeze-out process. As we showed in our simple approach
it is, in general, very important to take into account the possibility
that the cross sections exhibit non-trivial energy dependence. This
can lead to interesting phenomena connected with the existence of the
phase transition and even serve as a diagnose of its occurrence.

\begin{figure}[hb]
\label{distr}
\xslide{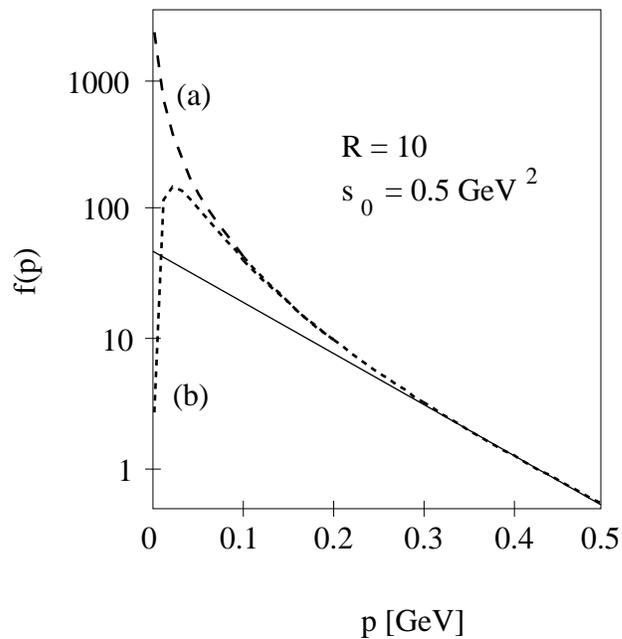}{10cm}{68}{155}{551}{693}
\caption{\small The distribution functions of the emitted pions (dashed
                lines) together with the thermal distribution (solid
                line) fitted to them in the high momentum region.
                The cases (a) and (b) correspond to the emission times
                calculated according to Eqs. (\ref{ema}) and (\ref{emb}),
                respectively.}
\end{figure}

\part{\bf SUMMARY}

\newpage
\begin{itemize}

\item[{\bf 1}] In general the vacuum correlation functions in the meson 
channels are screened and, due to Lorentz invariance, the screening
mass of a stable meson equals its dynamic mass. For unstable mesons
the screening mass equals the mass of the branch point, since the
asymptotic form of the correlation function picks out the lowest lying
singularity. As demonstrated by our work, these results are reproduced
by the NJL model: the pion screening mass is equal to its dynamic
mass, whereas the sigma screening mass is $2M_0$, where $M_0$ is the
constituent quark mass in vacuum.

At finite temperature and zero density, the correlation functions are
again exponentially damped. Moreover the screening and dynamic masses
differ --- this fact indicates that at finite temperature it is
impossible to obtain the exact information about time-like excitations
by studying space-like correlation functions.  The overall temperature
dependence of the screening masses in the NJL model agrees with the
corresponding lattice results.

We carefully explored the analytic structure of the correlation
functions. In the complex momentum plane they have two cuts (scalar
channel) or two cuts and a pole (pseudoscalar channel). At sufficiently
high temperature  the pion pole disappears and we have to deal, in
both channels, with two cuts. The contributions from these cuts show
large cancellations which lead to the exponential decay of the
correlation functions.

\item[{\bf 2}] We have demonstrated that the correlation functions at 
$T = 0$ and finite density differ qualitatively from those in vacuum
and those at $T>0$: they exhibit long ranged oscillations, of the
Friedel type, rather than exponential damping.  We found that the
appearance of the oscillations is connected with the existence of a
cut of finite range in the complex momentum plane. This cut is
responsible for the leading contribution to the correlation function
at large distances.  The length of the cut, which is proportional to
the Fermi momentum of the constituent quarks, is reflected in the
oscillation period at large distances $\delta r = \pi / p_F$.
Consequently this form of the correlation function is quite general
and is expected in all normal Fermi liquids. In particular, the
existence of the oscillations is independent of whether the basic
fermionic degrees of freedom are quarks or nucleons.  Therefore, we
feel the oscillatory behaviour of the correlation function at finite
density will not change qualitatively by confinement.

The fact that the correlation function exhibits Friedel-type
oscillations indicates again that in medium it is impossible to obtain
the information about time-like excitations by studying space-like
correlations. Their long distance behaviour is dominated by low lying
particle-hole excitations, and consequently not connected with the
dynamic mass.

\item[{\bf 3}] Our NJL studies concerning static meson correlation
functions were supplemented by the calculation within the perturbative
QCD. Keeping only the lowest order term we obtained compact analytic 
results. This work extends some of the earlier results, which were
concentrated only on the asymptotics of the correlation functions.

\item[{\bf 4}] Temperature dependence of the quark condensate was studied
in the NJL model with meson loops. Substantial differences were found
compared to the standard results obtained in the Hartree-Fock
approximation. In particular, we found finite slope of the condensate
vs. $T^2$ for $T \rightarrow 0$ in the chiral limit, faster melting of
the condensate, and a lower chiral restoration temperature.  The
importance of pions is connected with the smallness of their mass,
much smaller than the masses of other hadrons. In the particular case
of the NJL model, the pion mass is much smaller than the mass of the
non-confined constituent quarks.  The behaviour of $\langle
\overline{q} q \rangle$ is affected by the coupling of the quarks with
pions. This happens in such a way that the presence of pions leads to
the destruction of the condensate. Therefore, including meson loops we
speed up the decrease of the condensate with increasing temperature.
We demonstrated consistency of our results with the chiral
perturbation theory. Our calculation represents the first
self-consistent treatment of the NJL model with meson loops at finite
temperature.

\item[{\bf 5}] The chiral symmetry concepts have been
explicitly included in the construction of the transport equations for
quark matter. Our starting point was the chirally invariant Lagrangian
of the NJL model and we derived the transport equations via a spinor
decomposition of the Wigner function and a gradient expansion. Our
calculation was restricted to the mean-field approximation.  We have
taken into account the spin dynamics and discussed the possibility of
having a non-zero pseudoscalar condensate. In this aspect, our results
are a generalization of some earlier results.

The classical quark distribution functions satisfy the kinetic
equations of the standard form with the effective chirally invariant
mass $M^2(X) = \pi^2_{(0)}(X) + \sigma^2_{(0)}(X)$, where
$\pi_{(0)}(X)$ and $\sigma_{(0)}(X)$ are leading terms in the
classical expansion of the pseudoscalar and scalar condensates.
Furthermore, the angle $\Phi(X)$, defined through the relation
$\pi_{(0)}(X)/\sigma_{(0)}(X) = \hbox{tan}\Phi(X)$, must be a
constant. However, its value remains undefined, which reflects the
chiral symmetry of the problem.

The classical equation for the spin evolution, which has been derived
for the first time by us, is also invariant under chiral
transformations. However, its solutions are constrained by the
additional condition of the axial current conservation, which is not
simple to incorporate.

The inclusion of the quark mass term into the Lagrangian explicitly
breaks the chiral invariance. Consequently, we find that $\Phi(X)$ as
well as $\pi_{(0)}(X)$ must be zero.  Nevertheless, the general form
of the kinetic equations does not change and we can still use them
with the substitution $M(X)=\sigma_{(0)}(X)+m$, where $m$ is the
current quark mass. Moreover, in this case the requirement of the
axial current conservation reduces to a form familiar from PCAC.

\item[{\bf 6}] The chirally invariant kinetic theory for quark matter
was used to study large time-scale fluctuations of the quark
condensate. Our results showed that the general features of the
propagation of the excitations in the system were very similar to
those characterizing the wave propagation in a non-relativistic
Maxwellian plasma. The elementary fluctuations of the condensate
(Van Kampen modes) are not damped; they correspond to initial 
perturbations of the quark distribution function, which are strongly 
peaked for some given value of momentum. On the other hand, the 
superpositions of such elementary fluctuations are in practice always 
damped due to a phenomenon of destructive coherence (Landau damping).
The overall results of this study showed the stability of the in-medium
gap equation with respect to small perturbations of the quark
distribution functions.

\item[{\bf 7}] The system of kinetic equations describing hadronization
of the quark-antiquark plasma was analyzed. It was found the the
singularities in the elastic and hadronization cross sections, as
described by the NJL model in connection with the critical behaviour
at a second order phase transition, favor production of low-energetic
mesons. We found that this effect can give a contribution to the
observed low-$p_T$ enhancement of pions observed in the
ultra-relativistic heavy-ion collisions.

\end{itemize}

\newpage

\renewcommand\bibname{Bibliography \\
\bigskip
{\small \hspace{0.75cm} {\underline {\it General texts on QCD}}}
\vspace{-0.5cm} }

\end{document}